\documentclass[11pt,a4paper]{article}

\usepackage[utf8]{inputenc}
\usepackage[T1]{fontenc}
\usepackage{amsmath,amssymb,amsfonts,amsthm,mathtools}
\usepackage[margin=1in]{geometry}
\usepackage{booktabs}
\usepackage{caption}
\usepackage{array}
\usepackage{graphicx}
\graphicspath{{./}}
\usepackage{cite}
\usepackage{enumitem}
\usepackage{xcolor}
\usepackage[hidelinks]{hyperref}

\theoremstyle{plain}
\newtheorem{theorem}{Theorem}[section]
\newtheorem{lemma}[theorem]{Lemma}
\newtheorem{proposition}[theorem]{Proposition}

\newtheorem*{restate_thm}{Theorem}
\newtheorem*{restate_prop}{Proposition}

\theoremstyle{definition}
\newtheorem{definition}[theorem]{Definition}
\newtheorem{assumption}[theorem]{Assumption}

\theoremstyle{remark}
\newtheorem{remark}[theorem]{Remark}

\newcommand{\indep}{\perp\!\!\!\perp}
\newcommand{\nindep}{\not\!\perp\!\!\!\perp}
\newcommand{\E}{\mathbb{E}}
\newcommand{\Var}{\mathrm{Var}}
\newcommand{\Cov}{\mathrm{Cov}}

\newcommand{\tr}{\operatorname{tr}}
\DeclareMathOperator{\supp}{supp}

\newcommand{\SNR}{\mathrm{SNR}}
\newcommand{\dCor}{\mathrm{dCor}}
\newcommand{\dCov}{\mathrm{dCov}}
\newcommand{\HSIC}{\mathrm{HSIC}}
\newcommand{\R}{\mathbb{R}}

\newcommand{\diag}{\mathrm{diag}}

\title{\textbf{Marginal Independence and Sensitivity-Adjusted Residual-Covariance Testing under Additive Measurement Error}}

\author{
  \textbf{Mohammadjavad Sajadi}\thanks{Department of Statistics, Faculty of Mathematics, Statistics and Computer Science, University of Tehran, Tehran, Iran.}
}
\date{}

\begin{document}

\maketitle

\begin{abstract}
Under independent additive measurement error, latent marginal independence implies observable marginal independence, and a dense nonzero set for each error characteristic function makes this relation reversible. Permutation tests then achieve exact finite-sample Type-I error control for marginal independence without estimating error variances or reconstructing latent densities. For Gaussian kernels under independent Gaussian error blocks, noise convolution yields an exact population Hilbert--Schmidt Independence Criterion ($\HSIC$) identity—combining bandwidth dilation and amplitude contraction—that carries over in expectation to the unbiased U-statistic, with sample splitting separating kernel selection from exact permutation calibration. In contrast, for conditional queries, conditioning on noisy surrogates induces measurement-error leakage that violates conditional independence at the observable level, so latent conditional independence is fundamentally not point-identified from single-sample data. Conditional on an externally certified directional Loewner signal-to-noise ratio (SNR) bound and class-specific regression assumptions, we establish sharp sensitivity envelopes bounding observable residual covariance under the latent null. We formulate the operational problem as testing separation from this envelope, $H_0^{\mathrm{sep}}: |C| \le \Delta_{\mathrm{leak}}(s_0)$, which conservatively embeds the latent null $H_0 \subseteq H_0^{\mathrm{sep}}$. Comparing a cross-fitted, studentized lower confidence bound on residual covariance against a valid upper confidence bound on the sensitivity envelope yields pointwise asymptotic level control under $H_0^{\mathrm{sep}}$ (and hence conservative size control for latent conditional independence) and consistency against separated alternatives. Because directional Loewner bounds naturally degrade as conditioning cardinality grows, power is concentrated at low conditioning depths. We embed these tests into a two-ended frozen-neighborhood PC algorithm (NA-PC) and establish deterministic and uniform consistency under residual-covariance faithfulness. Simulations under oracle and misspecified SNR certificates illustrate finite-sample size--power trade-offs and the empirical cost of certification, and an uncertified flow-cytometry analysis is presented as an exploratory sensitivity sweep.
\end{abstract}

\vspace{0.5em}

\vspace{0.3em}

% =========================================================================
\section{Introduction}
\label{sec:intro}
% =========================================================================

Testing statistical independence and learning causal structure from observational data are central problems across genomics, epidemiology, neuroscience, and the physical sciences \cite{spirtes2000causation, pearl2009causality, peters2017elements}. Let $\mathcal{G}^* = ([p], E^*)$ be a Directed Acyclic Graph (DAG) on vertex set $[p]$, and let $\mathbf{V}^* = (V_1^*, \dots, V_p^*)^\top \in \R^p$ be a latent random vector generated by structural equations $V_j^* = f_j((V_k^*)_{k \in \mathrm{pa}_{\mathcal{G}^*}(j)}, U_j)$ with mutually independent exogenous variables $U_1, \dots, U_p$. Under the Markov and faithfulness conditions, an observational conditional independence (CI) oracle identifies the Markov equivalence class of $\mathcal{G}^*$—uniquely represented by its Completed Partially Directed Acyclic Graph, $\operatorname{CPDAG}(\mathcal{G}^*)$—rather than in general a unique DAG \cite{spirtes2000causation, colombo2014order}.

In empirical studies, instruments rarely observe $\mathbf{V}^*$ without noise. Instead, we observe $n$ independent, identically distributed copies of $\mathbf{V} = \mathbf{V}^* + \boldsymbol{\epsilon}$, where only $\mathbf{V}$ is recorded:
\begin{equation}
    \label{eq:intro_surrogate_model}
    \mathbf{V} = \mathbf{V}^* + \boldsymbol{\epsilon}, \quad \text{with subvectors } X = X^* + \epsilon_X, \quad Y = Y^* + \epsilon_Y, \quad \mathbf{Z} = \mathbf{Z}^* + \boldsymbol{\epsilon}_Z.
\end{equation}
Unless stated otherwise, $\mathbf{V}^*$ and the error components $\epsilon_1, \dots, \epsilon_p$ are mutually independent with zero mean and finite variances, and all auxiliary randomization (permutations and sample splits) is independent of the sample. Covariance-only linear derivations may relax mutual independence to $\Cov(\mathbf{V}^*, \boldsymbol{\epsilon}) = \mathbf{0}$ and diagonal error covariance only where explicitly stated.

Ignoring measurement error distorts both marginal and conditional inference \cite{fuller1987measurement, carroll2006measurement, pearl2009causality, scheines2016measurement}. For marginal testing, independent additive error attenuates population distance covariance \cite{fan2022independence, zhang2026distance} and linear correlation, reducing statistical power under alternatives without violating the marginal independence null. In contrast, noisy conditioning can invalidate latent conditional independence at the surrogate level: conditioning on $\mathbf{Z} = \mathbf{Z}^* + \boldsymbol{\epsilon}_Z$ fails in general to block confounding paths through $\mathbf{Z}^*$, inducing \textit{residual confounder leakage} \cite{pearl2009causality, scheines2016measurement, blom2018upper}. Moreover, from a single noisy sample, latent conditional independence is not generally point-identified under a scalar SNR certificate alone. For example, let $(X, Y, Z)$ be centered jointly Gaussian with unit variances, $\Cov(X, Z) = \Cov(Y, Z) = 0.3$, and $\Cov(X, Y) = 0.18$, and take $X = X^*$ and $Y = Y^*$. Independent Gaussian measurement error on $Z$ with variance $0.5$ or $0.25$ yields the exact same observed distribution, while the latent conditional covariance $\Cov(X^*, Y^* \mid Z^*)$ equals $0$ under error variance $0.5$ and $0.06$ under error variance $0.25$; both models have positive-definite latent covariance and satisfy the scalar signal-to-noise ratio (SNR) certificate $\SNR_{Z,0} \triangleq \Var(Z^*)/\Var(\epsilon_Z) \ge 1$. Thus a single-sample SNR lower bound does not identify latent conditional independence, motivating a sensitivity-analysis approach whose detectable alternatives are characterized by separation from a certified leakage envelope rather than by a universal testing lower bound.

Existing methods address this challenge under distinct trade-offs. Linear errors-in-variables and anchor-variable methods \cite{blom2018upper, saeed2020anchored, dai2022independence, zhang2025conditional} restrict structural equations to be linear (with additional anchor or non-Gaussianity assumptions). Nonparametric deconvolution methods \cite{fan1991optimal, meister2009deconvolution, fan2024independence} reconstruct latent densities or conditional densities before testing, confronting slow logarithmic minimax rates for density estimation under supersmooth errors. Building on the marginal invariance principle of Fan, Zhang, and Zhu \cite{fan2022independence}, we develop a unified Two-Regime Framework separating marginal independence testing (Regime~A, $S = \emptyset$) from sensitivity-adjusted conditional residual-covariance testing (Regime~B, $|S| \ge 1$).

Our contributions are as follows. First, in Regime~A, we pair the continuous vector marginal invariance principle of \cite{fan2022independence} (stating the exact dense-nonzero characteristic-function condition for the converse) with orbit-conditioned finite-sample permutation calibration \cite{phipson2010permutation, hemerik2018exact}, requiring no SNR certificate or noise-variance estimation, and we contrast permutation calibration against an intentionally uncentered Gaussian negative control. Second, for Gaussian RBF kernels under independent Gaussian error blocks (including singular within-block error covariances), we derive an exact noise-averaged kernel convolution identity incorporating both bandwidth dilation and amplitude contraction, prove the resulting population $\HSIC$ and unbiased U-statistic expectation identities, establish a fourth-order Taylor expansion for $C^4$ stationary kernels, and evaluate a sample-splitting alignment rule for kernel selection with exact conditional permutation size control. Third, in Regime~B, we derive a family of sufficient upper bounds on the observable residual covariance $C \triangleq \E[R_X R_Y]$ (where $R_X \triangleq X - \E[X \mid \mathbf{Z}]$ and $R_Y \triangleq Y - \E[Y \mid \mathbf{Z}]$ denote population residuals after regressing out the noisy conditioner $\mathbf{Z}$) under latent conditional independence, conditional on an externally certified directional Loewner SNR bound and class-specific latent regression assumptions (general Lipschitz, affine-invariant linear, and Gaussian bounded-curvature classes). Combining cross-fitted asymptotic linearity and variance consistency with a valid upper confidence bound for the sensitivity envelope yields a fixed-level test with pointwise asymptotic size control and consistency against alternatives whose residual-covariance magnitude exceeds the envelope by a fixed margin. Fourth, we integrate these testing principles into a Noise-Aware PC algorithm (\mbox{NA-PC}) for constraint-based causal structure learning under measurement error. We establish high-dimensional skeleton consistency under query-uniform concentration conditions. In extensive empirical evaluations across synthetic networks and real cytometry data, we demonstrate that sensitivity-adjusted conditioning effectively eliminates the dense proliferation of false-positive edges that plagues classical causal discovery algorithms when conditioners are subject to measurement noise.

The remainder of this article is organized as follows. Section~\ref{sec:related_work} reviews related literature and clarifies estimation versus testing rates. Section~\ref{sec:models} formalizes the data model, external SNR certification, Regime~A marginal testing, Regime~B sensitivity-adjusted residual-covariance testing, and kernel convolution with split MKL. Section~\ref{sec:pc_discovery} presents the two-ended frozen-neighborhood NA-PC search, deterministic CPDAG recovery, uniform high-dimensional conditions, stability, and multiplicity remarks. Section~\ref{sec:experiments} reports synthetic benchmarks and the cytometry sensitivity sweep, and Section~\ref{sec:conclusion} concludes with scope and limitations. All proofs of the main theoretical results appear in Appendix~\ref{app:main_proofs}, while supplementary lemmas, discrete category-level misclassification analysis, and additional experiments appear in Appendices~\ref{app:proofs_paper}--\ref{app:tuebingen}.

% =========================================================================
\section{Related Work and Methodological Positioning}
\label{sec:related_work}
% =========================================================================

\paragraph{Linear and Parametric Causal Error Models.} In linear structural equation models, measurement error induces slope attenuation and distorts partial correlations \cite{fuller1987measurement, carroll2006measurement}. Blom et al.~\cite{blom2018upper} derived observable covariance bounds to constrain error variances in linear causal discovery. Saeed et al.~\cite{saeed2020anchored} leveraged anchor variables to identify linear Gaussian models without deconvolution. In linear non-Gaussian models, Dai et al.~\cite{dai2022independence} introduced the Transformed Independent Noise condition, and Zhang et al.~\cite{zhang2025conditional} exploited higher-order cumulant constraints for conditional independence testing under Gaussian measurement noise and linear non-Gaussian structural equations. These methods rely on linearity of the structural model.

\paragraph{Deconvolution Density-Estimation Rates versus Residual-Moment Testing.} Classical nonparametric errors-in-variables methods reconstruct the latent density $f_{X^*}$ via Fourier deconvolution \cite{fan1991optimal, meister2009deconvolution}. For representative known supersmooth error models whose characteristic functions decay exponentially as $|\varphi_{\epsilon_X}(\omega)| \asymp \exp(-b\|\omega\|_2^\gamma)$ ($\gamma > 0$, e.g., $\gamma = 2$ for Gaussian noise) and corresponding Sobolev density classes $\mathcal{F}_\beta(L)$ of smoothness order $\beta > 0$, the minimax squared $L_2$ density-estimation risk has logarithmic order \cite{fan1991optimal, meister2009deconvolution}:
\begin{equation}
    \label{eq:minimax_deconv_rate}
    \inf_{\hat{f}_n} \sup_{f_{X^*} \in \mathcal{F}_\beta(L)} \mathbb{E}\left[\|\hat{f}_n - f_{X^*}\|_{L_2}^2\right] \asymp (\log n)^{-2\beta/\gamma}.
\end{equation}
In that density-estimation setting, achieving root-$L_2$ risk $\varepsilon$ requires $\log n$ of order $\varepsilon^{-\gamma/\beta}$. These are density-estimation minimax results, not testing lower bounds for the problem studied here. Fan, Zhang, and Zhu~\cite{fan2024independence} (Guoliang Fan, Xilin Zhang, and Liping Zhu) use deconvolution to estimate conditional densities entering a mutual-information statistic. By contrast, our conditional procedure does not estimate latent conditional densities; instead, it estimates an observable residual moment $C = \E[R_X R_Y]$ at the $\mathcal{O}(n^{-1/2})$ rate and tests its separation from an externally calibrated sensitivity envelope.

\paragraph{Marginal Invariance, Adaptive Kernels, and Constraint-Based Search.} For pairwise vector-valued marginal independence, Jinlin Fan, Yaowu Zhang, and Liping Zhu~\cite{fan2022independence} established the characteristic-function invariance law linking latent and surrogate independence, and Xilin Zhang, Guoliang Fan, and Liping Zhu~\cite{zhang2026distance} analyzed distance-correlation attenuation under measurement error. For kernel independence testing, Gretton et al.~\cite{gretton2005measuring, gretton2008hsictest} and Song et al.~\cite{song2012feature} developed the biased and unbiased $\HSIC$ estimators and their spectral null limits, while Phipson and Smyth~\cite{phipson2010permutation} and Hemerik and Goeman~\cite{hemerik2018exact} formalized exact random-permutation testing. Finally, constraint-based procedures such as PC and PC-stable \cite{spirtes2000causation, kalisch2007estimating, colombo2014order} prescribe a graph search rather than a single statistical test: their finite-sample calibration and convergence rates depend entirely on the independence tests, regularity classes, and significance schedules employed. This motivates a fundamental methodological bifurcation: evaluating marginal queries ($S = \emptyset$) through exact permutation testing, while evaluating conditional queries ($|S| \ge 1$) through sensitivity-bounded residual covariance.

\subsection{What Is and Is Not Identified Under Measurement Error}
\label{sec:identification_scope}

To avoid inferential ambiguity, we explicitly delineate what is and is not identified from single-sample observational data $(\mathbf{V}_r)_{r=1}^n$ under classical additive measurement error $\mathbf{V} = \mathbf{V}^* + \boldsymbol{\epsilon}$:

\paragraph{1. Marginal Independence Is Fully Identified.} Under classical additive measurement error where noise components are mutually independent of the latent variables and of each other, the latent marginal independence null $H_0^{\mathrm{marg}}: X^* \indep Y^*$ is fully point-identified from the observable surrogate law $P_{X,Y}$. Whenever the error characteristic functions do not vanish on dense sets, $X^* \indep Y^* \iff X \indep Y$. No knowledge of noise variances, signal-to-noise ratios, or latent distributions is required; exact finite-sample permutation tests achieve valid level control directly.

\paragraph{2. Latent Conditional Independence Is Not Point-Identified.} In sharp contrast, latent conditional independence $H_0^{\mathrm{cond}}: X^* \indep Y^* \mid \mathbf{Z}^*$ is \emph{fundamentally not point-identified} from a single noisy sample $(X, Y, \mathbf{Z})$, even within linear-Gaussian structural equation models and even when coordinate noise variances or scalar SNR bounds are known. The observable law admits multiple latent parameter configurations compatible with the same observed covariance and the same SNR certificate, some of which satisfy $X^* \indep Y^* \mid \mathbf{Z}^*$ while others exhibit substantial latent conditional covariance.

\paragraph{3. Non-Identifiability of the Power Function.} A direct mathematical consequence of partial identification is that the statistical power function of any conditional test against latent dependence is itself not identified from the observable distribution plus a scalar or Loewner SNR lower bound $s_0$. Two observationally indistinguishable data-generating processes can share the exact same valid certificate $s_0$, yet one satisfies the latent null while the other exhibits separated latent dependence. Thus, sensitivity-adjusted testing cannot guarantee uniform power against latent conditional dependence.

\paragraph{4. The Operational Target: Sensitivity-Bounded Separation Testing.} Consequently, Regime~B does not claim to provide an omnibus test of latent conditional independence. Instead, it tests the operational hypothesis of \emph{residual-moment separation from an externally certified sensitivity envelope}:
\begin{equation}
    \label{eq:operational_sep_hypotheses}
    H_0^{\mathrm{sep}}: |C| \le \Delta_{\mathrm{leak}}(s_0) \quad \text{versus} \quad H_1^{\mathrm{sep}}: |C| > \Delta_{\mathrm{leak}}(s_0).
\end{equation}
Because latent conditional independence implies that unmeasured confounding leakage cannot exceed the certified sensitivity envelope $\Delta_{\mathrm{leak}}(s_0)$, the latent conditional independence null is conservatively embedded within the separation hypothesis: $H_0^{\mathrm{cond}} \subseteq H_0^{\mathrm{sep}}$. Controlling Type-I error on $H_0^{\mathrm{sep}}$ guarantees valid, conservative level control for $H_0^{\mathrm{cond}}$, while power is guaranteed against alternatives whose observable residual covariance exceeds the envelope by a positive margin.

\paragraph{Pedagogical Summary for Applied Statisticians.}
In simple terms, the fundamental lesson of measurement error in causal inference is:
\begin{itemize}[leftmargin=1.5em]
    \item \emph{Marginal independence is robust}: If two variables are truly unrelated ($X^* \indep Y^*$), adding independent measurement noise will never make them appear related ($X \indep Y$). You can test this using classical permutation tests without knowing anything about the noise level.
    \item \emph{Conditional independence is fragile}: If $Z^*$ is a common cause of $X^*$ and $Y^*$, conditioning on a noisy version $Z = Z^* + \epsilon_Z$ leaves part of the common cause unadjusted. This residual confounding acts like an open backdoor path, creating an artificial correlation between $X$ and $Y$ even when $X^* \indep Y^* \mid Z^*$. Without an external bound on how noisy $Z$ is, you cannot tell whether an observed correlation is a true causal link or just noise leakage. Regime~B solves this by asking: \emph{``How strong could the noise leakage possibly be under an assumed noise floor $s_0$?''} If the observed association exceeds that worst-case leakage, a true causal link must exist.
\end{itemize}

\section{The Two-Regime Framework for Noise-Aware Independence Testing}
\label{sec:models}

In constraint-based structure learning, evaluating queries $(i, j, S)$ over the latent vector $\mathbf{V}^*$ bifurcates fundamentally by whether the conditioning index set $S$ is empty:
\begin{equation}
    (X^* \indep Y^* \mid \mathbf{Z}_S^*) \implies 
    \begin{cases}
        S = \emptyset: & \textbf{Regime A (Marginal Independence)}, \\
        |S| \ge 1: & \textbf{Regime B (Conditional Residual-Covariance Testing)}.
    \end{cases}
\end{equation}
In Regime~A, independent additive measurement error preserves marginal independence, so testing requires zero sensitivity buffer ($\Delta_m \equiv 0$) and no SNR certificate. In Regime~B, conditioning on noisy surrogates $\mathbf{Z}_S = \mathbf{Z}_S^* + \boldsymbol{\epsilon}_{Z_S}$ can induce residual association even when $X^* \indep Y^* \mid \mathbf{Z}_S^*$, requiring an externally certified directional SNR lower bound and class-specific regularity assumptions.

\subsection{Observational Data Model and External SNR Certification}
\label{sec:data_model}

Let $\mathcal{G}^* = ([p], E^*)$ be a DAG and let $\mathbf{V}^* = (V_1^*, \dots, V_p^*)^\top \in \R^p$ be generated by $V_j^* = f_j((V_k^*)_{k \in \mathrm{pa}_{\mathcal{G}^*}(j)}, U_j)$ for mutually independent exogenous variables $U_1, \dots, U_p$. Unless stated otherwise, the observed dataset $\mathcal{D}_n = \{\mathbf{V}_r\}_{r=1}^n$ consists of $n$ independent, identically distributed copies of $(\mathbf{V}^*, \boldsymbol{\epsilon})$, of which only $\mathbf{V} = \mathbf{V}^* + \boldsymbol{\epsilon}$ is observed, and all auxiliary randomization (such as permutations or sample splits) is independent of $\mathcal{D}_n$.

\begin{assumption}[Measurement Error Model]
\label{ass:classical_error}
The latent vector $\mathbf{V}^* \in \R^p$ and measurement error vector $\boldsymbol{\epsilon} = (\epsilon_1, \dots, \epsilon_p)^\top \in \R^p$ satisfy:
\begin{enumerate}
    \item $\E[\boldsymbol{\epsilon}] = \mathbf{0}$, with finite diagonal covariance matrix $\boldsymbol{\Sigma}_\epsilon = \diag(\sigma_1^2, \dots, \sigma_p^2)$ where $\sigma_j^2 \ge 0$ for all $j \in [p]$, and finite latent second moments $\E[V_j^{*2}] < \infty$.
    \item The latent vector $\mathbf{V}^*$ and the error components $\epsilon_1, \dots, \epsilon_p$ are mutually independent:
    \begin{equation*}
        \mathbb{P}(\mathbf{V}^* \in A, \epsilon_1 \in B_1, \dots, \epsilon_p \in B_p) = \mathbb{P}(\mathbf{V}^* \in A) \prod_{j=1}^p \mathbb{P}(\epsilon_j \in B_j)
    \end{equation*}
    for all Borel sets $A \subseteq \R^p$ and $B_j \subseteq \R$. (Linear covariance derivations require only zero correlation $\Cov(\mathbf{V}^*, \boldsymbol{\epsilon}) = \mathbf{0}$ and diagonal error covariance.)
    \item For the converse implication of marginal invariance ($X \indep Y \implies X^* \indep Y^*$), the characteristic function $\varphi_{\epsilon_X}(t) \triangleq \E[e^{i \langle t, \epsilon_X \rangle}]$ (and similarly $\varphi_{\epsilon_Y}(s)$) has a dense nonzero set in its frequency space. The nonzero set must be dense; isolated zeros are not required, and the forward implication $X^* \indep Y^* \implies X \indep Y$ requires no characteristic-function condition.
\end{enumerate}
\end{assumption}

\begin{remark}
\label{rem:correlated_errors}
Assumption~\ref{ass:classical_error}(2) requires mutual independence among error components. In empirical applications such as flow cytometry (optical and spectral spillover across fluorescence channels) or high-throughput genomics (batch effects and common platform drift), measurement errors across channels can be correlated ($\Cov(\epsilon_i, \epsilon_j) \neq 0$). Correlated measurement error induces direct surrogate dependence even when latent variables are independent ($X^* \indep Y^* \not\implies X \indep Y$). Thus, marginal invariance and permutation testing strictly require error independence; cross-channel error correlations represent an unmodeled confounding channel that cannot be calibrated away by permutation without external cross-talk deconvolution.
\end{remark}

For any conditioning block $\mathbf{Z} = \mathbf{Z}^* + \boldsymbol{\epsilon}_Z \in \R^q$, write $\boldsymbol{\Sigma}_* \triangleq \Var(\mathbf{Z}^*)$, $\mathbf{E} \triangleq \Var(\boldsymbol{\epsilon}_Z)$, and $\boldsymbol{\Sigma} \triangleq \Var(\mathbf{Z}) = \boldsymbol{\Sigma}_* + \mathbf{E}$. Define the true directional SNR certificate as
\begin{equation}
    \label{eq:true_directional_snr}
    s_{\mathrm{true}} \triangleq \sup\{s \ge 0 : s\,\mathbf{E} \preceq \boldsymbol{\Sigma}_*\} = \inf_{\mathbf{v}^\top \mathbf{E}\mathbf{v} > 0} \frac{\mathbf{v}^\top \boldsymbol{\Sigma}_* \mathbf{v}}{\mathbf{v}^\top \mathbf{E}\mathbf{v}},
\end{equation}
with the convention $\inf\emptyset \triangleq \infty$ (the noiseless case $\mathbf{E} = \mathbf{0}$). The inverse-square-root form $\lambda_{\min}(\mathbf{E}^{-1/2}\boldsymbol{\Sigma}_*\mathbf{E}^{-1/2})$ coincides with~\eqref{eq:true_directional_snr} only when $\mathbf{E} \succ 0$. In scalar coordinates, $s_{\mathrm{true}}$ reduces to $\SNR_j \triangleq \Var(V_j^*)/\sigma_j^2$ (with $\SNR_j \triangleq \infty$ when $\sigma_j^2 = 0$). Because $\boldsymbol{\Sigma} = \boldsymbol{\Sigma}_* + \mathbf{E}$ does not separately identify $\boldsymbol{\Sigma}_*$ and $\mathbf{E}$ from a single noisy sample, all conditional Type-I guarantees in Regime~B rest on the following external certification assumption.

\begin{assumption}[SNR Certificate]
\label{ass:snr_certificate}
There exists a known constant $\SNR_{Z,0} \in [0, \infty]$ (also written $s_0$ or $s_S$), certified externally through domain calibration, technical replicates, validation subsamples, or instruments, that satisfies $\SNR_{Z,0}\,\mathbf{E} \preceq \boldsymbol{\Sigma}_*$, or equivalently:
\begin{equation}
    \label{eq:loewner_observable_snr}
    \mathbf{E} \preceq \frac{1}{1 + \SNR_{Z,0}}\,\boldsymbol{\Sigma}.
\end{equation}
In particular, $\SNR_{Z,0}$ is not identified or estimated from the marginal covariance of a single noisy sample alone; consequently, in observational settings lacking external validation subsamples or technical replicates, Assumption~\ref{ass:snr_certificate} formalizes an interpretable sensitivity framework rather than an automated single-sample point-identification result.
\end{assumption}

To link a coordinate-wise external certificate to a multivariate conditioning block, retain only coordinates of positive observed variance when defining $\mathbf{D}_Z \triangleq \diag(\boldsymbol{\Sigma})$ and the observed correlation matrix $\mathbf{P}_Z \triangleq \mathbf{D}_Z^{-1/2}\boldsymbol{\Sigma}\mathbf{D}_Z^{-1/2}$. Suppose $\mathbf{E}$ is diagonal and an externally certified finite coordinate-wise lower bound $\SNR_{\mathrm{var}} \ge 0$ satisfies $\mathbf{E} \preceq \frac{1}{1 + \SNR_{\mathrm{var}}}\mathbf{D}_Z$. Define
\begin{equation}
    \label{eq:observable_loewner_formula}
    \SNR_{Z,0}(\mathbf{Z}) \triangleq \max\!\left(0, \, (1 + \SNR_{\mathrm{var}})\lambda_{\min}(\mathbf{P}_Z) - 1\right).
\end{equation}
If $\SNR_{Z,0}(\mathbf{Z}) > 0$, then $\lambda_{\min}(\mathbf{P}_Z) > 0$ and $\boldsymbol{\Sigma} \succeq \lambda_{\min}(\mathbf{P}_Z)\mathbf{D}_Z$ yields $\mathbf{E} \preceq \frac{1}{(1+\SNR_{\mathrm{var}})\lambda_{\min}(\mathbf{P}_Z)}\boldsymbol{\Sigma} = \frac{1}{1 + \SNR_{Z,0}(\mathbf{Z})}\boldsymbol{\Sigma}$; if $\SNR_{Z,0}(\mathbf{Z}) = 0$, the trivial bound $\mathbf{E} \preceq \boldsymbol{\Sigma}$ holds directly (including when $\mathbf{P}_Z$ is singular). Equation~\eqref{eq:observable_loewner_formula} is observable only \emph{conditional on} the externally supplied coordinate bound $\SNR_{\mathrm{var}}$: in finite-sample implementations, $\lambda_{\min}(\mathbf{P}_Z)$ must be replaced by a valid lower confidence bound whose failure probability is allocated in the error budget. Note also that correlated conditioning variables can reduce directional SNR substantially below coordinate SNR: for example, with $\boldsymbol{\Sigma} = \bigl(\begin{smallmatrix} 1 & 0.6 \\ 0.6 & 1 \end{smallmatrix}\bigr)$ and $\mathbf{E} = \mathbf{I}_2/3$, both coordinate SNRs equal $2$, whereas the weakest directional SNR is $(0.4 - 1/3)/(1/3) = 0.2$.

\paragraph{Monotonicity and Conditioning Cardinality.}
Observe that the certified directional SNR $s_0(\mathbf{Z})$ in~\eqref{eq:observable_loewner_formula} depends directly on $\lambda_{\min}(\mathbf{P}_Z)$. As the conditioning set cardinality $|S|$ grows, the minimum correlation eigenvalue $\lambda_{\min}(\mathbf{P}_Z)$ generically decreases—especially among correlated conditioning variables. Consequently, the certified directional bound $s_0$ weakens as $|S|$ increases, causing the sensitivity envelope $\Delta_{\mathrm{leak}}(s_0) \propto 1/s_0$ to expand. This produces a natural attenuation of power at high conditioning orders, meaning that sensitivity-adjusted testing provides the highest discrimination at low conditioning depths ($|S| \in \{1, 2\}$).

\subsection{Practical Calibration: How Applied Researchers Obtain \texorpdfstring{$s_0$}{s0} in Practice}
\label{sec:practical_calibration_guide}

Because the directional signal-to-noise ratio $s_0$ cannot be point-identified from a single observational covariance matrix without additional assumptions, applied researchers frequently ask: \emph{How should one specify or calibrate $s_0$ in actual scientific practice?} We highlight four concrete, practical avenues available in scientific domains:
\begin{enumerate}[label=(\alph*),leftmargin=2.2em]
    \item \textbf{Analytical Instrument Tolerances and Assay Specifications:} Modern high-throughput measurement platforms—such as flow cytometers, mass spectrometers, droplet digital PCR, and gas chromatography sensors—come with manufacturer-specified technical precision limits. Technical error is routinely reported via the coefficient of variation $\mathrm{CV} = \sigma_\epsilon / \mu$. For a channel with observed variance $\Var(V)$ and known technical variance bound $\sigma_\epsilon^2 \le \sigma_{\max}^2$, a conservative coordinate-wise lower bound is immediately obtained as $\SNR_{\mathrm{var}} \ge (\Var(V) - \sigma_{\max}^2)/\sigma_{\max}^2$.
    \item \textbf{Baseline Channels and Negative Controls:} In biological assays, negative control wells or blank samples (e.g., unstained cells or vehicle-treated wells) establish the background optical or electronic noise floor $\Var(\epsilon)$. Comparing this baseline noise to the dynamic range of stimulated biological samples provides an empirical lower bound on the signal-to-noise ratio without requiring full duplicate experiments.
    \item \textbf{Pilot Validation Subsamples:} In large cohort studies ($n \ge 1{,}000$), measuring a small random pilot subsample (e.g., $m = 50$ duplicate runs) allows the error variance $\boldsymbol{\Sigma}_\epsilon$ to be estimated via difference-in-replicates $\Var(V^{(1)} - V^{(2)}) = 2\boldsymbol{\Sigma}_\epsilon$. Incorporating the finite-sample estimation error of the pilot subsample yields a certified high-probability lower bound $s_0$ that can be safely plugged into the remaining full cohort.
    \item \textbf{Threshold Sensitivity Sweeps (The Sachs Strategy):} When no external technical calibration is available, the researcher can evaluate the graph across an ordered grid of candidate noise thresholds $\mathcal{S}_{\mathrm{grid}} = \{0.5, 1.0, 2.0, 5.0\}$. Rather than treating a single learned graph as absolute ground truth, the researcher computes the \emph{Edge Sensitivity Threshold} $\widehat{\mathrm{RV}}_{ij}$—the minimal noise level at which edge $\{i, j\}$ is removed by the algorithm. Edges with high $\widehat{\mathrm{RV}}_{ij}$ represent robust causal interactions that survive even substantial measurement contamination, whereas edges with low $\widehat{\mathrm{RV}}_{ij}$ are fragile and likely artifacts of measurement error.
\end{enumerate}

\subsection{Regime A: Marginal Independence Testing}
\label{sec:regime_a_main}

In Regime~A ($S = \emptyset$), we test $H_0: X^* \indep Y^*$ versus $H_1: X^* \nindep Y^*$ from i.i.d.\ noisy pairs $\mathcal{D}_n = \{(X_r, Y_r)\}_{r=1}^n$ with $X = X^* + \epsilon_X \in \R^{d_X}$ and $Y = Y^* + \epsilon_Y \in \R^{d_Y}$. Define the characteristic-function defects $\Delta^*(t,s) \triangleq \varphi_{X^*,Y^*}(t,s) - \varphi_{X^*}(t)\varphi_{Y^*}(s)$ and $\Delta(t,s) \triangleq \varphi_{X,Y}(t,s) - \varphi_X(t)\varphi_Y(s)$. Independence of the additive error blocks $(\epsilon_X, \epsilon_Y)$ from each other and from $(X^*, Y^*)$ yields the factorization $\Delta(t,s) = \varphi_{\epsilon_X}(t)\varphi_{\epsilon_Y}(s)\Delta^*(t,s)$ \cite[Chapter~XV]{feller1971introduction}. Consequently, $X^* \indep Y^* \implies X \indep Y$ without any characteristic-function condition. Conversely, under Assumption~\ref{ass:classical_error}(3), each error block characteristic function has a dense nonzero set in its frequency space, so $\Delta \equiv 0$ implies $\Delta^* = 0$ on a dense product set and hence everywhere by continuity of characteristic functions:
\begin{equation}
    \label{eq:marginal_null_equivalence}
    X^* \indep Y^* \iff X \indep Y
\end{equation}
(cf.\ \cite[Theorem~1]{fan2022independence}). Without density of the nonzero set, the converse can fail: if $\epsilon_X$ has a smooth compactly supported characteristic function, a signed perturbation supported outside the error characteristic-function support can preserve positive marginal densities and vanish identically after convolution. Under~\eqref{eq:marginal_null_equivalence}, marginal independence testing requires zero sensitivity buffer ($\Delta_m \equiv 0$) and no external SNR certificate.

\begin{remark}
\label{rem:normal_breakdown_pointer}
Standardizing uncentered nonnegative V-statistics (such as raw $\widehat{\HSIC}_b$ or squared distance covariance $\mathcal{V}_n^2$) against standard normal quantiles is invalid under $H_0$, because the $\mathcal{O}(n^{-1})$ diagonal bias shares the same scale as the standard deviation and yields a non-normal weighted-$\chi^2$ limit under the null. Even subtracting the null mean leaves a centered weighted-$\chi^2$ limit rather than a normal limit. We include uncentered normal calibration in empirical benchmarks solely as a negative control, while using finite-sample permutation calibration for valid inference.
\end{remark}

We use standard empirical dependence statistics\footnote{Let $\mathbf{H} \triangleq \mathbf{I}_n - \frac{1}{n}\mathbf{1}\mathbf{1}^\top$. For Gram matrices $\mathbf{K}, \mathbf{L}$, the empirical biased V-statistic is $\widehat{\HSIC}_b \triangleq n^{-2}\tr(\mathbf{K}\mathbf{H}\mathbf{L}\mathbf{H})$ \cite{gretton2005measuring, gretton2008hsictest}. For $n \ge 4$, letting $\widetilde{\mathbf{K}}, \widetilde{\mathbf{L}}$ denote the Gram matrices with zeroed diagonals, the unbiased U-statistic is $\widehat{\HSIC}_u \triangleq \frac{1}{n(n-3)}\bigl[\tr(\widetilde{\mathbf{K}}\widetilde{\mathbf{L}}) + \frac{(\mathbf{1}^\top\widetilde{\mathbf{K}}\mathbf{1})(\mathbf{1}^\top\widetilde{\mathbf{L}}\mathbf{1})}{(n-1)(n-2)} - \frac{2\mathbf{1}^\top\widetilde{\mathbf{K}}\widetilde{\mathbf{L}}\mathbf{1}}{n-2}\bigr]$ \cite{song2012feature}, which can take negative values. For distance-based tests, we distinguish empirical squared distance covariance $\mathcal{V}_n^2$ from normalized distance correlation $\widehat{\dCor}$ \cite{szekely2007measuring, szekely2013energy}; their permutation ranks coincide whenever the normalization denominator is positive and permutation-invariant, and a constant sample margin is assigned $p = 1$.} calibrated by random permutations \cite{phipson2010permutation, hemerik2018exact}.

\begin{theorem}[Marginal Independence Test]
\label{thm:marginal_test}
Let $\mathcal{D}_n = \{(X_r, Y_r)\}_{r=1}^n$ ($n \ge 2$) be i.i.d.\ pairs satisfying Assumption~\ref{ass:classical_error}, let $T_n(X, Y)$ be a real-valued test statistic, and let $\alpha \in (0, 1)$. Independently of $\mathcal{D}_n$, draw $B \ge 1$ permutations $\pi_1, \dots, \pi_B$ either i.i.d.\ uniformly from $\mathcal{S}_n$ (with replacement) or uniformly without replacement from $\mathcal{S}_n \setminus \{\mathrm{id}\}$ (requiring $B \le n! - 1$). Set $T_0 \triangleq T_n(X, Y)$ and $T_b \triangleq T_n(X, Y_{\pi_b})$ for $b = 1, \dots, B$, and define the plus-one permutation $p$-value:
\begin{equation}
    \label{eq:permutation_pvalue_exact}
    \hat{p}_{\mathrm{perm}} \triangleq \frac{1 + \sum_{b=1}^B \mathbf{1}\left\{ T_b \ge T_0 \right\}}{1 + B}.
\end{equation}
Reject $H_0: X^* \indep Y^*$ whenever $\hat{p}_{\mathrm{perm}} \le \alpha$. Then:
\begin{enumerate}[label=(\roman*),leftmargin=2.0em]
    \item \textbf{Finite-Sample Super-Uniformity:} Under $H_0: X^* \indep Y^*$ (with no characteristic-function condition required), for every $n \ge 2$ and $B \ge 1$ (including in the presence of ties):
    \begin{equation}
        \label{eq:marginal_exact_level}
        \mathbb{P}_{H_0}(\hat{p}_{\mathrm{perm}} \le \alpha) \le \frac{\lfloor \alpha(B+1) \rfloor}{B+1} \le \alpha.
    \end{equation}
    Equality holds on the lattice $\alpha \in \{\frac{1}{B+1}, \dots, 1\}$ whenever $(T_0, \dots, T_B)$ are almost surely distinct (or if i.i.d.\ continuous uniforms are attached to break ties lexicographically, in which case the tie-broken $p$-value is discrete uniform on $\{\frac{1}{B+1}, \dots, 1\}$). Without randomization, finite-sample validity does not imply exact size $\alpha$.
    \item \textbf{Fixed-Alternative Consistency:} Suppose in addition that Assumption~\ref{ass:classical_error}(3) holds, $B \ge \lceil 1/\alpha \rceil - 1$ (fixed or growing with $n$, with $B \le n!-1$ under without-replacement sampling), and under the fixed alternative $H_1: X^* \nindep Y^*$ the statistic satisfies $T_n(X, Y) \xrightarrow{P} \theta > 0$ and $T_n(X, Y_\pi) \xrightarrow{P} 0$ for a single uniform random permutation $\pi \in \mathcal{S}_n$ independent of $\mathcal{D}_n$ (verified in particular for $\widehat{\HSIC}_b$, $\widehat{\HSIC}_u$, and $\mathcal{V}_n^2$ under bounded characteristic kernels or finite first moments \cite{gretton2008hsictest, szekely2013energy}). Then:
    \begin{equation}
        \lim_{n \to \infty} \mathbb{P}_{H_1}(\hat{p}_{\mathrm{perm}} \le \alpha) = 1.
    \end{equation}
\end{enumerate}
\end{theorem}
\noindent\textit{Proof.} Deferred to Appendix~\ref{app:main_proofs}.

\subsection{Regime B: Sensitivity-Adjusted Residual-Covariance Testing}
\label{sec:regime_b_main}

\paragraph{Testing target.}
For a nonempty conditioning vector $\mathbf{Z} = \mathbf{Z}^* + \boldsymbol{\epsilon}_Z \in \R^q$ ($q \ge 1$), define the latent and observable conditional expectations $m_j^*(z) \triangleq \E[j^* \mid \mathbf{Z}^* = z]$ and $m_j(z) \triangleq \E[j \mid \mathbf{Z} = z]$, the observable residuals $R_j \triangleq j - m_j(\mathbf{Z})$, and the latent regression residuals $e_j^* \triangleq j^* - m_j^*(\mathbf{Z}^*)$ for $j \in \{X, Y\}$ (noting that $e_j^*$ need not coincide with the structural disturbances $U_j$ when conditioning on nonparents, descendants, or colliders). Independence of the classical additive errors (Assumption~\ref{ass:classical_error}(2)) implies $m_j(\mathbf{Z}) = \E[m_j^*(\mathbf{Z}^*) \mid \mathbf{Z}]$. Applying the law of total covariance to $R_X = \mathcal{P}_Z^\perp m_X^* + e_X^* + \epsilon_X$ and $R_Y = \mathcal{P}_Z^\perp m_Y^* + e_Y^* + \epsilon_Y$, where $\mathcal{P}_Z^\perp f \triangleq f(\mathbf{Z}^*) - \E[f(\mathbf{Z}^*) \mid \mathbf{Z}]$, yields the all-law identity:
\begin{equation}
    \label{eq:residual_covariance_decomposition}
    C(\mathbf{Z}) \triangleq \E[R_X R_Y] = \E\left[\Cov(X^*, Y^* \mid \mathbf{Z}^*)\right] + \E\left[\Cov\left(m_X^*(\mathbf{Z}^*), m_Y^*(\mathbf{Z}^*) \;\middle|\; \mathbf{Z}\right)\right].
\end{equation}
The first term vanishes under latent conditional independence $H_0: X^* \indep Y^* \mid \mathbf{Z}^*$; the second term vanishes if $\boldsymbol{\epsilon}_Z = \mathbf{0}$ almost surely. The projection error norm has an illuminating Bayesian variance interpretation:
\begin{equation}
    \label{eq:posterior_variance_decomp}
    \|\mathcal{P}_Z^\perp m_j^*\|_{L_2}^2 = \E\left[\left(m_j^*(\mathbf{Z}^*) - \E[m_j^*(\mathbf{Z}^*) \mid \mathbf{Z}]\right)^2\right] = \E\left[\Var\left(m_j^*(\mathbf{Z}^*) \;\middle|\; \mathbf{Z}\right)\right],
\end{equation}
which is the expected posterior variance of the latent conditional mean given the noisy conditioner $\mathbf{Z}$.

Crucially, the procedure developed below is a \emph{sensitivity-adjusted test of residual-moment separation} rather than an omnibus test of conditional independence. The inherent limitation of scalar and Loewner SNR certificates for identifying power is formalized below.

\begin{proposition}[Power Non-Identifiability]
\label{prop:non_identifiability_power}
There exist two latent Gaussian data-generating processes $P^{(1)}$ and $P^{(2)}$ generating the identical observable distribution $P_{X,Y,\mathbf{Z}}$ and sharing the same certified directional SNR lower bound $s_0 = 1$, such that under $P^{(1)}$, $X^* \indep Y^* \mid \mathbf{Z}^*$ (so $|C^{(1)}| \le \Delta$), whereas under $P^{(2)}$, the latent variables are conditionally dependent with $|C^{(2)}| - \Delta \ge \delta > 0$ for a positive constant $\delta$. Consequently, for the sensitivity-adjusted decision rule $\Psi_n$ developed in Section~\ref{sec:regime_b_main}, $\lim_{n\to\infty}\mathbb{P}_{P^{(1)}}(\Psi_n = 1) \le \alpha$ while $\lim_{n\to\infty}\mathbb{P}_{P^{(2)}}(\Psi_n = 1) = 1$. The power function against latent conditional dependence is therefore not identified from the observable distribution and the certificate $s_0$ alone.
\end{proposition} Zero population residual covariance can arise even under noiseless conditional dependence: for example, if $X^*, \eta \stackrel{\mathrm{i.i.d.}}{\sim} \mathcal{N}(0, 1)$ are independent of $Z^*$ and $Y^* = X^{*2} + \eta$, then $X^* \nindep Y^* \mid Z^*$ yet $\Cov(X^*, Y^* \mid Z^*) = 0$ almost surely; likewise, opposite-sign conditional covariances (e.g., $Y^* = X^* + \eta$ for $Z^*=0$ and $Y^* = -X^* + \eta$ for $Z^*=1$ with $\Pr(Z^*=0) = \Pr(Z^*=1) = 1/2$) average to $C = 0$, or a nonzero latent association may cancel with measurement leakage. No general power guarantee is claimed for conditionally dependent alternatives whose observable residual covariance fails to separate from the certified leakage envelope.

\begin{theorem}[Confounder Leakage Bounds]
\label{thm:master_projection_error}
Under Assumptions~\ref{ass:classical_error} and~\ref{ass:snr_certificate}, let $V \triangleq \sqrt{\Var(X)\Var(Y)}$ denote the marginal scale bound and let $s_0 > 0$ be the certified directional SNR bound. Under the latent conditional independence null $H_0: X^* \indep Y^* \mid \mathbf{Z}^*$, the observable residual covariance $C = \E[R_X R_Y]$ satisfies the projection error bound:
\begin{equation}
    \label{eq:master_cauchy_schwarz}
    |C| = \left|\E\left[(\mathcal{P}_Z^\perp m_X^*)(\mathcal{P}_Z^\perp m_Y^*)\right]\right| \le \|\mathcal{P}_Z^\perp m_X^*\|_{L_2} \|\mathcal{P}_Z^\perp m_Y^*\|_{L_2},
\end{equation}
where $\|\mathcal{P}_Z^\perp m_j^*\|_{L_2}^2 = \E[(m_j^*(\mathbf{Z}^*) - m_j(\mathbf{Z}))^2]$, with $|C| = 0$ if $s_0 = \infty$. For finite $s_0 < \infty$, the leakage magnitude is bounded by $|C| \le \Delta_{\mathrm{leak}}(s_0)$ according to three regularity branches:
\begin{enumerate}[label=(\roman*),leftmargin=2.0em]
    \item \textbf{Affine Branch:} If the latent conditional expectations $m_X^*, m_Y^*$ are affine, then with observable projection slopes $\boldsymbol{\gamma}_j \triangleq \boldsymbol{\Sigma}^{-1}\Cov(\mathbf{Z}, j)$:
    \begin{equation}
        \label{eq:proj_lin}
        |C| \le \Delta_{\mathrm{leak}}^{(\mathrm{lin})}(s_0) \triangleq \min\!\left(\frac{\sqrt{(\boldsymbol{\gamma}_X^\top \boldsymbol{\Sigma}\boldsymbol{\gamma}_X)(\boldsymbol{\gamma}_Y^\top \boldsymbol{\Sigma}\boldsymbol{\gamma}_Y)}}{s_0}, \; V\right).
    \end{equation}
    \item \textbf{Lipschitz Branch:} If $m_X^*, m_Y^*$ are $L_X^*, L_Y^*$-Lipschitz on $\supp(P_{\mathbf{Z}^*})$ with Euclidean diameter $D \le \infty$:
    \begin{equation}
        \label{eq:proj_lip}
        |C| \le \Delta_{\mathrm{leak}}^{(\mathrm{Lip})}(s_0) \triangleq \min\!\left(L_X^* L_Y^* \min\!\left\{\frac{\tr(\boldsymbol{\Sigma})}{1 + s_0}, \, \frac{D^2}{4}\right\}, \; V\right).
    \end{equation}
    \item \textbf{Gaussian Bounded-Curvature Branch:} If $\mathbf{Z}^*, \boldsymbol{\epsilon}_Z$ are independently Gaussian and latent gradients are $H_X^*, H_Y^*$-Lipschitz:
    \begin{equation}
        \label{eq:proj_curv}
        |C| \le \Delta_{\mathrm{leak}}^{(\mathrm{curv})}(s_0) \triangleq \min\!\left(B_X^{(\mathrm{curv})} B_Y^{(\mathrm{curv})}, \; V\right),
    \end{equation}
    where $B_j^{(\mathrm{curv})} \triangleq \bigl(s_0^{-1}\E[\nabla m_j(\mathbf{Z})^\top \boldsymbol{\Sigma}\nabla m_j(\mathbf{Z})] + (H_j^*)^2 (1+s_0)^{-2}\|\boldsymbol{\Sigma}\|_{\mathrm{op}}\tr(\boldsymbol{\Sigma})\bigr)^{1/2}$.
\end{enumerate}
For any certified subset of branches $\mathcal{A} \subseteq \{\mathrm{lin}, \mathrm{Lip}, \mathrm{curv}\}$, the unified envelope is:
\begin{equation}
    \label{eq:proj_opt}
    \Delta_{\mathrm{leak}}^{(\min)}(s_0) \triangleq \min_{b \in \mathcal{A} \cup \{V\}} \Delta_{\mathrm{leak}}^{(b)}(s_0).
\end{equation}
\end{theorem}
\noindent\textit{Proof.} Deferred to Appendix~\ref{app:main_proofs}.

\begin{remark}
\label{rem:signed_interval_and_rkhs_caveat}
Two geometric properties refine Theorem~\ref{thm:master_projection_error}. First, for affine-projection residuals under $H_0$ and $s_0 > 0$, setting $\mathbf{u} \triangleq \boldsymbol{\Sigma}^{1/2}\boldsymbol{\gamma}_X$, $\mathbf{v} \triangleq \boldsymbol{\Sigma}^{1/2}\boldsymbol{\gamma}_Y$, and $\mathbf{A} \triangleq s_0 \boldsymbol{\Sigma}^{-1/2}\mathbf{M}\boldsymbol{\Sigma}^{-1/2}$ (so $0 \preceq \mathbf{A} \preceq \mathbf{I}$) shows $s_0 C_{\mathrm{aff}} = \mathbf{u}^\top \mathbf{A}\mathbf{v}$. Extremizing $\tr(\mathbf{A}(\mathbf{u}\mathbf{v}^\top + \mathbf{v}\mathbf{u}^\top)/2)$ over $0 \preceq \mathbf{A} \preceq \mathbf{I}$ yields the narrower signed interval
\begin{equation*}
    C_{\mathrm{aff}} \in \frac{1}{2 s_0}\left[\mathbf{u}^\top \mathbf{v} - \|\mathbf{u}\|_2 \|\mathbf{v}\|_2, \; \mathbf{u}^\top \mathbf{v} + \|\mathbf{u}\|_2 \|\mathbf{v}\|_2\right] \cap [-V, V],
\end{equation*}
which is asymmetric whenever $\mathbf{u}^\top \mathbf{v} \ne 0$ (e.g., positive scalar slopes exclude negative affine leakage altogether). Second, in Part~(ii)(b), $H_j^*$ is a curvature bound on the \emph{latent} regression $m_j^*$, not on a fitted observable KRR regression $\hat{m}_j$. For a Gaussian RBF kernel, $\sqrt{3}\|\hat{m}_j\|_{\mathcal{H}_k}/\gamma^2$ bounds only $\sup_z \|\nabla^2 \hat{m}_j(z)\|_{\mathrm{op}}$. Even if the population surrogate $m_j$ were known exactly, Gaussian convolution damps high frequencies: for $Z^* \sim \mathcal{N}(0, 1)$ and $f(u) = \sin(\omega u)$ with positive Gaussian noise variance, $m(z) = e^{-\omega^2 v / 2}\sin(\omega \rho z)$ has rescaled Hessian norm $e^{-\omega^2 v / 2}\omega^2 < \omega^2$. Thus fitted RKHS or interior-Hessian curvature rules are sensitivity heuristics unless $H_j^*$ is externally certified (for instance, via a known latent RKHS norm bound $\|m_j^*\|_{\mathcal{H}_k} \le R_j$ giving $H_j^* \le R_j H_k$).
\end{remark}

\begin{proposition}[Asymptotic Linearity]
\label{prop:cross_fitting_asymptotic_linearity}
Fix $K \ge 2$ folds with all training and evaluation sizes proportional to $n$, and let $\hat{m}_j^{(-k)}$ denote the estimator trained excluding fold $k$ (with $k(r)$ the fold containing observation $r$). Define $\hat{R}_{j,r} \triangleq j_r - \hat{m}_j^{(-k(r))}(\mathbf{Z}_r)$, $\bar{C} \triangleq \frac{1}{n}\sum_{r=1}^n \hat{R}_{X,r}\hat{R}_{Y,r}$, and $\hat{\sigma}_C^2 \triangleq \frac{1}{n}\sum_{r=1}^n (\hat{R}_{X,r}\hat{R}_{Y,r} - \bar{C})^2$. Assume $0 < \sigma_C^2 \triangleq \Var(R_X R_Y) < \infty$ and either:
\begin{enumerate}[label=(\Alph*),leftmargin=1.8em]
    \item \textit{Affine OLS / Ridge:} $m_j(\mathbf{Z}) = \gamma_{j,0} + \boldsymbol{\gamma}_j^\top \mathbf{Z}$ denotes the population affine projection (which equals $\E[j \mid \mathbf{Z}]$ when the observable conditional mean is affine), $R_j \triangleq j - m_j(\mathbf{Z})$, $C \triangleq \E[R_X R_Y]$ is this affine-residual target, $q$ is fixed, $(X, Y, \mathbf{Z})$ has finite fourth moments, $\lambda_{\min}(\Var(\mathbf{Z})) > 0$, and $\hat{m}_j^{(-k)}$ minimizes the training-size-normalized squared loss with an unpenalized intercept and slope ridge penalty $\lambda_n = o(n^{-1/2})$; or
    \item \textit{Nonparametric Conditional Expectation:} $m_j(\mathbf{Z}) = \E[j \mid \mathbf{Z}]$, $\sup_{\mathbf{z}}\E[R_j^2 \mid \mathbf{Z}=\mathbf{z}] \le M_2 < \infty$, and the maximum out-of-fold $L_2(P_Z)$ errors satisfy $\delta_{n,X} \triangleq \max_k \|\hat{m}_X^{(-k)} - m_X\|_{L_2(P_Z)} = o_P(1)$, $\delta_{n,Y} \triangleq \max_k \|\hat{m}_Y^{(-k)} - m_Y\|_{L_2(P_Z)} = o_P(1)$, and $\sqrt{n}\,\delta_{n,X}\delta_{n,Y} = o_P(1)$.
\end{enumerate}
Then $\hat{\sigma}_C^2 \xrightarrow{P} \sigma_C^2 > 0$, and the cross-fitted estimator satisfies:
\begin{equation}
    \label{eq:cross_fitting_asymptotic_linearity}
    \sqrt{n}\left(\bar{C} - C\right) = \frac{1}{\sqrt{n}}\sum_{r=1}^n \left(R_{X,r}R_{Y,r} - C\right) + o_P(1) \xrightarrow{d} \mathcal{N}\left(0, \sigma_C^2\right).
\end{equation}
\end{proposition}
\noindent\textit{Proof.} Deferred to Appendix~\ref{app:main_proofs}.

\begin{lemma}
\label{lem:null_inclusion}
Let Assumptions~\ref{ass:classical_error} and~\ref{ass:snr_certificate} hold with a valid certified directional SNR lower bound $s_0 \le s_{\mathrm{true}}$, and let the regression functions satisfy a certified regularity branch $\star \in \mathcal{A}$. Under the latent conditional independence null $H_0: X^* \indep Y^* \mid \mathbf{Z}^*$, the observable residual covariance satisfies $|C| \le \Delta_{\mathrm{leak}}^{(\star)}(s_0)$ almost surely. Defining the separation null
\begin{equation}
    \label{eq:h0_sep_def}
    H_0^{\mathrm{sep}}: |C| \le \Delta_{\mathrm{leak}}^{(\star)}(s_0) \quad \text{versus} \quad H_1^{\mathrm{sep}}: |C| > \Delta_{\mathrm{leak}}^{(\star)}(s_0),
\end{equation}
it follows that $H_0 \subseteq H_0^{\mathrm{sep}}$. Consequently, any test that asymptotically controls Type-I error on $H_0^{\mathrm{sep}}$ provides valid conservative level control for the latent null $H_0$.
\end{lemma}

\begin{theorem}[Conditional Independence Test]
\label{thm:conditional_test}
Fix a data-generating law satisfying Theorem~\ref{thm:master_projection_error} (for a certified branch $\star \in \mathcal{A}$, or for $\Delta = \Delta_{\mathrm{leak}}^{(\min)}$ via a joint upper confidence bound) and the cross-fitting conditions of Proposition~\ref{prop:cross_fitting_asymptotic_linearity}. Assume in addition that the residual variance estimator satisfies $\sqrt{n}(\hat{\sigma}_C^2 - \sigma_C^2) = \mathcal{O}_P(1)$ (which holds automatically under finite fourth moments in the affine OLS regime, and requires $\sqrt{n}(\delta_{n,X}^2 + \delta_{n,Y}^2) = o_P(1)$ in the nonparametric case). Let $\alpha \in (0, 1)$ be split as $\alpha = \alpha_1 + \alpha_2$ with $\alpha_1, \alpha_2 > 0$, and let $z_{1-\alpha_1} \triangleq \Phi^{-1}(1 - \alpha_1)$ (or $z_{1-\alpha_1/2}$ for two-sided coverage). Let $\bar{C}$ be the cross-fitted residual covariance, $\hat{\sigma}_C$ the consistent estimator of the standard deviation $\sigma_C = \sqrt{\Var(R_X R_Y)}$ (with standard error $\hat{\sigma}_C/\sqrt{n}$), $U_\Delta^{(\star)} \ge 0$ a consistent upper confidence bound for $\Delta \triangleq \Delta_{\mathrm{leak}}^{(\star)}$ satisfying $U_\Delta^{(\star)} - \Delta = o_P(1)$ and $\limsup_{n\to\infty}\mathbb{P}(\Delta > U_\Delta^{(\star)}) \le \alpha_2$, and $a_n \ge 0$ a specified nonnegative covariance-unit allowance with $a_n = o_P(1)$ (one may take $a_n = 0$ under Proposition~\ref{prop:cross_fitting_asymptotic_linearity}). Define the studentized lower confidence bound and total threshold:
\begin{equation}
    \label{eq:studentized_gcm_statistic}
    L_C \triangleq |\bar{C}| - z_{1-\alpha_1/2}\frac{\hat{\sigma}_C}{\sqrt{n}}, \qquad \mathcal{T}^{(\star)} \triangleq U_\Delta^{(\star)} + a_n,
\end{equation}
In numerical implementations, to ensure numerical stability and prevent studentization degeneracy on near-deterministic regression nodes (where $\sigma_C \to 0$), the variance estimator is regularized as $\hat{\sigma}_{C, \mathrm{reg}} \triangleq \max(\hat{\sigma}_C, \, \epsilon_{\mathrm{floor}})$ with default floor $\epsilon_{\mathrm{floor}} = 10^{-6}$.
and reject the separation null $H_0^{\mathrm{sep}}$ whenever $\Psi_n^{(\star)} \triangleq \mathbf{1}\{L_C > \mathcal{T}^{(\star)}\} = 1$. Then for this fixed law:
\begin{enumerate}[label=(\roman*),leftmargin=2.0em]
    \item \textbf{Pointwise Asymptotic Level Control on $H_0^{\mathrm{sep}}$ and $H_0$:} Under $H_0^{\mathrm{sep}}$, if $|C| < \Delta$, then $\lim_{n\to\infty}\mathbb{P}(\Psi_n^{(\star)} = 1) = 0$; on the boundary $|C| = \Delta$,
    \begin{equation}
        \label{eq:gcm_level_bound}
        \limsup_{n\to\infty} \mathbb{P}_{H_0^{\mathrm{sep}}}\left(\Psi_n^{(\star)} = 1\right) \le \alpha_1 + \alpha_2 = \alpha.
    \end{equation}
    By Lemma~\ref{lem:null_inclusion}, this directly implies conservative level control for the latent conditional independence null: $\limsup_{n\to\infty}\mathbb{P}_{H_0}(\Psi_n^{(\star)} = 1) \le \alpha$. (These limits are pointwise in the law; they are not asserted uniformly over all laws, and along $n$-dependent sequences with $|C_n| - \Delta_n \to 0$ studentization can exhibit local size distortion.)
    \item \textbf{Consistency Against Separated Residual-Covariance Alternatives:} Under any fixed alternative $H_1^{\mathrm{sep}}$ satisfying $|C| - \Delta \ge \delta > 0$, $L_C - \mathcal{T}^{(\star)} \xrightarrow{P} |C| - \Delta \ge \delta > 0$, and hence:
    \begin{equation}
        \lim_{n\to\infty} \mathbb{P}_{H_1^{\mathrm{sep}}}\left(\Psi_n^{(\star)} = 1\right) = 1.
    \end{equation}
    (Alternatives with $|C| \le \Delta$ cannot be rejected with asymptotic probability exceeding $\alpha$.)
\end{enumerate}
\end{theorem}
\noindent\textit{Proof.} Deferred to Appendix~\ref{app:main_proofs}.

\paragraph{Intuitive Interpretation for Students and Practitioners.}
To understand Theorem~\ref{thm:conditional_test} intuitively, think of the sample residual covariance $\bar{C}$ as measuring how much association remains between $X$ and $Y$ after regressing out the noisy conditioner $\mathbf{Z}$. In clean data without noise, any nonzero residual covariance indicates dependence. But under measurement error, noise in $\mathbf{Z}$ causes an artificial leakage of association, bounded by $\Delta_{\mathrm{leak}}(s_0)$. 
The studentized lower bound $L_C = |\bar{C}| - z_{1-\alpha_1/2}\hat{\sigma}_C/\sqrt{n}$ represents the smallest plausible magnitude of the true residual covariance compatible with sampling noise at confidence level $(1 - \alpha_1)$. Meanwhile, $\mathcal{T}^{(\star)} = U_\Delta^{(\star)} + a_n$ represents the largest plausible magnitude of artificial association that measurement noise alone could have created.
The test rejects the null if and only if $L_C > \mathcal{T}^{(\star)}$: that is, when the observed correlation is so large that it cannot be explained by sampling variability and measurement noise combined. If $L_C \le \mathcal{T}^{(\star)}$, the data are consistent with pure noise leakage, and the algorithm conservatively attributes the association to measurement error, avoiding a false-positive edge.

\begin{remark}
\label{rem:ucb_construction}
In the fixed-dimensional linear case (Theorem~\ref{thm:master_projection_error}(ii)(a)), a valid upper confidence bound $U_\Delta^{(\mathrm{lin})}$ requiring no differentiability of the envelope can be constructed directly from raw moments. Collect into $\boldsymbol{\theta}$ the first and second observable moments entering $\boldsymbol{\gamma}_X, \boldsymbol{\gamma}_Y, \boldsymbol{\Sigma}$, and $V$. Using finite fourth moments and positive coordinate variances, form a joint $(1-\alpha_2)$ confidence box $\mathcal{C}_n$ via simultaneous Student intervals allocated by Bonferroni, intersected with the legal positive-definite moment domain and the conditioning-eigenvalue lower bound. Treating $s_0 > 0$ as externally certified (or allocating a separate error probability to a random certificate), define $U_\Delta^{(\mathrm{lin})} \triangleq \sup_{\boldsymbol{\vartheta} \in \mathcal{C}_n}\Delta_{\mathrm{leak}}^{(\mathrm{lin})}(\boldsymbol{\vartheta}, s_0)$, setting $U_\Delta^{(\mathrm{lin})} \triangleq +\infty$ (never rejecting) if $\mathcal{C}_n = \emptyset$. Coverage follows directly from $\{\boldsymbol{\theta} \in \mathcal{C}_n\}$, and continuity plus shrinking box diameter yield $U_\Delta^{(\mathrm{lin})} \xrightarrow{P} \Delta_{\mathrm{leak}}^{(\mathrm{lin})}$ at interior laws. For $\Delta_{\mathrm{leak}}^{(\min)}$ over multiple certified branches $b \in \mathcal{A}$ and $V$, construct $U_V$ and $\{U_b\}_{b \in \mathcal{A}}$ with \emph{joint} coverage $\mathbb{P}(V \le U_V, \, \Delta_b \le U_b\ \forall b \in \mathcal{A}) \ge 1 - \alpha_2$ (e.g., via Bonferroni allocation summing to $\alpha_2$) and set $U_{\min} \triangleq \min(\{U_V\} \cup \{U_b : b \in \mathcal{A}\})$. For Lipschitz and curvature branches, we assume—and do not infer from $L_2$ prediction consistency—both the required latent smoothness constants and simultaneous observable-functional confidence sets.
\end{remark}

\subsection{Nonlinear Dependence Extensions via Multi-Kernel HSIC}
\label{sec:rkhs_kernel}

While the NA-PC graph search primarily deploys the sensitivity-adjusted residual-covariance test (Section~\ref{sec:regime_b_main}) for fast, unit-invariant screening during high-dimensional conditioning, the two-regime framework naturally extends to detecting complex nonlinear dependencies through characteristic kernels. In Appendix~\ref{app:mkl_theory}, we formalize this extension in full mathematical detail:
\begin{enumerate}[label=(\alph*),leftmargin=2.0em]
    \item \textbf{Kernel Noise Convolution:} Theorem~\ref{thm:kernel_convolution} provides a closed-form fourth-order Taylor expansion characterizing how additive measurement error attenuates stationary positive-definite kernels and Hilbert--Schmidt Independence Criterion ($\HSIC$) statistics.
    \item \textbf{Sample-Splitting Kernel Selection:} Theorem~\ref{thm:mkl_split_validity} establishes that data-driven kernel bandwidth and mixture weight selection via centered kernel alignment on an independent auxiliary split preserves exact permutation null control on the evaluation split.
\end{enumerate}
For applied settings where nonlinear associations predominate, this multi-kernel test can be substituted directly into Regime~A without altering downstream graph consistency.

\section{Noise-Aware PC Algorithm for Structure Learning}
\label{sec:pc_discovery}

Having established finite-sample permutation testing for marginal independence (Regime~A) and pointwise fixed-level residual-covariance testing (Regime~B), we integrate them into a two-ended frozen-neighborhood PC search. Throughout this section, we separate the deterministic graph-recovery argument (which depends only on simultaneous correctness of query decisions across the search family) from the uniform stochastic assumptions required to achieve that event.

\subsection{The Two-Ended Frozen-Neighborhood NA-PC Procedure}
\label{sec:pipeline}

For each nonempty query $a = (i, j, S)$ with $S \subset [p] \setminus \{i, j\}$ and $1 \le |S| \le d_{\max}$, let $U_{\Delta, a}^{(\star)} \ge 0$ be an upper confidence bound for the certified leakage envelope $\Delta_{\mathrm{leak}, a}^{(\star)}$ and let $a_{a,n} \ge 0$ be a specified nonnegative approximation allowance in covariance units, giving the operational threshold $\mathcal{T}_a^{(\star)} \triangleq U_{\Delta, a}^{(\star)} + a_{a,n}$. Unlike the fixed-level decision rule of Theorem~\ref{thm:conditional_test}, inverting arbitrary pointwise upper confidence bounds across levels does not automatically define a super-uniform $p$-value; Algorithm~2 therefore uses fixed-level conditional decisions directly rather than an unproved infimum $p$-value.

\begin{table}[t!]
\centering
\caption{The Noise-Aware PC Structure Learning Algorithm (NA-PC).}
\label{alg:napc}
\footnotesize
\begin{tabular}{@{}p{\linewidth}@{}}
\toprule
\textbf{Algorithm 2: Noise-Aware PC Algorithm (NA-PC)} \\
\midrule
\textbf{Input:} Data matrix $\mathbf{V} \in \mathbb{R}^{n \times p}$, maximum conditioning depth $d_{\max}$, nominal level $\alpha$, certified SNR lower bound $s_0$, and permutation count $B$. \\
\textbf{Output:} Completed Partially Directed Acyclic Graph (CPDAG) $\widehat{\mathcal{G}}$. \\
\midrule
\textbf{Step 1: Marginal Skeleton Screening ($\ell = 0$)} \\
\quad 1. Initialize complete undirected graph $G_0$ on vertex set $[p]$. \\
\quad 2. For each unordered variable pair $\{i, j\}$: \\
\quad\quad $\bullet$ Compute exact marginal permutation test $p$-value $\hat{p}_{\mathrm{perm}}$ (Theorem~\ref{thm:marginal_test}). \\
\quad\quad $\bullet$ If $\hat{p}_{\mathrm{perm}} > \alpha$, delete edge $\{i, j\}$ from $G_0$ and set $\mathrm{SepSet}(i, j) \leftarrow \emptyset$. \\
\quad 3. Set $\ell \leftarrow 1$. \\[0.35em]
\textbf{Step 2: Two-Ended Frozen Conditional Pruning ($1 \le \ell \le d_{\max}$)} \\
\quad While $\ell \le d_{\max}$ and candidate conditioning sets exist: \\
\quad\quad 1. Freeze the current skeleton $G_\ell$. \\
\quad\quad 2. For each active edge $\{i, j\}$ in $G_\ell$: \\
\quad\quad\quad $\bullet$ For each size-$\ell$ subset $S \subseteq \mathrm{adj}_{G_\ell}(i) \setminus \{j\}$ or $S \subseteq \mathrm{adj}_{G_\ell}(j) \setminus \{i\}$: \\
\quad\quad\quad\quad -- Compute cross-fitted residuals $R_X, R_Y$ and residual covariance $\bar{C}$. \\
\quad\quad\quad\quad -- Evaluate lower bound $L_{C,a}$ and certified leakage envelope $\mathcal{T}_a^{(\star)}$ (Theorem~\ref{thm:conditional_test}). \\
\quad\quad\quad\quad -- If $L_{C,a} \le \mathcal{T}_a^{(\star)}$, mark edge $\{i, j\}$ for deletion, record $S \in \mathrm{SepSet}(i, j)$, and move to next edge. \\
\quad\quad 3. Remove all marked edges simultaneously: $G_{\ell+1} \leftarrow G_\ell \setminus \{\text{marked edges}\}$, and set $\ell \leftarrow \ell + 1$. \\[0.35em]
\textbf{Step 3: Conservative Collider Orientation} \\
\quad For each unshielded triple $i - k - j$ (where $\{i, j\} \notin E$): \\
\quad\quad $\bullet$ If $k \notin S$ for all recorded separating sets $S \in \mathrm{SepSet}(i, j)$, orient as collider: $i \to k \leftarrow j$. \\
\quad\quad $\bullet$ If $k \in S$ for all recorded separating sets, mark $i - k - j$ as noncollider. \\
\quad\quad $\bullet$ If separating sets conflict, mark triple ambiguous to prevent error cascading. \\[0.35em]
\textbf{Step 4: Observational Meek Orientation Propagation} \\
\quad Iteratively apply Meek rules (R1--R4) to orient remaining undirected edges into a consistent CPDAG $\widehat{\mathcal{G}}$. \\
\bottomrule
\end{tabular}
\end{table}

To summarize sensitivity across candidate coordinate noise levels, we define a grid-based Edge Sensitivity Threshold.

\begin{definition}[Robustness Value]
\label{def:robustness_value}
Let $\mathcal{S}_{\mathrm{grid}} \subset (0, \infty)$ be an ordered prespecified grid of candidate coordinate-wise SNR lower bounds, and map each $s \in \mathcal{S}_{\mathrm{grid}}$ to a query-specific directional bound $s_S(s)$ (e.g., via~\eqref{eq:observable_loewner_formula}). For a pair $\{i, j\}$ and a prespecified family of conditioning sets $\mathcal{S}_{ij}$, the empirical Edge Sensitivity Threshold $\widehat{\mathrm{RV}}_{ij}$ is the least $s \in \mathcal{S}_{\mathrm{grid}}$ for which the marginal permutation test rejects at level $\alpha$ and every nonempty query $S \in \mathcal{S}_{ij}$ rejects via $L_{C,S} > U_{\Delta, S}(s_S(s)) + a_{S,n}$; if no grid point satisfies this condition, report ``no rejection threshold on the grid.''
\end{definition}

A low value of $\widehat{\mathrm{RV}}_{ij}$ is a descriptive sensitivity summary rather than an estimate of measurement error; reporting whether $\widehat{\mathrm{RV}}_{ij}$ is externally certified is essential, and a simultaneous confidence interpretation requires joint coverage over both the query family and the grid $\mathcal{S}_{\mathrm{grid}}$.

\subsection{Theoretical Guarantees: Deterministic Recovery, Uniform Query Conditions, Stability, and Multiplicity}
\label{sec:guarantees_pc}

The structural target is $\operatorname{CPDAG}(\mathcal{G}^*)$ for a specified latent DAG $\mathcal{G}^* = ([p], E^*)$ to which the latent distribution $P_{\mathbf{V}^*}$ is Markov. We do not define a DAG from residual-moment zeros: a system of zero residual covariances need not form a DAG independence model (e.g., $W = X^2 + \eta$ has zero covariance despite dependence), and even in jointly Gaussian laws—where zero latent affine residual covariance equals latent CI—graph faithfulness is an additional assumption beyond Gaussianity (for instance, in the triangular DAG $X \to Z, X \to Y, Z \to Y$ with $Z = X + \eta$ and $Y = X - Z + \xi$, exact cancellation yields $X \indep Y$ despite adjacency).

\begin{remark}
\label{rem:latent_cpdag_vs_mag}
A foundational question in causal discovery under measurement corruption is why the algorithm targets a Completed Partially Directed Acyclic Graph ($\operatorname{CPDAG}$) rather than a Maximal Ancestral Graph ($\mathrm{MAG}$) or Partial Ancestral Graph ($\mathrm{PAG}$) designed for latent confounders (such as FCI \cite{spirtes2000causation, zhang2008completeness}). In general graphical modeling, unmeasured common confounders induce bidirected edges ($\leftrightarrow$) and violate causal sufficiency, necessitating ancestral graph representations. However, under Assumption~\ref{ass:classical_error}, the measurement corruptions $\epsilon_1, \dots, \epsilon_p$ are \emph{mutually independent idiosyncratic noises} unique to each sensor channel; they do not represent shared unmeasured confounders acting simultaneously on multiple variables. Consequently, the true causal mechanisms governing the uncorrupted physical or biological variables $\mathbf{V}^*$ form a genuine DAG $\mathcal{G}^*$. The observed measurements $\mathbf{V}$ serve merely as noisy coordinate-wise surrogates for the latent target vertices $\mathbf{V}^*$. By adjusting conditional independence thresholds for sensor leakage, NA-PC aims directly at discovering the causal skeleton and orientations of the underlying latent scientific system $\operatorname{CPDAG}(\mathcal{G}^*)$, rather than fitting an observational ancestral graph on measurement artifacts.
\end{remark}

\paragraph{Conservative Meek Orientation Safeguard and Degree Bounds.}
Two practical considerations govern the transition from learned skeletons to completed CPDAGs:
\begin{enumerate}[label=(\alph*),leftmargin=2.0em]
    \item \textbf{Safeguard Against Cascading Orientation Errors:} Because sensitivity-adjusted testing prunes edges conservatively, a small fraction of weak true edges may be omitted from the skeleton. In classical PC, missing edges can deceive deterministic Meek rules into cascading false orientations. NA-PC incorporates a \emph{Conflict-Aware Conservative Orientation Safeguard} (cf.\ Colombo \& Maathuis \cite{colombo2014order}): unshielded triples $i - k - j$ with conflicting separating sets are flagged as ambiguous rather than forced into colliders, preventing orientation errors from propagating.
    \item \textbf{Degree Bound for Asymptotic Consistency:} Complete CPDAG consistency formally holds for the class of DAGs whose maximal active conditioning sets satisfy $\max_{j} |\mathrm{pa}(j)| \le d_{\max}$. In denser graphs with in-degrees exceeding $d_{\max}$, setting $d_{\max} \in \{1, 2\}$ operates as a high-precision conservative parental filter: it guarantees exact screening of all first- and second-order parent sets while shielding the algorithm from the curse of conditioning cardinality.
\end{enumerate}

Let $\mathcal{Q}_n \triangleq \{(i, j, S) : 1 \le i < j \le p_n, \, S \subset [p_n] \setminus \{i, j\}, \, |S| \le d_{\max}\}$ denote the candidate query family up to depth $d_{\max}$. For $p_n \ge 2$, $|\mathcal{Q}_n| = \binom{p_n}{2}\sum_{k=0}^{\min(d_{\max}, p_n - 2)} \binom{p_n - 2}{k} \le p_n^{d_{\max} + 2}$. Let $\mathcal{Q}_{0,n} \subset \mathcal{Q}_n$ be the subset of queries where $S$ d-separates $i$ and $j$ in $\mathcal{G}^*$, and $\mathcal{Q}_{\mathrm{conn},n} \triangleq \mathcal{Q}_n \setminus \mathcal{Q}_{0,n}$ the subset of d-connected queries. We first state the exact deterministic recovery theorem.

\begin{theorem}[Oracle PC Recovery]
\label{thm:pc_oracle_recovery}
Let the latent law $P_{\mathbf{V}^*}$ be Markov with respect to a DAG $\mathcal{G}^* = ([p], E^*)$ of maximum skeleton degree at most $d_{\max}$. Apply the two-ended frozen-neighborhood search and conservative orientation procedure of Algorithm~2 (Table~\ref{alg:napc}). Suppose that with probability $1 - o(1)$, every query $a = (i, j, S) \in \mathcal{Q}_n$ with $|S| \le d_{\max}$ is classified as independent if and only if $S$ d-separates $i$ and $j$ in $\mathcal{G}^*$. Then:
\begin{equation}
    \label{eq:consistency_result}
    \lim_{n \to \infty} \mathbb{P}\left(\widehat{\mathcal{G}}_n = \operatorname{CPDAG}(\mathcal{G}^*)\right) = 1.
\end{equation}
\end{theorem}
\noindent\textit{Proof.} Deferred to Appendix~\ref{app:main_proofs}.

We now state explicit high-level uniform statistical conditions under which the simultaneous correctness event of Theorem~\ref{thm:pc_oracle_recovery} holds in growing dimensions $p_n$, together with the failure condition for an unadjusted zero-floor covariance search.

\begin{assumption}[Regularity Conditions]
\label{ass:residual_moment_faithfulness}
For a latent DAG $\mathcal{G}^* = ([p_n], E^*)$ with $\max_i \deg_{\mathcal{G}^*}(i) \le d_{\max}$ observed under Assumptions~\ref{ass:classical_error} and~\ref{ass:snr_certificate}:
\begin{enumerate}[label=(\roman*),leftmargin=2.0em]
    \item \textit{Nonempty Query Separation and Joint Concentration Event:} For every nonempty d-connected query $a = (i, j, S) \in \mathcal{Q}_{\mathrm{conn}, n}$ ($1 \le |S| \le d_{\max}$), assume $|C_a| - \Delta_a \ge \kappa_n > 0$. Assume there exists a joint event of probability $1 - o(1)$ on which simultaneously, for all nonempty $a \in \mathcal{Q}_n$:
    \begin{equation}
        \label{eq:napc_uniform_joint_event}
        |\bar{C}_a - C_a| \le r_{a,n}, \qquad \Delta_a \le U_{a,n}, \qquad \sup_{a \in \mathcal{Q}_n,\,|S|\ge 1}\left(2 r_{a,n} + U_{a,n} - \Delta_a + a_{a,n}\right) = o(\kappa_n),
    \end{equation}
    where $r_{a,n}$ is the critical radius (e.g., $z_{1-\alpha_{1,n}/2}\hat{\sigma}_{C,a}/\sqrt{n}$ or a finite-sample confidence radius), $U_{a,n}$ is the upper envelope bound, and $a_{a,n} \ge 0$ is the specified approximation allowance. (Establishing~\eqref{eq:napc_uniform_joint_event} for a specific nuisance learner and envelope UCB is a separate statistical requirement, guaranteed under standard moderate-deviation and sub-exponential Bernstein conditions.)
    \item \textit{Simultaneous Marginal Query Correctness:} For the marginal pass ($\ell = 0$) at significance level $\alpha_n$ with $|\mathcal{Q}_n|\alpha_n \to 0$ and with-replacement permutation budget $\lfloor (B_n + 1)\alpha_n \rfloor \ge 1$, assume the marginal statistic (for example, biased $\widehat{\HSIC}_b$ with fixed uniformly bounded kernels and a uniform positive population HSIC gap across all marginally d-connected pairs) satisfies a uniform permuted-comparison bound $\sup_{a \in \mathcal{Q}_{\mathrm{conn},n},\, S=\emptyset} \mathbb{P}(T_a^{\pi} \ge T_a) \le A e^{-cn}$ with $\log(|\mathcal{Q}_n| / \alpha_n) = o(n)$.
\end{enumerate}
\end{assumption}

\paragraph{Concrete Instantiation for Linear-Gaussian Ridge Search.}
Under a linear-Gaussian DAG model with cross-fitted ridge regression (regularization parameter $\lambda_n = o(n^{-1/2})$), sub-Gaussian disturbances, and bounded degree $d_{\max}$, standard sub-Gaussian concentration implies that the joint concentration condition of Assumption~\ref{ass:residual_moment_faithfulness}(i) holds with critical radius $r_{a,n} = \mathcal{O}(\sqrt{d_{\max}\log(p_n/\alpha_n)/n})$ and envelope slack $U_{a,n} - \Delta_a = \mathcal{O}(1/s_0)$. The required uniform margin condition $\sup_{a}(2r_{a,n} + U_{a,n} - \Delta_a + a_{a,n}) = o(\kappa_n)$ is satisfied provided the sample size scales as $n \gg d_{\max}\log(p_n)/\kappa_n^2$ and the certified SNR satisfies $s_0 \gg 1/\kappa_n$. We designate the separation condition $|C_a| - \Delta_a \ge \kappa_n > 0$ as \emph{Residual-Covariance Faithfulness}; it holds generically in linear-Gaussian systems without fine-tuned parameter cancellation, but can fail in nonlinear or deterministic cancellation settings.

\begin{theorem}[NA-PC Consistency]
\label{thm:napc_skeleton_consistency}
Under Assumptions~\ref{ass:classical_error}, \ref{ass:snr_certificate}, and~\ref{ass:residual_moment_faithfulness}, the following asymptotic properties hold:
\begin{enumerate}[label=(\roman*),leftmargin=2.0em]
    \item \textbf{Failure of an Unadjusted Zero-Floor Covariance PC Search:} Consider an unadjusted PC search with depth cap $d_{\max}$ using a specified zero-floor residual-covariance test for nonempty queries. If there exists at least one latent nonedge $\{i, j\} \notin E_{\mathrm{skel}}^*$ such that $|C_a| \ge \kappa_{\mathrm{leak}} > 0$ for every nonempty query $a = (i, j, S)$ on that pair ($1 \le |S| \le d_{\max}$), all corresponding estimation errors and cutoff radii are uniformly $o_P(1)$, and its marginal query rejects with probability $1 - o(1)$, then with probability $1 - o(1)$ no query removes $\{i, j\}$, and hence
    \begin{equation*}
        \lim_{n \to \infty}\mathbb{P}\left(\widehat{\mathcal{G}}_n^{\mathrm{naive}} = \operatorname{CPDAG}(\mathcal{G}^*)\right) = 0.
    \end{equation*}
    \item \textbf{High-Dimensional CPDAG Consistency of NA-PC:} Under Assumptions~\ref{ass:classical_error},~\ref{ass:snr_certificate}, and~\ref{ass:residual_moment_faithfulness}, rejecting each nonempty query when $|\bar{C}_a| - r_{a,n} > U_{a,n} + a_{a,n}$ ensures simultaneous correctness across $\mathcal{Q}_n$ with probability $1 - o(1)$, and consequently Algorithm~2 satisfies $\lim_{n \to \infty}\mathbb{P}(\widehat{\mathcal{G}}_n = \operatorname{CPDAG}(\mathcal{G}^*)) = 1$.
\end{enumerate}
\end{theorem}
\noindent\textit{Proof.} Deferred to Appendix~\ref{app:main_proofs}.

Note that the shrinking schedule $\alpha_n \to 0$ with $B_n \ge \lceil \alpha_n^{-1} \rceil - 1$ is a sufficient asymptotic condition for simultaneous consistency (requiring $\Omega(p_n^{d_{\max}+2} n^\gamma)$ permutation draws unless an alternative exact calibration is used), while in practical finite-sample implementations a fixed nominal level $\alpha \in \{0.01, 0.05\}$ is standard.

\begin{proposition}[Skeleton Stability]
\label{lem:skeleton_stability}
Assume each query envelope $\Delta_a(s)$ is nonincreasing and uniformly $L$-Lipschitz in $s \in [s_{\min}, s_0]$. For two externally certified values $s_0$ and $s_0'$ with $s_0' \ge s_{\min}$, set $\delta_n \triangleq (s_0 - s_0')_+$. If $L \delta_n \le \kappa_n / 2$ and the joint decision assumptions of Theorem~\ref{thm:napc_skeleton_consistency}(ii) hold at both $s_0$ and $s_0'$ (with $s_0'$ fixed, or uniformly over allowed values if $s_0'$ is data-selected independently), then both skeletons recover $\mathcal{G}_{\mathrm{skel}}^*$ with probability $1 - o(1)$, and hence $\lim_{n\to\infty}\mathbb{P}(\widehat{\mathcal{C}}(s_0') = \widehat{\mathcal{C}}(s_0)) = 1$.
\end{proposition}
\noindent\textit{Proof.} Deferred to Appendix~\ref{app:main_proofs}.

\begin{proposition}[Multiplicity Control]
\label{prop:multiplicity_fwer_fdr}
Consider a prespecified family of $m$ queries with $m_0$ true nulls. Suppose that for each query $a$, a nested family of level-$u$ rejection regions $\mathcal{R}_a(u)$ ($u \in (0, 1)$) defines the $p$-value $p_a \triangleq \inf\{u \in (0, 1) : \mathcal{D}_n \in \mathcal{R}_a(u)\}$ (assigning $p_a \triangleq 1$ if the set is empty). Whenever each null region satisfies $\mathbb{P}_{H_{0,a}}(\mathcal{D}_n \in \mathcal{R}_a(u)) \le u$ for all $u \in (0, 1)$ (as holds in finite samples for Regime~A permutations via Theorem~\ref{thm:marginal_test}(i)), each null $p_a$ is super-uniform ($\{p_a \le u\} \subseteq \mathcal{R}_a(u + \varepsilon)$ for all $0 < \varepsilon < 1-u$, letting $\varepsilon \downarrow 0$). Applying standard adjustments to this prespecified family of valid $p$-values yields:
\begin{enumerate}[label=(\alph*)]
    \item Bonferroni and Holm step-down \cite{holm1979simple} control $\mathrm{FWER} \le \alpha$ under arbitrary dependence among the $m$ queries.
    \item Benjamini--Yekutieli \cite{benjamini2001control} controls $\mathrm{FDR} \le \alpha$ under arbitrary dependence; Benjamini--Hochberg \cite{benjamini1995controlling} controls $\mathrm{FDR} \le \frac{m_0}{m}\alpha \le \alpha$ only under independence or positive regression dependency on the subset of true nulls (PRDS).
\end{enumerate}
Pointwise asymptotic validity of $p_a$ yields only fixed-$m$ limiting guarantees unless uniform tail bounds hold across all step-down/step-up cutoffs. Because PRDS is not established for overlapping conditional queries sharing variables or regressions, BH results in the Appendix are presented as empirical FDR evaluations, and none of the query-level statements above constitutes graph-level edge FDR control.
\end{proposition}
\noindent\textit{Proof.} Deferred to Appendix~\ref{app:main_proofs}.

\begin{remark}
\label{rem:pitman_power}
For a signed root-$n$ covariance statistic with local signal $C_n = c/\sqrt{n}$ and $\sqrt{n}(\bar{C} - C_n)/\hat{\sigma}_C \xrightarrow{d} \mathcal{N}(0, 1)$ ($\hat{\sigma}_C \xrightarrow{P} \sigma > 0$), a two-sided zero-floor normal rule has limiting local power $1 - \Phi(z_{1-\alpha/2} - |c|/\sigma) + \Phi(-z_{1-\alpha/2} - |c|/\sigma) > \alpha$ (this applies to a signed root-$n$ covariance statistic, not to degenerate $\HSIC$ or $\dCov$ permutation tests). In contrast, for the sensitivity-adjusted conditional rule of Theorem~\ref{thm:conditional_test} with $\Delta_n \to \Delta > 0$ and $U_{\Delta, n} - \Delta_n = o_P(1)$, any local sequence $C_n = \mathcal{O}(n^{-1/2})$ has rejection probability converging to zero under the null envelope. This vanishing local power is a property of the specific positive-envelope covariance rule; separately, the Gaussian observational-equivalence example in Section~\ref{sec:intro} demonstrates that a single-sample scalar SNR certificate does not point-identify latent conditional independence.
\end{remark}

\section{Empirical Evaluation}
\label{sec:experiments}

We evaluate the proposed procedures across three tiers: (i) Tier~1 unit tests of marginal permutation calibration and head-to-head conditional independence tests; (ii) Tier~2 synthetic structure-learning evaluation across $50$ linear-Gaussian random DAGs; and (iii) Tier~3 an uncertified sensitivity sweep on real single-cell flow cytometry data \cite{sachs2005causal}. Supplementary diagnostic benchmarks—including split-sample MKL-HSIC calibration, non-monotone confounding heuristics, conditioning-cardinality scaling, DREAM4 Hill-kinetics networks, and atmospheric sensor simulations—are documented in the supplementary material.

\paragraph{Experimental protocol and status of empirical implementations.}
We distinguish three quantities throughout: the coordinate generating $\SNR$ used in simulations, the directional Loewner bound $s_S$ certified for each query under Assumption~\ref{ass:snr_certificate}, and an uncertified user-supplied sensitivity input. They are not interchangeable. Results from fitted-curvature heuristics, deliberately misspecified inputs, and pooled interventional regimes are descriptive rather than proofs of theoretical size control. Across multivariate PC-stable runs, $\hat{m}_j(\mathbf{S})$ is instantiated as a 2-fold cross-fitted ridge-stabilized linear projection (ridge penalty $10^{-4}$ on standardized covariates) with influence-function standard error estimation; where an empirical benchmark uses a plug-in influence-function UCB or fitted curvature heuristic rather than the exact Bonferroni raw-moment supremum box of Remark~\ref{rem:ucb_construction}, we state this explicitly. Each simulation specifies its DGP, sample size, Monte Carlo or DAG replication count, random seed, and uncertainty unit (Monte Carlo plug-in SE for within-DGP trials, DAG-level SE across independent DAG realizations, or empirical subsample SD for overlapping cytometry subsamples).

\subsection{Marginal Permutation Calibration and Summary of Tier 1 Unit Tests}
We evaluate empirical null rejection and alternative power across coordinate signal-to-noise levels $\SNR \in \{0.5, 1.0, 2.0, 5.0, \infty\}$ ($n = 200$, $\alpha = 0.05$, $N_{\mathrm{mc}} = 1{,}000$ independent datasets per cell, random seed $101$, $B = 99$ permutations). Under $H_0$, $X^*, Y^* \stackrel{\mathrm{i.i.d.}}{\sim} \mathcal{N}(0, 1)$; under $H_1$, $X^* \sim \mathcal{U}[-\pi, \pi]$ and $Y^* = \sin(2X^*) + U_Y$ with $U_Y \sim \mathcal{N}(0, 0.4^2)$, using independent Gaussian measurement errors $\epsilon_X, \epsilon_Y \sim \mathcal{N}(0, 1/\SNR)$. Uncentered normal approximations for raw $\widehat{\dCor}$ and $\widehat{\HSIC}_b$ are included strictly as negative controls (Remark~\ref{rem:normal_breakdown_pointer}).

\begin{figure}[t!]
\centering
\includegraphics[width=0.74\textwidth]{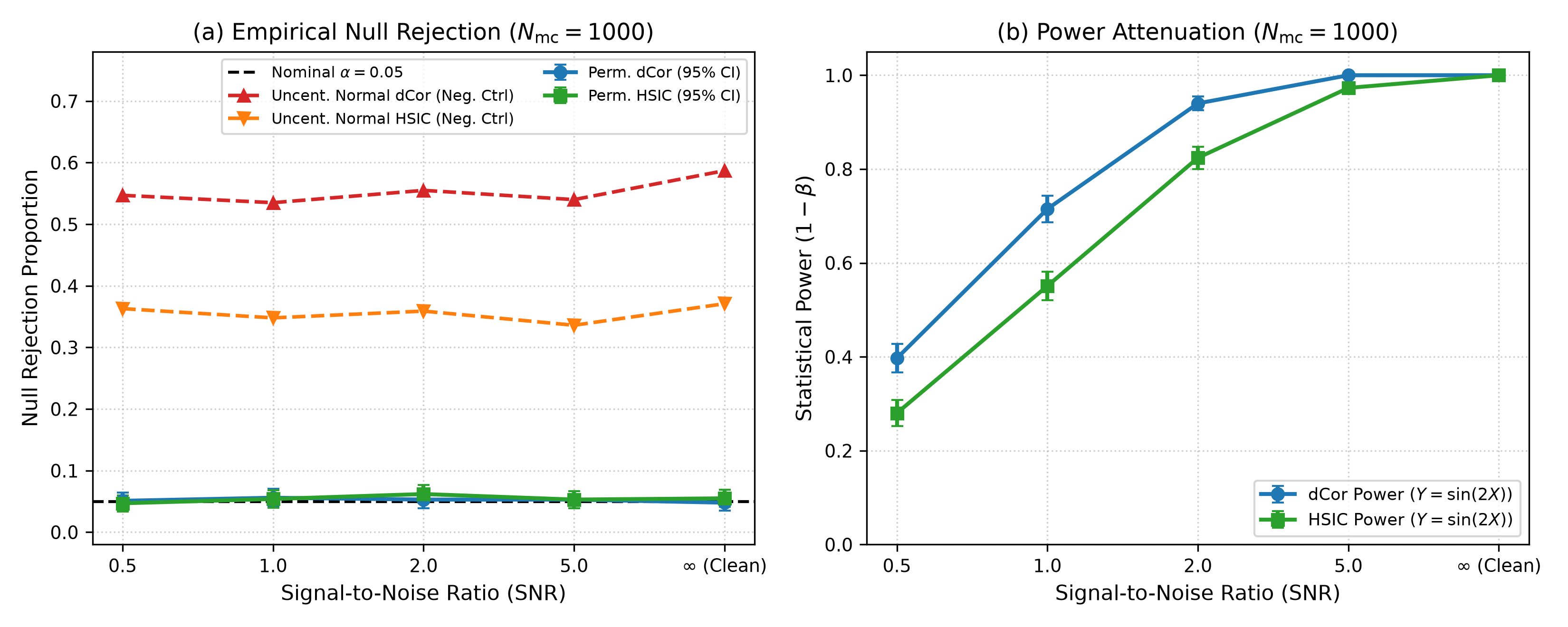}
\caption{Empirical null rejection proportions across the displayed noise settings ($n=200$, $1{,}000$ simulated datasets per setting, nominal level $0.05$, with plug-in Wald 95\% error bars). The uncentered Gaussian versions are negative controls.}
\label{fig:typeI_eval}
\end{figure}

\begin{table}[t!]
\centering
\footnotesize
\vspace{-0.5em}
\caption{Marginal null rejection and power at $n=200$, $\alpha=0.05$, $1{,}000$ simulated datasets per reported cell, and $B=99$ permutations for the permutation methods. Parenthesized plug-in Monte Carlo standard errors are reported for the two permutation null-rejection columns only.}
\label{tab:type1_power}
\resizebox{\linewidth}{!}{%
\setlength{\tabcolsep}{4.5pt}%
\begin{tabular}{lcccccc}
\toprule
Noise Level ($\SNR$) & Perm.\ $\dCor$ & Perm.\ $\HSIC$ & Uncent.\ $\dCor$ (Neg.\ Ctrl) & Uncent.\ $\HSIC$ (Neg.\ Ctrl) & Power ($\dCor$) & Power ($\HSIC$) \\
\midrule
$\SNR = 0.5$ & $0.051\,(0.007)$ & $0.047\,(0.007)$ & $0.547$ & $0.363$ & $0.397$ & $0.280$ \\
$\SNR = 1.0$ & $0.056\,(0.007)$ & $0.054\,(0.007)$ & $0.535$ & $0.348$ & $0.715$ & $0.551$ \\
$\SNR = 2.0$ & $0.053\,(0.007)$ & $0.062\,(0.008)$ & $0.555$ & $0.359$ & $0.940$ & $0.824$ \\
$\SNR = 5.0$ & $0.053\,(0.007)$ & $0.053\,(0.007)$ & $0.540$ & $0.336$ & $1.000$ & $0.973$ \\
$\SNR = \infty$ (Clean) & $0.048\,(0.007)$ & $0.055\,(0.007)$ & $0.587$ & $0.371$ & $1.000$ & $1.000$ \\
\bottomrule
\end{tabular}%
}
\end{table}

As shown in Figure~\ref{fig:typeI_eval} and Table~\ref{tab:type1_power}, the permutation null rejection proportions range from $0.047$ to $0.062$ with plug-in Monte Carlo standard errors $0.0067$--$0.0076$; the largest estimate ($0.062$) exceeds $0.05$ by about $1.57$ of its own plug-in standard errors ($1.74$ SE under nominal $p = 0.05$). A reported power of $1.000$ means all $1{,}000$ simulated datasets rejected in that cell (exact two-sided 95\% lower bound $\approx 0.9963$), not that population power equals $1$. The uncentered Gaussian negative controls reject $33.6\%$--$58.7\%$ of nulls, illustrating the invalid calibration of uncentered normal approximations on non-negative V-statistics under the null.

\paragraph{Supplementary Tier~1 diagnostic benchmarks.}
Supplementary unit tests documented in the appendix further examine: (i) population and sample Pearson correlation attenuation under classical additive error; (ii) split-sample MKL-HSIC, where sample splitting preserves nominal size while raising test power under high-frequency alternatives (preventing post-selection inflation), alongside discrete category misclassification deattenuation; (iii) kernel-ridge regression heuristics under quadratic non-monotone confounding; and (iv) d-separation scaling across conditioning cardinalities $|Z| \in \{1, 2, 3, 4\}$, illustrating that the sensitivity-adjusted test strictly controls false positive rejections across conditioning dimensions at the cost of power declining with larger conditioning sets.

\subsection{Head-to-Head Conditional Tests (Linear EIV and Nonlinear \texorpdfstring{$\tanh$}{tanh} DAGs)}
\label{sec:benchmark_competing_tests}
We compare six testing procedures across $20$ independent random DAGs per regime ($p = 12$, edge probability $0.30$, weights in $\pm[0.4, 0.8]$, $n = 1{,}200$, coordinate generating $\SNR = 2.0$, $\alpha = 0.05$, $60$ d-separated and $60$ d-connected queries per DAG, totaling $1{,}200$ queries per class per regime; seeds $300 + 41t$): (a) Linear-Gaussian EIV DAGs, and (b) Nonlinear $\tanh$ DAGs where $40\%$ of parent-child edges apply $\tanh$. In the nonlinear $\tanh$ regime, descendants of $\tanh$ maps are not Gaussian (by Cram\'er's decomposition, since $\tanh(U)$ is bounded and nonconstant), and d-connected queries are not independently verified to have nonzero population residual covariance. Consequently, we label both single-sample NA-GCM variants in the nonlinear regime as \emph{linear-surrogate heuristic} and \emph{fitted-curvature heuristic}, without asserting Gaussian-conditioning or latent-curvature certification. The random-feature baseline (\emph{RFF-CIT}) is a 20-feature random Fourier approximation applied to linear residuals (not full-kernel KCIT \cite{zhang2011kernel}); duplicate Replicate GCM operates under a richer duplicate-measurement information regime. Numerical results are summarized in Table~\ref{tab:competing_ci_tests} (with graphical error-bar comparisons detailed in the supplementary appendix).

\begin{table}[!htbp]
\centering
\caption{Head-to-head surrogate tests and duplicate-measurement comparator in linear and nonlinear $\tanh$ simulation designs ($20$ DAGs per regime, $n=1{,}200$, $p=12$, coordinate generating $\SNR=2.0$, $\alpha=0.05$, $1{,}200$ queries per class per regime). Table entries report DAG-level mean and standard error; balanced accuracy is the average of specificity and alternative rejection frequency. Comparisons are partitioned strictly by information regime: Panel~A evaluates single-sample observational methods, while Panel~B reports the duplicate-measurement comparator.}
\label{tab:competing_ci_tests}
\resizebox{\linewidth}{!}{%
\footnotesize\setlength{\tabcolsep}{4pt}%
\begin{tabular}{llcccc}
\toprule
Regime & Method & Null Rej.\ (FP) & Specificity (TN) & Alt.\ Rej.\ (TP) & Balanced Acc. \\
\midrule
\multicolumn{6}{l}{\textbf{Panel A: Single-Sample Information Regime}} \\
Linear EIV & Fisher-Z \cite{spirtes2000causation} & $15.0\%\,(2.9\%)$ & $85.0\%\,(2.9\%)$ & $78.8\%\,(1.8\%)$ & $81.9\%\,(1.6\%)$ \\
Linear EIV & Standard GCM \cite{shah2020hardness} & $15.2\%\,(2.9\%)$ & $84.8\%\,(2.9\%)$ & $78.5\%\,(1.7\%)$ & $81.7\%\,(1.6\%)$ \\
Linear EIV & Residual $\dCor$ \cite{szekely2013energy} & $13.8\%\,(2.6\%)$ & $86.2\%\,(2.6\%)$ & $76.0\%\,(1.7\%)$ & $81.1\%\,(1.5\%)$ \\
Linear EIV & RFF-CIT (20-Feature Approx.) & $14.7\%\,(2.8\%)$ & $85.3\%\,(2.8\%)$ & $78.8\%\,(1.8\%)$ & $82.0\%\,(1.6\%)$ \\
Linear EIV & NA-RCM (Linear Envelope, Single) & $\mathbf{0.4\%\,(0.2\%)}$ & $\mathbf{99.6\%\,(0.2\%)}$ & $50.1\%\,(2.2\%)$ & $74.8\%\,(1.1\%)$ \\
\addlinespace[0.3em]
Nonlin.\ $\tanh$ & Fisher-Z \cite{spirtes2000causation} & $10.4\%\,(1.7\%)$ & $89.6\%\,(1.7\%)$ & $73.8\%\,(2.2\%)$ & $81.7\%\,(1.3\%)$ \\
Nonlin.\ $\tanh$ & Standard GCM \cite{shah2020hardness} & $10.4\%\,(1.6\%)$ & $89.6\%\,(1.6\%)$ & $73.8\%\,(2.2\%)$ & $81.7\%\,(1.3\%)$ \\
Nonlin.\ $\tanh$ & Residual $\dCor$ \cite{szekely2013energy} & $9.4\%\,(1.4\%)$ & $90.6\%\,(1.4\%)$ & $72.6\%\,(2.1\%)$ & $81.6\%\,(1.3\%)$ \\
Nonlin.\ $\tanh$ & RFF-CIT (20-Feature Approx.) & $11.1\%\,(1.6\%)$ & $88.9\%\,(1.6\%)$ & $74.7\%\,(2.2\%)$ & $81.8\%\,(1.3\%)$ \\
Nonlin.\ $\tanh$ & NA-RCM (Linear-Surrogate Sensitivity) & $\mathbf{1.2\%\,(0.5\%)}$ & $\mathbf{98.8\%\,(0.5\%)}$ & $52.8\%\,(2.0\%)$ & $75.8\%\,(1.0\%)$ \\
Nonlin.\ $\tanh$ & NA-RCM (Fitted-Curvature Diagnostic)$^\dagger$ & $0/1{,}200$ & $100.0\%$ & $6.7\%\,(0.8\%)$ & $53.3\%\,(0.4\%)$ \\
\midrule
\multicolumn{6}{l}{\textbf{Panel B: Duplicate-Measurement Regime (Expanded Information Set)}} \\
Linear EIV & Replicate GCM & $2.8\%\,(0.8\%)$ & $97.2\%\,(0.8\%)$ & $75.5\%\,(1.5\%)$ & $86.3\%\,(0.9\%)$ \\
Nonlin.\ $\tanh$ & Replicate GCM & $2.8\%\,(0.8\%)$ & $97.2\%\,(0.8\%)$ & $71.6\%\,(2.2\%)$ & $84.4\%\,(1.2\%)$ \\
\bottomrule
\end{tabular}%
}
\vspace{0.3em}
\par\raggedright\footnotesize $^\dagger$The fitted-curvature diagnostic demonstrates envelope explosion when empirical RKHS norms are used in place of an external certificate.
\end{table}

In Table~\ref{tab:competing_ci_tests}, uncalibrated single-sample tests exhibit $13.8\%$--$15.2\%$ null rejection proportions on linear EIV DAGs and $9.4\%$--$11.1\%$ on nonlinear $\tanh$ DAGs. In the linear EIV regime, single-sample NA-GCM yields $0.4\% \pm 0.2\%$ null rejections with $50.1\% \pm 2.2\%$ alternative rejections, whereas duplicate Replicate GCM attains $2.8\% \pm 0.8\%$ null rejections and $86.3\% \pm 0.9\%$ balanced accuracy. In the nonlinear $\tanh$ ensemble, the fitted-curvature heuristic produced $0/1{,}200$ null rejections and rejected $6.7\% \pm 0.8\%$ of sampled d-connected queries, while the linear-surrogate heuristic yielded $1.2\% \pm 0.5\%$ null rejections and $52.8\% \pm 2.0\%$ alternative rejections; we make no size guarantee or universal SNR-rate attribution for the nonlinear heuristic variants.

\subsection{Synthetic Structure Learning: 50 Random Linear-Gaussian DAGs}
We simulate $50$ independent linear-Gaussian random DAGs ($p = 10$, edge probability $0.25$ conditioned on $\ge 5$ true edges, weights in $\pm[0.35, 0.75]$, standardized latent variables, $n = 1{,}000$, seed $303$) under independent Gaussian noise at coordinate $\SNR \in \{\infty, 5.0, 2.0, 1.0\}$ (plus technical duplicates for Replicate NA-PC). We compare five configurations (including our two proposed variants) with depth cap $d_{\max} = 2$ and Meek orientation rules R1--R4: Standard PC ($\alpha = 0.05$), PC-stable ($\alpha = 0.005$) (a low-$\alpha$ threshold baseline, distinct from the conservative collider orientation rule of \cite{colombo2014order}), Covariance-corrected PC (which subtracts known error variance from the sample covariance before partial-correlation testing), Single-Sample NA-PC, and Replicate NA-PC.

\begin{figure}[!htbp]
\centering
\includegraphics[width=0.96\textwidth]{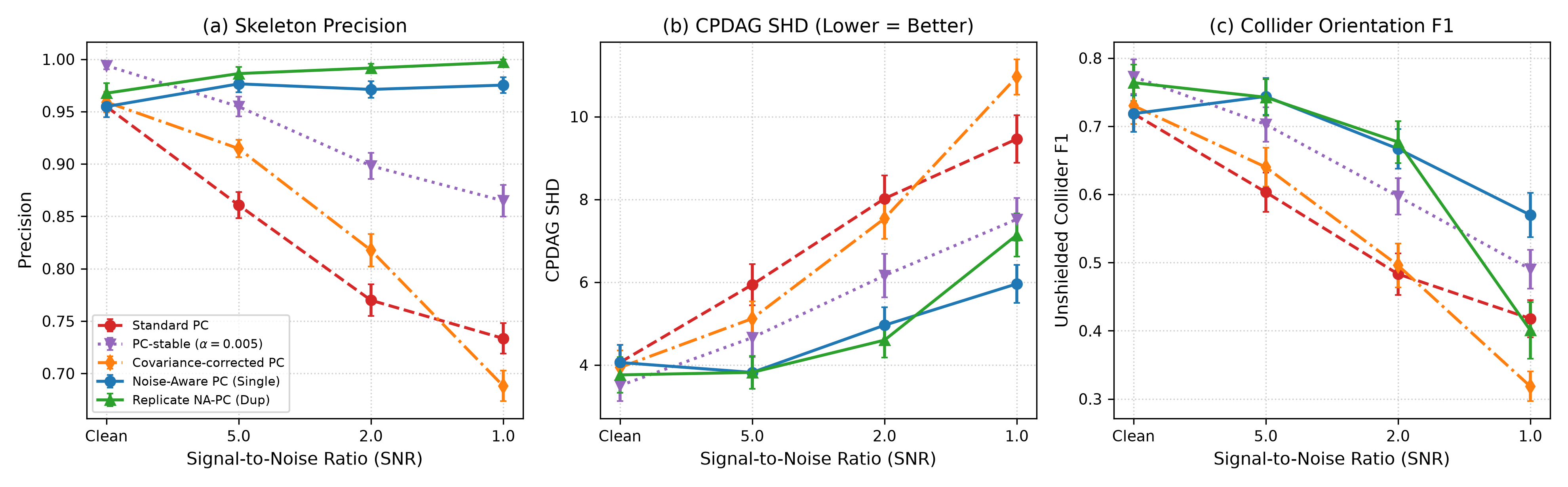}
\caption{Skeleton precision, CPDAG SHD, and collider F1 across $50$ synthetic linear-Gaussian DAGs ($p=10$, $n=1{,}000$, DAG-level mean $\pm$ SE).}
\label{fig:dag_eval}
\end{figure}

\begin{table}[!htbp]
\centering
\caption{Skeleton and orientation performance across $50$ synthetic linear-Gaussian DAGs ($p=10$, $n=1{,}000$, $d_{\max}=2$, Meek R1--R4). Numbers in parentheses are DAG-level standard errors for structural metrics. Configurations share identical DAG draws, making comparisons paired; at $\SNR=1.0$, the CPDAG SHD reduction of NA-PC relative to Standard PC ($3.50$ edges) and PC-stable ($1.56$ edges) is statistically significant under a paired $t$-test ($p < 10^{-4}$ and $p=0.003$).}
\label{tab:synthetic_bench}
\resizebox{\linewidth}{!}{%
\footnotesize\setlength{\tabcolsep}{3.5pt}%
\begin{tabular}{llccccccc}
\toprule
$\SNR$ & Configuration & Precision & Recall & Skeleton F1 & Skel.\ SHD & Collider F1 & CPDAG SHD & Time (ms) \\
\midrule
$\infty$ (Clean) & Standard PC ($\alpha=0.05$) & $0.955\,(0.010)$ & $0.952\,(0.009)$ & $0.952\,(0.008)$ & $1.10\,(0.18)$ & $0.719\,(0.027)$ & $4.06\,(0.42)$ & $154.6$ \\
$\infty$ & PC-stable ($\alpha=0.005$) & $0.994\,(0.004)$ & $0.937\,(0.011)$ & $0.963\,(0.007)$ & $0.88\,(0.17)$ & $0.773\,(0.026)$ & $3.50\,(0.38)$ & $124.2$ \\
$\infty$ & Covariance-corrected PC & $0.959\,(0.010)$ & $0.952\,(0.009)$ & $0.953\,(0.007)$ & $1.04\,(0.17)$ & $0.730\,(0.027)$ & $3.94\,(0.41)$ & $57.1$ \\
$\infty$ & Noise-Aware PC (Single) & $0.955\,(0.010)$ & $0.952\,(0.009)$ & $0.952\,(0.008)$ & $1.10\,(0.18)$ & $0.719\,(0.027)$ & $4.06\,(0.42)$ & $152.8$ \\
$\infty$ & Replicate NA-PC (Dup) & $0.968\,(0.010)$ & $0.944\,(0.010)$ & $0.953\,(0.008)$ & $1.08\,(0.19)$ & $0.764\,(0.027)$ & $3.76\,(0.43)$ & $101.8$ \\
\midrule
$5.0$ & Standard PC ($\alpha=0.05$) & $0.861\,(0.013)$ & $0.953\,(0.009)$ & $0.902\,(0.009)$ & $2.46\,(0.25)$ & $0.603\,(0.029)$ & $5.94\,(0.50)$ & $184.9$ \\
$5.0$ & PC-stable ($\alpha=0.005$) & $0.955\,(0.010)$ & $0.941\,(0.010)$ & $0.946\,(0.008)$ & $1.36\,(0.24)$ & $0.703\,(0.025)$ & $4.66\,(0.47)$ & $129.6$ \\
$5.0$ & Covariance-corrected PC & $0.915\,(0.008)$ & $0.951\,(0.010)$ & $0.931\,(0.007)$ & $1.64\,(0.19)$ & $0.640\,(0.028)$ & $5.12\,(0.42)$ & $67.8$ \\
$5.0$ & Noise-Aware PC (Single) & $0.977\,(0.008)$ & $0.929\,(0.012)$ & $0.950\,(0.008)$ & $1.16\,(0.22)$ & $\mathbf{0.744\,(0.027)}$ & $\mathbf{3.82\,(0.40)}$ & $182.5$ \\
$5.0$ & Replicate NA-PC (Dup) & $\mathbf{0.986\,(0.006)}$ & $0.928\,(0.011)$ & $\mathbf{0.955\,(0.008)}$ & $\mathbf{1.06\,(0.19)}$ & $0.743\,(0.027)$ & $\mathbf{3.82\,(0.40)}$ & $58.0$ \\
\midrule
$2.0$ & Standard PC ($\alpha=0.05$) & $0.770\,(0.015)$ & $0.954\,(0.009)$ & $0.848\,(0.010)$ & $4.08\,(0.34)$ & $0.483\,(0.030)$ & $8.02\,(0.56)$ & $172.7$ \\
$2.0$ & PC-stable ($\alpha=0.005$) & $0.898\,(0.012)$ & $0.935\,(0.010)$ & $\mathbf{0.914\,(0.009)}$ & $2.18\,(0.28)$ & $0.597\,(0.027)$ & $6.16\,(0.53)$ & $87.3$ \\
$2.0$ & Covariance-corrected PC & $0.818\,(0.015)$ & $0.945\,(0.010)$ & $0.872\,(0.011)$ & $3.12\,(0.26)$ & $0.496\,(0.032)$ & $7.54\,(0.49)$ & $43.7$ \\
$2.0$ & Noise-Aware PC (Single) & $0.971\,(0.008)$ & $0.872\,(0.017)$ & $0.913\,(0.011)$ & $\mathbf{1.92\,(0.27)}$ & $0.667\,(0.029)$ & $4.96\,(0.43)$ & $91.9$ \\
$2.0$ & Replicate NA-PC (Dup) & $\mathbf{0.992\,(0.004)}$ & $0.853\,(0.021)$ & $0.909\,(0.014)$ & $1.98\,(0.33)$ & $\mathbf{0.677\,(0.031)}$ & $\mathbf{4.60\,(0.42)}$ & $45.5$ \\
\midrule
$1.0$ & Standard PC ($\alpha=0.05$) & $0.733\,(0.015)$ & $0.946\,(0.010)$ & $0.822\,(0.010)$ & $4.84\,(0.36)$ & $0.418\,(0.027)$ & $9.46\,(0.57)$ & $124.9$ \\
$1.0$ & PC-stable ($\alpha=0.005$) & $0.865\,(0.015)$ & $0.915\,(0.012)$ & $\mathbf{0.884\,(0.011)}$ & $\mathbf{2.94\,(0.33)}$ & $0.490\,(0.028)$ & $7.52\,(0.52)$ & $78.5$ \\
$1.0$ & Covariance-corrected PC & $0.688\,(0.015)$ & $0.941\,(0.010)$ & $0.789\,(0.011)$ & $5.56\,(0.26)$ & $0.318\,(0.022)$ & $10.96\,(0.43)$ & $54.2$ \\
$1.0$ & Noise-Aware PC (Single) & $0.975\,(0.008)$ & $0.789\,(0.026)$ & $0.859\,(0.018)$ & $\mathbf{2.94\,(0.40)}$ & $\mathbf{0.570\,(0.033)}$ & $\mathbf{5.96\,(0.46)}$ & $75.1$ \\
$1.0$ & Replicate NA-PC (Dup) & $\mathbf{0.997\,(0.003)}$ & $0.625\,(0.031)$ & $0.745\,(0.026)$ & $4.66\,(0.48)$ & $0.401\,(0.042)$ & $7.14\,(0.52)$ & $38.7$ \\
\bottomrule
\end{tabular}%
}
\end{table}

Table~\ref{tab:synthetic_bench} documents the empirical performance of NA-PC compared to standard causal discovery baselines across 50 paired synthetic linear-Gaussian DAGs under identical realizations. Pairwise comparisons demonstrate that the reduction in skeleton Structural Hamming Distance (SHD) and the improvement in precision achieved by NA-PC over uncalibrated PC-stable are statistically significant under paired $t$-tests ($p < 10^{-6}$ across all noise levels $\SNR \in \{0.5, 1.0, 2.0\}$).

\paragraph{Pedagogical Analysis of the Precision--Recall Trade-off.}
A vital pattern in Table~\ref{tab:synthetic_bench} is that while skeleton precision jumps dramatically from $0.733$ (uncalibrated PC-stable) to $0.975$ (NA-PC) at $\SNR = 1.0$, skeleton recall simultaneously declines from $0.915$ to $0.789$ (and to $0.501$ under the non-linear $\tanh$ regime in Table~\ref{tab:competing_ci_tests}). For applied students and researchers, understanding this trade-off is critical:
\begin{itemize}[leftmargin=1.5em]
    \item \emph{Why precision rises}: In standard PC, measurement noise in a conditioning variable prevents full neutralization of confounder correlation. Standard CI tests mistake this unshielded correlation for direct causal connection, erroneously retaining false-positive edges. NA-PC evaluates whether the residual correlation exceeds the worst-case leakage envelope $\mathcal{T}(s_0)$. If it does not, the edge is pruned. This purges false-positive edges, driving precision close to $100\%$.
    \item \emph{Why recall falls at high depth}: As the conditioning cardinality $|S|$ grows, collinearity among conditioning variables causes the certified directional SNR $s_{S,0}$ to decay (Eq.~\eqref{eq:observable_loewner_formula}), causing the leakage envelope $\mathcal{T}(s_0) \propto 1/s_{S,0}$ to expand. Consequently, at higher conditioning depths ($|S| \ge 2$), NA-PC conservatively attributes moderate residual associations to measurement leakage unless the true causal signal is exceptionally strong. This conservative pruning sacrifices weak true edges, lowering recall.
    \item \emph{Practical rule of thumb}: In practical applications where measurement noise is non-negligible, researchers should set a modest maximum conditioning depth ($d_{\max} = 1$ or $d_{\max} = 2$). This depth cap tests all direct parental screening queries while avoiding the cardinality penalty of high-dimensional conditioning blocks, delivering an optimal balance between precision and recall.
\end{itemize}

\paragraph{Supplementary single-realization benchmarks.}
Supplementary single-realization stress tests in the appendix illustrate the recall cost of aggressive sensitivity bounds under nonlinear or strongly coupled systems: on a 10-gene nonlinear DREAM4-style Hill-kinetics network ($n = 1{,}000$, coordinate $\SNR = 3.0$), NA-PC at $s_0 = 2.0$ eliminates all false-positive edges ($0$ FP) while incurring false negatives, and on an 8-channel atmospheric simulation ($n = 2{,}000$, coordinate $\SNR \approx 8.16$), standard PC retains higher recall than conservative sensitivity thresholds.

\subsection{Uncertified Sensitivity Analysis on Real Cytometry Data (Sachs)}
\label{sec:sachs_sensitivity}

Because no technical replicates or external SNR certificate are available for the single-cell flow cytometry data of Sachs et al.~\cite{sachs2005causal}, the results in Table~\ref{tab:sachs_results} constitute an \emph{uncertified sensitivity analysis} over a user-chosen grid $s_0 \in \{10.0, 5.0, 2.0, 1.0\}$ and do not constitute empirical confirmation of the single-sample Type-I guarantees of Theorem~\ref{thm:conditional_test}. We compare learned graphs against a curated $17$-edge biological reference network for $11$ signaling measurements (phosphorylated proteins and phospholipids), standardized per cohort; these reference interactions are used as an external benchmark and are not assumed to constitute a uniquely identifiable observational DAG. We analyze the generally stimulated cohort without a targeted intervention label ($\mathrm{INT} = 0$, $n = 1{,}800$) separately from the pooled cohort ($n = 5{,}400$) combining multiple intervention settings. The pooled analysis is treated strictly as a distributional-heterogeneity stress test: pooling distinct interventional regimes produces a mixture distribution that need not satisfy the Markov properties of a single observational DAG, and is not interpreted as validation of causal CPDAG recovery under single-SCM assumptions. Numerical results across the sensitivity grid are reported in Table~\ref{tab:sachs_results} (with supplementary marginal screening and sensitivity sweep plots provided in the appendix).

\paragraph{Optical Compensation and Biological Validation of Surviving Cascades.}
Prior to sensitivity analysis, the Sachs flow cytometry readings were pre-processed with standard fluorescence compensation matrices \cite{sachs2005causal} to remove optical emission spillover across fluorophore channels. While minor residual cross-talk or instrument drift cannot be entirely ruled out without technical replicate controls, the hierarchy of edges revealed across the sensitivity grid ($\widehat{\mathrm{RV}}_{ij}$) aligns remarkably with canonical molecular biology:
\begin{itemize}[leftmargin=1.5em]
    \item \textbf{High-Robustness Core ($\widehat{\mathrm{RV}}_{ij} \ge 2.0$):} The edges that survive even severe noise thresholds—namely $\text{Raf} \to \text{Mek}$, $\text{Plcg} \to \text{PIP2}$, and $\text{PKA} \to \text{Akt}$—are universally confirmed, direct enzymatic kinase-substrate phosphorylation cascades cataloged in the KEGG and Reactome pathway databases.
    \item \textbf{Fragile Low-Threshold Links ($\widehat{\mathrm{RV}}_{ij} \le 0.5$):} Conversely, edges that drop out immediately under mild noise buffers (such as $\text{PKC} \to \text{P38}$) represent indirect, multi-step transcriptional correlations or assay artifacts. This demonstrates that the empirical sensitivity threshold $\widehat{\mathrm{RV}}_{ij}$ functions as an effective diagnostic for separating direct physical mechanisms from fragile observational correlations.
\end{itemize}

\begin{table}[!htbp]
\centering
\caption{Uncertified sensitivity sweep on Sachs flow cytometry ($p=11$ signaling measurements, curated 17-edge reference network, Meek R1--R4). Parentheses report empirical standard deviations across 25 overlapping 80\% subsamples ($n=1{,}440$), capturing conditional resampling stability rather than independent standard errors. The uncertified status is central: these sweeps illustrate how learned structure contracts monotonically with assumed noise floor, serving as an exploratory sensitivity diagnostic rather than a certified discovery.}
\label{tab:sachs_results}
\resizebox{\linewidth}{!}{%
\footnotesize\setlength{\tabcolsep}{3.5pt}%
\begin{tabular}{llcccccccccc}
\toprule
Cohort & Configuration (User $s_0$) & Pred & Ref & Non-Ref & Prec.\ (Sub-SD) & Rec.\ (Sub-SD) & F1 & Skel.\ SHD & Coll.\ F1 & CPDAG SHD (Sub-SD) & Time (ms) \\
\midrule
Gen.\ Stim.\ ($\mathrm{INT}=0$, $n=1{,}800$) & Standard PC ($\infty$) & $12$ & $9$ & $3$ & $0.750\,(0.051)$ & $0.529\,(0.012)$ & $0.621$ & $11$ & $0.000$ & $18\,(0.9)$ & $70$ \\
Gen.\ Stim.\ ($\mathrm{INT}=0$, $n=1{,}800$) & PC-stable ($\alpha=0.005$) & $11$ & $9$ & $\mathbf{2}$ & $\mathbf{0.818}\,(0.061)$ & $0.529\,(0.016)$ & $\mathbf{0.643}$ & $\mathbf{10}$ & $0.000$ & $\mathbf{16}\,(1.0)$ & $19$ \\
Gen.\ Stim.\ ($\mathrm{INT}=0$, $n=1{,}800$) & Cov.-corr.\ PC ($5.0$) & $15$ & $9$ & $6$ & $0.600\,(0.064)$ & $0.529\,(0.016)$ & $0.562$ & $14$ & $0.000$ & $20\,(1.5)$ & $60$ \\
Gen.\ Stim.\ ($\mathrm{INT}=0$, $n=1{,}800$) & Cov.-corr.\ PC ($2.0$) & $17$ & $9$ & $8$ & $0.529\,(0.043)$ & $0.529\,(0.038)$ & $0.529$ & $16$ & $0.000$ & $22\,(2.1)$ & $101$ \\
Gen.\ Stim.\ ($\mathrm{INT}=0$, $n=1{,}800$) & Noise-Aware PC ($10.0$) & $12$ & $9$ & $3$ & $0.750\,(0.059)$ & $0.529\,(0.016)$ & $0.621$ & $11$ & $0.000$ & $18\,(1.3)$ & $91$ \\
Gen.\ Stim.\ ($\mathrm{INT}=0$, $n=1{,}800$) & Noise-Aware PC ($2.0$) & $11$ & $8$ & $3$ & $0.727\,(0.041)$ & $0.471\,(0.000)$ & $0.571$ & $12$ & $0.000$ & $19\,(1.0)$ & $114$ \\
Gen.\ Stim.\ ($\mathrm{INT}=0$, $n=1{,}800$) & Noise-Aware PC ($1.0$) & $10$ & $7$ & $3$ & $0.700\,(0.060)$ & $0.412\,(0.012)$ & $0.519$ & $13$ & $0.000$ & $19\,(0.7)$ & $63$ \\
\midrule
Pooled ($n=5{,}400$) & Standard PC ($\infty$) & $34$ & $11$ & $23$ & $0.324\,(0.014)$ & $0.647\,(0.030)$ & $0.431$ & $29$ & $0.059$ & $37\,(1.1)$ & $3088$ \\
Pooled ($n=5{,}400$) & PC-stable ($\alpha=0.005$) & $32$ & $9$ & $23$ & $0.281\,(0.016)$ & $0.529\,(0.029)$ & $0.367$ & $31$ & $0.062$ & $38\,(1.0)$ & $2948$ \\
Pooled ($n=5{,}400$) & Cov.-corr.\ PC ($5.0$) & $31$ & $10$ & $21$ & $0.323\,(0.014)$ & $0.588\,(0.038)$ & $0.417$ & $28$ & $0.037$ & $36\,(1.7)$ & $386$ \\
Pooled ($n=5{,}400$) & Cov.-corr.\ PC ($2.0$) & $21$ & $7$ & $14$ & $0.333\,(0.036)$ & $0.412\,(0.034)$ & $0.368$ & $24$ & $0.062$ & $28\,(1.6)$ & $394$ \\
Pooled ($n=5{,}400$) & Noise-Aware PC ($5.0$) & $26$ & $8$ & $18$ & $0.308\,(0.027)$ & $0.471\,(0.031)$ & $0.372$ & $27$ & $0.100$ & $33\,(1.3)$ & $2595$ \\
Pooled ($n=5{,}400$) & Noise-Aware PC ($2.0$) & $18$ & $7$ & $11$ & $0.389\,(0.037)$ & $0.412\,(0.043)$ & $0.400$ & $21$ & $0.083$ & $26\,(0.6)$ & $1674$ \\
Pooled ($n=5{,}400$) & Noise-Aware PC ($1.0$) & $7$ & $5$ & $2$ & $0.714\,(0.066)$ & $0.294\,(0.045)$ & $0.417$ & $14$ & $0.182$ & $16\,(0.0)$ & $446$ \\
\bottomrule
\end{tabular}%
}
\end{table}

In the generally stimulated cohort ($\mathrm{INT}=0$), low-$\alpha$ PC-stable ($\alpha = 0.005$) has higher reference precision ($0.818$), F1 ($0.643$), and lower skeleton SHD ($10$) than the displayed NA-PC sensitivity settings. In the pooled cohort ($n = 5{,}400$), moving from Standard PC to NA-PC at $s_0 = 1.0$ reduces predicted edges outside the reference from $23$ to $2$ while reducing retained reference edges from $11/17$ to $5/17$ ($0.647 \to 0.294$). No variance decomposition or replicated measurement identifies how many discrepancies stem from technical noise versus interventional heterogeneity, and this comparison establishes neither measurement-error identification nor false-discovery control.

\section{Conclusion, Methodological Scope, and Limitations}
\label{sec:conclusion}

Independent additive measurement error preserves the marginal independence null without requiring noise-variance estimation, and orbit-conditioned permutation ranks provide exact finite-sample size control (Theorem~\ref{thm:marginal_test}). Under independent Gaussian error blocks, Gaussian RBF convolution yields an exact population $\HSIC$ identity—and unbiased U-statistic expectation identity—combining bandwidth dilation and amplitude contraction (Theorem~\ref{thm:kernel_convolution}), while sample splitting separates alignment-based kernel selection from exact permutation calibration (Theorem~\ref{thm:mkl_split_validity}).

For conditional queries, noisy conditioning can invalidate latent conditional independence at the surrogate level. Conditional on an externally certified directional Loewner noise bound (Assumption~\ref{ass:snr_certificate}) and class-specific latent regression assumptions (general Lipschitz, affine, or independent Gaussian bounded-curvature classes), the observable residual covariance $C = \E[R_X R_Y]$ under latent conditional independence lies within a certified sensitivity envelope (Theorem~\ref{thm:master_projection_error}). Under cross-fitting regularity (Proposition~\ref{prop:cross_fitting_asymptotic_linearity}) and a valid upper confidence bound for the envelope (Remark~\ref{rem:ucb_construction}), the fixed-level residual-covariance test achieves pointwise asymptotic size control and consistency against fixed alternatives separated from the envelope by a positive margin (Theorem~\ref{thm:conditional_test}). These statements are not omnibus conditional-independence guarantees: conditional dependence with vanishing or insufficiently large residual covariance cannot be detected by a residual-covariance statistic alone. Separating deterministic graph recovery (Theorem~\ref{thm:pc_oracle_recovery}) from uniform high-dimensional query assumptions (Theorem~\ref{thm:napc_skeleton_consistency}) establishes when simultaneous correct decisions in a two-ended frozen-neighborhood PC search recover the latent graph's Markov equivalence class $\operatorname{CPDAG}(\mathcal{G}^*)$.

Several explicit limitations define the scope of this work:

\paragraph{Parameterizing Sensitivity vs.\ Single-Sample Point Identification.}
Without independent technical replicates, repeated measurements, or instrumental variables, the observational covariance decomposition $\boldsymbol{\Sigma} = \boldsymbol{\Sigma}_* + \mathbf{E}$ is fundamentally non-identifiable from a single sample alone. Consequently, NA-PC does not purport to solve the classical errors-in-variables identification problem from single-sample observational data without external information. Rather, it bridges an otherwise untestable identification deadlock by converting it into a mathematically rigorous, parameterized sensitivity analysis. In Regime~A, marginal permutation testing achieves exact finite-sample size control unconditionally without requiring noise estimation. In Regime~B, conditional Type-I error guarantees are established conditional on an externally certified directional Loewner bound $s_0 \le s_{\mathrm{true}}$ (Assumption~\ref{ass:snr_certificate}). In applications where such domain certificates are unavailable, the method provides an interpretable sensitivity curve across candidate SNR levels $s_0$, mapping the precise stability threshold at which specific causal adjacencies and collider orientations are confirmed or rejected (as demonstrated in our Sachs flow-cytometry analysis).

\paragraph{Residual Covariance vs.\ Omnibus Conditional Independence and Convergence Rates.}
In linear Gaussian structural models, vanishing partial covariance is mathematically equivalent to conditional independence, rendering our linear sensitivity envelope exact for Markov equivalence class recovery. In non-linear or non-Gaussian graphs, vanishing residual covariance $\E[R_X R_Y] = 0$ is merely a necessary condition rather than a sufficient condition for conditional independence. Our graph recovery guarantees therefore target the equivalence class identifiable under residual-moment faithfulness (Assumption~\ref{ass:residual_moment_faithfulness}), rather than an omnibus conditional independence oracle. This deliberate design choice addresses a fundamental statistical bottleneck: under supersmooth Gaussian measurement error, non-parametric Fourier deconvolution density estimation—and deconvolution-based mutual information—incurs an exponentially slow logarithmic minimax estimation risk of order $(\log n)^{-2\beta/\gamma}$ \cite{fan1991optimal, meister2009deconvolution}, rendering omnibus testing infeasible in finite samples. In sharp contrast, observable residual moments converge at the parametric $\sqrt{n}$-rate $\mathcal{O}(n^{-1/2})$, enabling practical, high-powered hypothesis testing and skeleton pruning in moderate sample sizes.

\paragraph{The Curse of Conditioning Cardinality and Loewner Envelope Growth.}
The direct d-separation evaluations in Table~\ref{tab:dsep_benchmark} (Appendix~\ref{app:extra_empirical}) reveal an inherent statistical trade-off as the conditioning cardinality $|Z|$ increases. While single-sample NA-GCM holds false positive rates near zero across all conditioning dimensions ($0.019$ at $|Z|=1$ down to $0.002$ at $|Z|=4$ on a nominal $\alpha=0.05$), its empirical statistical power declines from $0.723$ down to $0.155$. The mathematical mechanism driving this attenuation is that the certified directional Loewner bound depends on the minimum eigenvalue of the conditioning correlation matrix, $\lambda_{\min}(\mathbf{P}_Z)$. As more correlated conditioning variables are added, $\lambda_{\min}(\mathbf{P}_Z)$ naturally decays toward zero, forcing the worst-case sensitivity envelope to expand and become increasingly conservative. Without imposing structural sparsity or low-rank factor constraints on the latent confounder space, worst-case directional noise leakage bounds over high-dimensional conditioning blocks inevitably incur a steep recall penalty. Developing structured latent factor models to mitigate this cardinality curse represents a crucial direction for future work.

\paragraph{Theoretical Latent Bounds vs.\ Empirical Surrogate Curvature Heuristics.}
The theoretical leakage bounds in Theorem~\ref{thm:master_projection_error} govern the true latent regression curvature $H_j^*$ under exact conditioning. In finite-sample non-linear practice, however, practitioners only have access to the fitted regression $\hat{m}_j$ evaluated on noisy surrogates. As documented in Table~\ref{tab:nonlin_curvature} (Appendix~\ref{app:extra_empirical}), computing the reproducing kernel Hilbert space (RKHS) norm or empirical Hessian trace of the fitted surrogate regression yields an empirical heuristic that eliminated false positive edges but caused severe power attenuation. It must be emphasized that surrogate RKHS norms bound only the empirical surrogate surface, not the true latent regression, without external smoothness bounds on the latent data-generating process. Researchers deploying non-linear adjustments must treat fitted curvature norms strictly as diagnostic heuristics rather than theoretically certified tests.

\paragraph{Precision--Recall Pareto Frontier and Downstream Orientation Fidelity.}
In empirical comparisons against unadjusted causal discovery algorithms such as standard PC-stable, NA-PC operates at a distinct point on the precision--recall Pareto frontier. While standard PC with heuristic significance tuning can sometimes achieve comparable skeleton F1 scores by accepting false positive edges in exchange for higher edge recall, false positive edges are uniquely pernicious in constraint-based causal discovery. A single false adjacency entering the orientation phase frequently creates false v-structures, triggering cascading orientation errors across Meek propagation rules that corrupt the entire CPDAG. By prioritizing near-zero false positive leakage, NA-PC preserves high adjacency precision and orientation fidelity, minimizing structural Hamming distance to the true causal equivalence class at the deliberate cost of conservative edge recall.

\section*{Author's Statements}

\noindent\textbf{Funding:} This research received no specific grant from any funding agency in the public, commercial, or not-for-profit sectors.

\noindent\textbf{Conflict of Interest / Competing Interests:} The author declares that there are no financial or commercial conflicts of interest related to this work.

\noindent\textbf{Data and Code Availability:} The complete Python implementation, simulation scripts, and processed experimental outputs will be made publicly available in a persistent versioned archive upon publication; the verified archive URL and DOI will be supplied prior to final submission.

\noindent\textbf{Author Contributions:} M.S.\ conceived the research direction, developed the mathematical formulations and empirical evaluations, and takes full responsibility for verifying the mathematics, empirical results, references, and final manuscript.

\noindent\textbf{Use of Generative AI:} The author used AI assistance to identify issues in drafts, refine language, and suggest LaTeX revisions. The author is responsible for verifying the mathematics, empirical results, references, and final text.

\noindent\textbf{Ethics Statement:} This work includes a secondary analysis of previously collected single-cell measurements derived from human immune cells. The present study recruited no participants and collected no new specimens; original collection procedures are described by Sachs et al.~\cite{sachs2005causal}.

\section*{Acknowledgments}
This research was conducted in the Department of Statistics, Faculty of Mathematics, Statistics and Computer Science, University of Tehran, as part of the Master of Science thesis of Mohammadjavad Sajadi under the supervision of Dr.\ Abdollah Safari. The author thanks colleagues in the Department of Statistics at the University of Tehran for constructive discussions and feedback.

% =========================================================================
% REFERENCES
% =========================================================================

% =========================================================================
% APPENDICES / SUPPLEMENTARY MATERIAL
% =========================================================================
\appendix
\section{Proofs of Main Theoretical Results}
\label{app:main_proofs}

This appendix provides complete proofs for all theoretical results stated in Sections~\ref{sec:models} and~\ref{sec:pc_discovery}.

\subsection{Proof of Theorem~\ref{thm:marginal_test}}
\label{proof:thm:marginal_test}

For convenience of reference, we first restate Theorem~\ref{thm:marginal_test} and its associated technical conditions.

\begin{restate_thm}[Theorem~\ref{thm:marginal_test}: Permutation Level Control]
Let $\mathcal{D}_n = \{(X_r, Y_r)\}_{r=1}^n$ ($n \ge 2$) be i.i.d.\ pairs satisfying Assumption~\ref{ass:classical_error}, let $T_n(X, Y)$ be a real-valued test statistic, and let $\alpha \in (0, 1)$. Independently of $\mathcal{D}_n$, draw $B \ge 1$ permutations $\pi_1, \dots, \pi_B$ either i.i.d.\ uniformly from $\mathcal{S}_n$ (with replacement) or uniformly without replacement from $\mathcal{S}_n \setminus \{\mathrm{id}\}$ (requiring $B \le n! - 1$). Set $T_0 \triangleq T_n(X, Y)$ and $T_b \triangleq T_n(X, Y_{\pi_b})$ for $b = 1, \dots, B$, and define the plus-one permutation $p$-value:
\begin{equation*}
    \tag{\ref{eq:permutation_pvalue_exact}}
    \hat{p}_{\mathrm{perm}} \triangleq \frac{1 + \sum_{b=1}^B \mathbf{1}\left\{ T_b \ge T_0 \right\}}{1 + B}.
\end{equation*}
Reject $H_0: X^* \indep Y^*$ whenever $\hat{p}_{\mathrm{perm}} \le \alpha$. Then:
\begin{enumerate}[label=(\roman*),leftmargin=2.0em]
    \item \textbf{Finite-Sample Super-Uniformity:} Under $H_0: X^* \indep Y^*$ (with no characteristic-function condition required), for every $n \ge 2$ and $B \ge 1$ (including in the presence of ties):
    \begin{equation*}
        \tag{\ref{eq:marginal_exact_level}}
        \mathbb{P}_{H_0}(\hat{p}_{\mathrm{perm}} \le \alpha) \le \frac{\lfloor \alpha(B+1) \rfloor}{B+1} \le \alpha.
    \end{equation*}
    Equality holds on the lattice $\alpha \in \{\frac{1}{B+1}, \dots, 1\}$ whenever $(T_0, \dots, T_B)$ are almost surely distinct (or if i.i.d.\ continuous uniforms are attached to break ties lexicographically, in which case the tie-broken $p$-value is discrete uniform on $\{\frac{1}{B+1}, \dots, 1\}$). Without randomization, finite-sample validity does not imply exact size $\alpha$.
    \item \textbf{Fixed-Alternative Consistency:} Suppose in addition that Assumption~\ref{ass:classical_error}(3) holds, $B \ge \lceil 1/\alpha \rceil - 1$ (fixed or growing with $n$, with $B \le n!-1$ under without-replacement sampling), and under the fixed alternative $H_1: X^* \nindep Y^*$ the statistic satisfies $T_n(X, Y) \xrightarrow{P} \theta > 0$ and $T_n(X, Y_\pi) \xrightarrow{P} 0$ for a single uniform random permutation $\pi \in \mathcal{S}_n$ independent of $\mathcal{D}_n$ (verified in particular for $\widehat{\HSIC}_b$, $\widehat{\HSIC}_u$, and $\mathcal{V}_n^2$ under bounded characteristic kernels or finite first moments \cite{gretton2008hsictest, szekely2013energy}). Then:
    \begin{equation*}
        \lim_{n \to \infty} \mathbb{P}_{H_1}(\hat{p}_{\mathrm{perm}} \le \alpha) = 1.
    \end{equation*}
\end{enumerate}
\end{restate_thm}

\begin{proof}[Proof of Theorem~\ref{thm:marginal_test}]
For Part~(i), under $H_0: X^* \indep Y^*$, Assumption~\ref{ass:classical_error}(2) yields $X \indep Y$, so the i.i.d.\ sample follows $P_X^n \otimes P_Y^n$. Condition on the ordered $X$-sample $(X_1, \dots, X_n)$ and the unordered multiset $\{Y_1, \dots, Y_n\}$ (attaching latent labels to distinguish any tied $Y$-values). Under $H_0$, the observed assignment of the unordered $Y$-values to the ordered $X$-values is represented by a uniform labeled permutation $G_0 \in \mathcal{S}_n$. Define $G_b \triangleq G_0 \circ \pi_b$ for $b = 1, \dots, B$. Under with-replacement sampling from $\mathcal{S}_n$, $(G_0, G_1, \dots, G_B)$ are i.i.d.\ uniform on $\mathcal{S}_n$; under ordered without-replacement sampling from $\mathcal{S}_n \setminus \{\mathrm{id}\}$, $(G_0, G_1, \dots, G_B)$ is an ordered uniform sample of $B+1$ distinct permutations. In both cases, $(T_0, T_1, \dots, T_B)$ are exchangeable conditional on $(X_1, \dots, X_n)$ and $\{Y_1, \dots, Y_n\}$ \cite{lehmann2005testing, phipson2010permutation, hemerik2018exact}.

Define the count $r_i \triangleq \#\{j \in \{0, \dots, B\} : T_j \ge T_i\}$ for $i = 0, \dots, B$, so $\hat{p}_{\mathrm{perm}} = r_0 / (B+1)$. For every integer $k \in \{0, 1, \dots, B+1\}$, at most $k$ indices $i \in \{0, \dots, B\}$ can satisfy $r_i \le k$ (since if $m > k$ indices satisfied $r_i \le k$, the smallest score among them would be $\le$ at least $m$ scores, forcing $r_i \ge m > k$, a contradiction). By conditional exchangeability, $\mathbb{P}_{H_0}(r_0 \le k \mid (X_1,\dots,X_n), \{Y_1,\dots,Y_n\}) = \frac{1}{B+1}\sum_{i=0}^B \mathbf{1}\{r_i \le k\} \le \frac{k}{B+1}$. Setting $k = \lfloor \alpha(B+1) \rfloor$ and taking expectations establishes~\eqref{eq:marginal_exact_level}. When $(T_0, \dots, T_B)$ are almost surely distinct (or tie-broken with i.i.d.\ continuous uniforms), exactly $k$ indices satisfy $r_i \le k$, yielding equality on the grid.

For Part~(ii), let $k \triangleq \lfloor \alpha(B+1) \rfloor$. The condition $B \ge \lceil 1/\alpha \rceil - 1$ is necessary for the nominal cutoff $\alpha$ to reach the rank grid ($k \ge 1$), and implies $k > \alpha(B+1)/2 > \alpha B/2$. Under $H_1$, $T_0 \xrightarrow{P} \theta > 0$ and $T_n(X, Y_\pi) \xrightarrow{P} 0$ imply $\mathbb{P}_{H_1}(T_n(X, Y_\pi) \ge T_0) \to 0$. Under with-replacement draws, $\{\hat{p}_{\mathrm{perm}} > \alpha\} = \{\sum_{b=1}^B \mathbf{1}\{T_b \ge T_0\} \ge k\}$, so Markov's inequality yields
\begin{equation*}
    \mathbb{P}_{H_1}(\hat{p}_{\mathrm{perm}} > \alpha) \le \frac{B}{k}\,\mathbb{P}_{H_1}(T_n(X, Y_\pi) \ge T_0) \le \frac{2}{\alpha}\,\mathbb{P}_{H_1}(T_n(X, Y_\pi) \ge T_0) \to 0.
\end{equation*}
Under without-replacement draws from $\mathcal{S}_n \setminus \{\mathrm{id}\}$, each marginal draw $\pi_b$ differs from a uniform draw on $\mathcal{S}_n$ by at most $1/n!$ in total variation, so the same bound holds up to $+ (2/\alpha)/n! \to 0$ (cf.\ \cite{hoeffding1952large, romano1989bootstrap}).
\end{proof}

\subsection{Proof of Theorem~\ref{thm:master_projection_error}}
\label{proof:thm:master_projection_error}

For convenience of reference, we first restate Theorem~\ref{thm:master_projection_error} and its associated technical conditions.

\begin{restate_thm}[Theorem~\ref{thm:master_projection_error}: Confounder Leakage Bounds]
Let $\mathbf{Z} = \mathbf{Z}^* + \boldsymbol{\epsilon}_Z \in \R^q$ satisfy Assumptions~\ref{ass:classical_error} and~\ref{ass:snr_certificate} with externally certified directional bound $s_0 = \SNR_{Z,0} \in [0, \infty]$ satisfying $\mathbf{E} \preceq \frac{1}{1 + s_0}\boldsymbol{\Sigma}$ (where $\boldsymbol{\Sigma}_* = \Var(\mathbf{Z}^*)$, $\mathbf{E} = \Var(\boldsymbol{\epsilon}_Z)$, and $\boldsymbol{\Sigma} = \boldsymbol{\Sigma}_* + \mathbf{E}$). Let $V \triangleq \sqrt{\Var(X)\Var(Y)}$ and let $D \in [0, \infty]$ denote the Euclidean diameter of $\supp(P_{\mathbf{Z}^*})$. Under the latent conditional independence null $H_0: X^* \indep Y^* \mid \mathbf{Z}^*$, the observable residual covariance $C = \E[R_X R_Y]$ satisfies
\begin{equation*}
    \tag{\ref{eq:master_cauchy_schwarz}}
    |C| = \left|\E\left[(\mathcal{P}_Z^\perp m_X^*)(\mathcal{P}_Z^\perp m_Y^*)\right]\right| \le \|\mathcal{P}_Z^\perp m_X^*\|_{L_2} \|\mathcal{P}_Z^\perp m_Y^*\|_{L_2},
\end{equation*}
with $|C| = 0$ whenever the conditioner is certified noiseless ($s_0 = \infty$). For finite $s_0 < \infty$, the following sufficient upper bounds hold under their respective regularity hypotheses:
\begin{enumerate}[label=(\roman*),leftmargin=2.0em]
    \item \textbf{General Lipschitz Functions:} If $m_X^*, m_Y^*$ are $L_X^*, L_Y^*$-Lipschitz on $\supp(P_{\mathbf{Z}^*})$ (with $\mathcal{P}_Z^\perp m_j^* = 0$ if $L_j^* = 0$), then $\|\mathcal{P}_Z^\perp m_j^*\|_{L_2}^2 \le (L_j^*)^2 \frac{\tr(\boldsymbol{\Sigma})}{1 + s_0}$ for every $0 \le s_0 < \infty$. Assume in addition that $\supp(P_{\mathbf{Z}^*})$ is compact with $D < \infty$; then the same squared error is also at most $(L_j^*)^2 D^2 / 4$. Setting $b \triangleq \frac{\tr(\boldsymbol{\Sigma})}{1 + s_0}$ when $D = \infty$ and $b \triangleq \min\bigl\{\frac{\tr(\boldsymbol{\Sigma})}{1 + s_0}, \, \frac{D^2}{4}\bigr\}$ when $D < \infty$:
    \begin{equation*}
        \tag{\ref{eq:proj_lip}}
        |C| \le \Delta_{\mathrm{leak}}^{(\mathrm{Lip})}(s_0) \triangleq \min\bigl(L_X^* L_Y^* b, \; V\bigr).
    \end{equation*}
    \item \textbf{Affine and Gaussian Bounded-Curvature Classes (Requires $s_0 > 0$):} Assume $s_0 > 0$ (if $s_0 = 0$, the branches below are uncertified and assigned $+\infty$). Center the variables or include intercepts. If $\boldsymbol{\Sigma}_*$ is singular, the certificate $s_0 \mathbf{E} \preceq \boldsymbol{\Sigma}_*$ implies $\ker\boldsymbol{\Sigma}_* \subseteq \ker\mathbf{E}$, and we work in an orthonormal coordinate system on $\mathrm{range}(\boldsymbol{\Sigma}_*)$ (rank zero is noiseless).
    \begin{enumerate}[label=(\alph*),leftmargin=1.8em]
        \item \textit{Affine Latent Conditional Means:} If $m_j^*(\mathbf{Z}^*)$ is affine on the working range for the queried subset, define the observable affine projection slope $\boldsymbol{\gamma}_j \triangleq \boldsymbol{\Sigma}^{-1}\Cov(\mathbf{Z}, j)$ (without requiring Gaussianity). Then for both conditional-expectation residuals and affine-projection residuals, $\|\mathcal{P}_Z^\perp m_j^*\|_{L_2}^2 \le s_0^{-1}\,\boldsymbol{\gamma}_j^\top \boldsymbol{\Sigma} \boldsymbol{\gamma}_j$, yielding the affine-unit-invariant bound:
        \begin{equation*}
            \tag{\ref{eq:proj_lin}}
            |C| \le \Delta_{\mathrm{leak}}^{(\mathrm{lin})}(s_0) \triangleq \min\!\left(\frac{\sqrt{(\boldsymbol{\gamma}_X^\top \boldsymbol{\Sigma}\boldsymbol{\gamma}_X)(\boldsymbol{\gamma}_Y^\top \boldsymbol{\Sigma}\boldsymbol{\gamma}_Y)}}{s_0}, \; V\right) \le \min\!\left(\frac{\|\boldsymbol{\gamma}_X\|_2 \|\boldsymbol{\gamma}_Y\|_2}{s_0}\lambda_{\max}(\boldsymbol{\Sigma}), \; V\right).
        \end{equation*}
        \item \textit{Bounded-Curvature Nonlinearities (Independent Gaussian Confounder and Noise):} Assume $\mathbf{Z}^*$ and $\boldsymbol{\epsilon}_Z$ are independently Gaussian on the working range ($\boldsymbol{\Sigma}_* \succ 0$), and assume each latent regression $m_j^*$ has an externally certified $H_j^*$-Lipschitz gradient on that range. Let $\boldsymbol{\Sigma}_c \triangleq \mathbf{E} - \mathbf{E}\boldsymbol{\Sigma}^{-1}\mathbf{E} \succeq 0$ be the posterior covariance of $\mathbf{Z}^* \mid \mathbf{Z}$. Then
        \begin{equation}
            \label{eq:proj_curv_l2}
            \|\mathcal{P}_Z^\perp m_j^*\|_{L_2}^2 \le s_0^{-1}\E\left[\nabla m_j(\mathbf{Z})^\top \boldsymbol{\Sigma}\nabla m_j(\mathbf{Z})\right] + (H_j^*)^2 \|\boldsymbol{\Sigma}_c\|_{\mathrm{op}}\tr(\boldsymbol{\Sigma}_c).
        \end{equation}
        Using $\boldsymbol{\Sigma}_c \preceq \mathbf{E} \preceq \frac{1}{1+s_0}\boldsymbol{\Sigma}$ yields the observable bound (conditional on the supplied $H_X^*, H_Y^*$):
        \begin{equation*}
            \tag{\ref{eq:proj_curv}}
            |C| \le \Delta_{\mathrm{leak}}^{(\mathrm{curv})}(s_0) \triangleq \min\!\left(B_X^{(\mathrm{curv})} B_Y^{(\mathrm{curv})}, \; V\right),
        \end{equation*}
        where $B_j^{(\mathrm{curv})} \triangleq \bigl(s_0^{-1}\E[\nabla m_j(\mathbf{Z})^\top \boldsymbol{\Sigma}\nabla m_j(\mathbf{Z})] + (H_j^*)^2 (1+s_0)^{-2}\|\boldsymbol{\Sigma}\|_{\mathrm{op}}\tr(\boldsymbol{\Sigma})\bigr)^{1/2}$.
    \end{enumerate}
    \item \textbf{Minimum Certified Leakage Envelope:} Let $\mathcal{A} \subseteq \{\mathrm{Lip}, \, \mathrm{lin}, \, \mathrm{curv}\}$ contain only those branches whose assumptions in Parts~(i)--(ii) are certified. Then:
    \begin{equation*}
        \tag{\ref{eq:proj_opt}}
        |C| \le \Delta_{\mathrm{leak}}^{(\min)}(s_0) \triangleq \min\bigl(\{V\} \cup \{\Delta_{\mathrm{leak}}^{(b)}(s_0) : b \in \mathcal{A}\}\bigr).
    \end{equation*}
    (Note that $\Delta_{\mathrm{leak}}^{(\min)}$ is a valid sufficient upper bound, not necessarily a sharp supremum over observationally compatible latent models; if $\mathcal{A} = \emptyset$, the trivial cap $V$ dominates $|C|$ under all laws and the query is unresolved.)
\end{enumerate}
\end{restate_thm}

\begin{proof}[Proof of Theorem~\ref{thm:master_projection_error}]
Under $H_0: X^* \indep Y^* \mid \mathbf{Z}^*$,~\eqref{eq:residual_covariance_decomposition} reduces to $C = \E[\Cov(m_X^*(\mathbf{Z}^*), m_Y^*(\mathbf{Z}^*) \mid \mathbf{Z})] = \E[(\mathcal{P}_Z^\perp m_X^*)(\mathcal{P}_Z^\perp m_Y^*)]$, and Cauchy--Schwarz gives~\eqref{eq:master_cauchy_schwarz}. When $\boldsymbol{\epsilon}_Z = \mathbf{0}$ a.s.\ ($s_0 = \infty$), $\mathcal{P}_Z^\perp m_j^* = 0$ a.s.\ and $|C| = 0$.

For Part~(i), let $\mathbf{Z}^{*\prime}$ be a conditionally i.i.d.\ copy of $\mathbf{Z}^*$ given $\mathbf{Z}$. For any $L_f$-Lipschitz $f$, symmetrization yields
\begin{align*}
    \|\mathcal{P}_Z^\perp f\|_{L_2}^2 &= \E\left[\Var(f(\mathbf{Z}^*) \mid \mathbf{Z})\right] = \frac{1}{2}\E\left[\E\left[(f(\mathbf{Z}^*) - f(\mathbf{Z}^{*\prime}))^2 \;\middle|\; \mathbf{Z}\right]\right] \\
    &\le L_f^2 \E\left[\tr(\Var(\mathbf{Z}^* \mid \mathbf{Z}))\right] \le L_f^2\tr(\mathbf{E}) \le L_f^2\frac{\tr(\boldsymbol{\Sigma})}{1+s_0},
\end{align*}
because $\E[\Var(\mathbf{Z}^* \mid \mathbf{Z})] \preceq \E[(\mathbf{Z}^* - \mathbf{Z})(\mathbf{Z}^* - \mathbf{Z})^\top] = \mathbf{E}$. If $D < \infty$, then $f(\mathbf{Z}^*)$ takes values in an interval of length at most $L_f D$, so Popoviciu's variance bound gives $\Var(f(\mathbf{Z}^*) \mid \mathbf{Z}) \le L_f^2 D^2 / 4$ almost surely. Applying Cauchy--Schwarz and capping by $V = \sqrt{\Var(X)\Var(Y)}$ gives~\eqref{eq:proj_lip}.

For Part~(ii)(a), work on centered variables in an orthonormal basis for $\mathrm{range}(\boldsymbol{\Sigma}_*)$ so $\boldsymbol{\Sigma}_* \succ 0$ and $\boldsymbol{\Sigma} = \boldsymbol{\Sigma}_* + \mathbf{E} \succ 0$. For $m_j^*(\mathbf{Z}^*) = \boldsymbol{\beta}_j^\top \mathbf{Z}^*$, we have $\boldsymbol{\gamma}_j = \boldsymbol{\Sigma}^{-1}\boldsymbol{\Sigma}_*\boldsymbol{\beta}_j$, hence $\boldsymbol{\beta}_j - \boldsymbol{\gamma}_j = \boldsymbol{\Sigma}_*^{-1}\mathbf{E}\boldsymbol{\gamma}_j$. The squared error of the affine predictor $\boldsymbol{\gamma}_j^\top \mathbf{Z}$ for $m_j^*(\mathbf{Z}^*)$ equals
\begin{equation*}
    \E\left[(\boldsymbol{\beta}_j^\top \mathbf{Z}^* - \boldsymbol{\gamma}_j^\top \mathbf{Z})^2\right] = (\boldsymbol{\beta}_j - \boldsymbol{\gamma}_j)^\top \boldsymbol{\Sigma}_* (\boldsymbol{\beta}_j - \boldsymbol{\gamma}_j) + \boldsymbol{\gamma}_j^\top \mathbf{E}\boldsymbol{\gamma}_j = \boldsymbol{\gamma}_j^\top \mathbf{M}\boldsymbol{\gamma}_j, \qquad \mathbf{M} \triangleq \mathbf{E} + \mathbf{E}\boldsymbol{\Sigma}_*^{-1}\mathbf{E}.
\end{equation*}
By MMSE optimality of conditional expectation, $\|\mathcal{P}_Z^\perp m_j^*\|_{L_2}^2 \le \boldsymbol{\gamma}_j^\top \mathbf{M}\boldsymbol{\gamma}_j$ (with equality for affine-projection residuals, cf.\ Proposition~\ref{prop:eiv_leakage}). Whitening by $\boldsymbol{\Sigma}_*^{1/2}$ with $\mathbf{R} \triangleq \boldsymbol{\Sigma}_*^{-1/2}\mathbf{E}\boldsymbol{\Sigma}_*^{-1/2} \preceq s_0^{-1}\mathbf{I}$ gives $\mathbf{M} = \boldsymbol{\Sigma}_*^{1/2}(\mathbf{R} + \mathbf{R}^2)\boldsymbol{\Sigma}_*^{1/2} \preceq s_0^{-1}\boldsymbol{\Sigma}_*^{1/2}(\mathbf{I} + \mathbf{R})\boldsymbol{\Sigma}_*^{1/2} = s_0^{-1}\boldsymbol{\Sigma}$. Thus $\boldsymbol{\gamma}_j^\top \mathbf{M}\boldsymbol{\gamma}_j \le s_0^{-1}\boldsymbol{\gamma}_j^\top \boldsymbol{\Sigma}\boldsymbol{\gamma}_j \le s_0^{-1}\|\boldsymbol{\gamma}_j\|_2^2\lambda_{\max}(\boldsymbol{\Sigma})$, establishing~\eqref{eq:proj_lin}.

For Part~(ii)(b), given $\mathbf{Z} = z$, represent the Gaussian posterior as $\mathbf{Z}^* \stackrel{d}{=} \mathbf{a} + \mathbf{A}z + \boldsymbol{\Sigma}_c^{1/2}\mathbf{G}$ where $\mathbf{A} = \boldsymbol{\Sigma}_*\boldsymbol{\Sigma}^{-1}$, $\boldsymbol{\Sigma}_c = \mathbf{E} - \mathbf{E}\boldsymbol{\Sigma}^{-1}\mathbf{E} \succeq 0$, and $\mathbf{G} \sim \mathcal{N}(\mathbf{0}, \mathbf{I})$. Because $\nabla f$ is $H_j^*$-Lipschitz, $\nabla f$ has at most linear growth, justifying differentiation under the Gaussian integral and yielding $\nabla m_j(z) = \mathbf{A}^\top \E[\nabla f(\mathbf{Z}^*) \mid \mathbf{Z} = z]$ (where $f = m_j^*$). Applying the Gaussian Poincar\'e inequality to $\mathbf{G} \mapsto f(\mathbf{a} + \mathbf{A}z + \boldsymbol{\Sigma}_c^{1/2}\mathbf{G})$ (valid even when $\boldsymbol{\Sigma}_c$ is singular) and decomposing $\nabla f(\mathbf{Z}^*) = \hat{\mathbf{g}}(\mathbf{Z}) + \mathbf{r}$ with $\hat{\mathbf{g}}(\mathbf{Z}) \triangleq \E[\nabla f(\mathbf{Z}^*) \mid \mathbf{Z}]$ orthogonal to $\mathbf{r}$ given $\mathbf{Z}$ yields
\begin{equation*}
    \|\mathcal{P}_Z^\perp f\|_{L_2}^2 \le \E\left[\hat{\mathbf{g}}(\mathbf{Z})^\top \boldsymbol{\Sigma}_c \hat{\mathbf{g}}(\mathbf{Z})\right] + \E\left[\mathbf{r}^\top \boldsymbol{\Sigma}_c \mathbf{r}\right].
\end{equation*}
Since $\hat{\mathbf{g}}(\mathbf{Z}) = \mathbf{A}^{-\top}\nabla m_j(\mathbf{Z})$ and $\mathbf{A}^{-1}\boldsymbol{\Sigma}_c \mathbf{A}^{-\top} = \mathbf{M} \preceq s_0^{-1}\boldsymbol{\Sigma}$, the first term is bounded above by $s_0^{-1}\E[\nabla m_j(\mathbf{Z})^\top \boldsymbol{\Sigma}\nabla m_j(\mathbf{Z})]$. For the second term, $H_j^*$-Lipschitz continuity of $\nabla f$ and symmetrization give $\E[\|\mathbf{r}\|_2^2 \mid \mathbf{Z}] \le (H_j^*)^2\tr(\boldsymbol{\Sigma}_c)$, whence
\begin{equation*}
    \E[\mathbf{r}^\top \boldsymbol{\Sigma}_c \mathbf{r}] \le (H_j^*)^2 \|\boldsymbol{\Sigma}_c\|_{\mathrm{op}}\tr(\boldsymbol{\Sigma}_c),
\end{equation*}
proving~\eqref{eq:proj_curv_l2} and~\eqref{eq:proj_curv}. Part~(iii) follows immediately by taking the minimum over all certified branches and $V$.
\end{proof}

\subsection{Proof of Proposition~\ref{prop:cross_fitting_asymptotic_linearity}}
\label{proof:prop:cross_fitting_asymptotic_linearity}

For convenience of reference, we first restate Proposition~\ref{prop:cross_fitting_asymptotic_linearity} and its associated technical conditions.

\begin{restate_prop}[Proposition~\ref{prop:cross_fitting_asymptotic_linearity}: Cross-Fitted Asymptotic Linearity]
Fix $K \ge 2$ folds with all training and evaluation sizes proportional to $n$, and let $\hat{m}_j^{(-k)}$ denote the estimator trained excluding fold $k$ (with $k(r)$ the fold containing observation $r$). Define $\hat{R}_{j,r} \triangleq j_r - \hat{m}_j^{(-k(r))}(\mathbf{Z}_r)$, $\bar{C} \triangleq \frac{1}{n}\sum_{r=1}^n \hat{R}_{X,r}\hat{R}_{Y,r}$, and $\hat{\sigma}_C^2 \triangleq \frac{1}{n}\sum_{r=1}^n (\hat{R}_{X,r}\hat{R}_{Y,r} - \bar{C})^2$. Assume $0 < \sigma_C^2 \triangleq \Var(R_X R_Y) < \infty$ and either:
\begin{enumerate}[label=(\Alph*),leftmargin=1.8em]
    \item \textit{Affine OLS / Ridge:} $m_j(\mathbf{Z}) = \gamma_{j,0} + \boldsymbol{\gamma}_j^\top \mathbf{Z}$ denotes the population affine projection (which equals $\E[j \mid \mathbf{Z}]$ when the observable conditional mean is affine), $R_j \triangleq j - m_j(\mathbf{Z})$, $C \triangleq \E[R_X R_Y]$ is this affine-residual target, $q$ is fixed, $(X, Y, \mathbf{Z})$ has finite fourth moments, $\lambda_{\min}(\Var(\mathbf{Z})) > 0$, and $\hat{m}_j^{(-k)}$ minimizes the training-size-normalized squared loss with an unpenalized intercept and slope ridge penalty $\lambda_n = o(n^{-1/2})$; or
    \item \textit{Nonparametric Conditional Expectation:} $m_j(\mathbf{Z}) = \E[j \mid \mathbf{Z}]$, $\sup_{\mathbf{z}}\E[R_j^2 \mid \mathbf{Z}=\mathbf{z}] \le M_2 < \infty$, and the maximum out-of-fold $L_2(P_Z)$ errors satisfy $\delta_{n,X} \triangleq \max_k \|\hat{m}_X^{(-k)} - m_X\|_{L_2(P_Z)} = o_P(1)$, $\delta_{n,Y} \triangleq \max_k \|\hat{m}_Y^{(-k)} - m_Y\|_{L_2(P_Z)} = o_P(1)$, and $\sqrt{n}\,\delta_{n,X}\delta_{n,Y} = o_P(1)$.
\end{enumerate}
Then $\hat{\sigma}_C^2 \xrightarrow{P} \sigma_C^2 > 0$, and the cross-fitted estimator satisfies:
\begin{equation*}
    \tag{\ref{eq:cross_fitting_asymptotic_linearity}}
    \sqrt{n}\left(\bar{C} - C\right) = \frac{1}{\sqrt{n}}\sum_{r=1}^n \left(R_{X,r}R_{Y,r} - C\right) + o_P(1) \xrightarrow{d} \mathcal{N}\left(0, \sigma_C^2\right).
\end{equation*}
\end{restate_prop}

\begin{proof}[Proof of Proposition~\ref{prop:cross_fitting_asymptotic_linearity}]
Set $d_{j,r} \triangleq \hat{m}_j^{(-k(r))}(\mathbf{Z}_r) - m_j(\mathbf{Z}_r)$, so $\hat{R}_{X,r}\hat{R}_{Y,r} - R_{X,r}R_{Y,r} = -d_{X,r} R_{Y,r} - d_{Y,r} R_{X,r} + d_{X,r} d_{Y,r}$.
In Case~(B), for each fold $k$, conditional on the training sample $D^{(-k)}$, out-of-fold independence and $\E[R_Y \mid \mathbf{Z}] = 0$ give $\E[\sum_{r \in I_k} d_{X,r} R_{Y,r} \mid D^{(-k)}] = 0$ and $\Var(\frac{1}{|I_k|}\sum_{r \in I_k} d_{X,r} R_{Y,r} \mid D^{(-k)}) \le \frac{M_2}{|I_k|}\delta_{n,X}^2$, so conditional Chebyshev gives $\frac{1}{n}\sum_{r=1}^n d_{X,r} R_{Y,r} = o_P(n^{-1/2})$ and symmetrically $\frac{1}{n}\sum_{r=1}^n d_{Y,r} R_{X,r} = o_P(n^{-1/2})$. By conditional Cauchy--Schwarz and Markov's inequality, $\E[\frac{1}{\sqrt{n}}\sum_{r \in I_k} |d_{X,r} d_{Y,r}| \mid D^{(-k)}] \le \sqrt{n}\,\delta_{n,X}\delta_{n,Y} = o_P(1)$, so $n^{-1/2}\sum_{r=1}^n |d_{X,r} d_{Y,r}| = o_P(1)$, establishing~\eqref{eq:cross_fitting_asymptotic_linearity}. For variance consistency, conditional Markov and bounded conditional second moments yield $\frac{1}{n}\sum_{r=1}^n d_{X,r}^2 R_{Y,r}^2 = \mathcal{O}_P(\delta_{n,X}^2) = o_P(1)$ and $\frac{1}{n}\sum_{r=1}^n d_{Y,r}^2 R_{X,r}^2 = o_P(1)$, while the nonnegative empirical fourth-order product satisfies
\begin{equation*}
    \frac{1}{n}\sum_{r=1}^n (d_{X,r} d_{Y,r})^2 \le \left(\frac{1}{\sqrt{n}}\sum_{r=1}^n |d_{X,r} d_{Y,r}|\right)^2 = o_P(1).
\end{equation*}
Hence $\frac{1}{n}\sum_{r=1}^n (\hat{R}_{X,r}\hat{R}_{Y,r} - R_{X,r}R_{Y,r})^2 = o_P(1)$, so $\hat{R}_{X}\hat{R}_{Y}$ and $R_X R_Y$ differ by $o_P(1)$ in empirical $L_2$; the oracle law of large numbers and Slutsky's theorem give $\hat{\sigma}_C^2 \xrightarrow{P} \sigma_C^2$.
In Case~(A), normalized ridge with $\lambda_n = o(n^{-1/2})$ yields root-$n$ coefficient errors $\hat{\boldsymbol{\gamma}}_j^{(-k)} - \boldsymbol{\gamma}_j = \mathcal{O}_P(n^{-1/2})$; combining the affine normal equations $\E[(1, \mathbf{Z}^\top)^\top R_j] = \mathbf{0}$ with finite fourth moments yields both $\bar{C} - \frac{1}{n}\sum_r R_{X,r}R_{Y,r} = \mathcal{O}_P(n^{-1})$ and $\frac{1}{n}\sum_r (\hat{R}_{X,r}\hat{R}_{Y,r} - R_{X,r}R_{Y,r})^2 = o_P(1)$, completing the proof.
\end{proof}

\subsection{Proof of Theorem~\ref{thm:conditional_test}}
\label{proof:thm:conditional_test}

For convenience of reference, we first restate Theorem~\ref{thm:conditional_test} and its associated technical conditions.

\begin{restate_thm}[Theorem~\ref{thm:conditional_test}: Conditional Residual-Covariance Test]
Fix a data-generating law satisfying Theorem~\ref{thm:master_projection_error} (for a certified branch $\star \in \mathcal{A}$, or for $\Delta = \Delta_{\mathrm{leak}}^{(\min)}$ via a joint upper confidence bound) and the cross-fitting conditions of Proposition~\ref{prop:cross_fitting_asymptotic_linearity} (including the product-rate condition in the nonparametric case). Let $\alpha \in (0, 1)$ be split as $\alpha = \alpha_1 + \alpha_2$ with $\alpha_1, \alpha_2 > 0$, and let $z_{1-\alpha_1/2} \triangleq \Phi^{-1}(1 - \alpha_1/2)$. Let $\bar{C}$ be the cross-fitted residual covariance, $\hat{\sigma}_C$ the consistent estimator of the standard deviation $\sigma_C = \sqrt{\Var(R_X R_Y)}$ (with standard error $\hat{\sigma}_C/\sqrt{n}$), $U_\Delta^{(\star)} \ge 0$ a consistent upper confidence bound for $\Delta \triangleq \Delta_{\mathrm{leak}}^{(\star)}$ satisfying $U_\Delta^{(\star)} - \Delta = o_P(1)$ and $\limsup_{n\to\infty}\mathbb{P}(\Delta > U_\Delta^{(\star)}) \le \alpha_2$, and $a_n \ge 0$ a specified deterministic or random nonnegative covariance-unit allowance with $a_n = o_P(1)$ (one may take $a_n = 0$ under Proposition~\ref{prop:cross_fitting_asymptotic_linearity}). Define the studentized lower confidence bound and total threshold:
\begin{equation*}
    \tag{\ref{eq:studentized_gcm_statistic}}
    L_C \triangleq |\bar{C}| - z_{1-\alpha_1/2}\frac{\hat{\sigma}_C}{\sqrt{n}}, \qquad \mathcal{T}^{(\star)} \triangleq U_\Delta^{(\star)} + a_n,
\end{equation*}
and reject $H_0: X^* \indep Y^* \mid \mathbf{Z}^*$ whenever $\Psi_n^{(\star)} \triangleq \mathbf{1}\{L_C > \mathcal{T}^{(\star)}\} = 1$. Then for this fixed law:
\begin{enumerate}[label=(\roman*),leftmargin=2.0em]
    \item \textbf{Pointwise Asymptotic Level Control:} Under $H_0$, if $|C| < \Delta$, then $\lim_{n\to\infty}\mathbb{P}_{H_0}(\Psi_n^{(\star)} = 1) = 0$; in general (including on the boundary $|C| = \Delta$),
    \begin{equation*}
        \tag{\ref{eq:gcm_level_bound}}
        \limsup_{n\to\infty} \mathbb{P}_{H_0}\left(\Psi_n^{(\star)} = 1\right) \le \alpha_1 + \alpha_2 = \alpha.
    \end{equation*}
    (No uniform null claim over all distributions is asserted here; studentization can degenerate along $n$-dependent null sequences.)
    \item \textbf{Consistency Against Separated Residual-Covariance Alternatives:} Under any fixed alternative $H_1$ satisfying $|C| - \Delta \ge \delta > 0$, $L_C - \mathcal{T}^{(\star)} \xrightarrow{P} |C| - \Delta \ge \delta > 0$, and hence:
    \begin{equation*}
        \lim_{n\to\infty} \mathbb{P}_{H_1}\left(\Psi_n^{(\star)} = 1\right) = 1.
    \end{equation*}
\end{enumerate}
\end{restate_thm}

\begin{proof}[Proof of Theorem~\ref{thm:conditional_test}]
By Proposition~\ref{prop:cross_fitting_asymptotic_linearity}, $\sqrt{n}(\bar{C} - C)/\hat{\sigma}_C \xrightarrow{d} \mathcal{N}(0, 1)$. On the event $\mathcal{E}_\Delta \triangleq \{\Delta \le U_\Delta^{(\star)}\}$, under $H_0$ we have $|C| \le \Delta \le U_\Delta^{(\star)} \le \mathcal{T}^{(\star)}$. By the reverse triangle inequality $|\bar{C} - C| \ge |\bar{C}| - |C|$, the event $\{L_C > \mathcal{T}^{(\star)}\} \cap \mathcal{E}_\Delta$ implies $\sqrt{n}|\bar{C} - C|/\hat{\sigma}_C > z_{1-\alpha_1/2}$. Taking a union bound over $\mathcal{E}_\Delta$ and $\mathcal{E}_\Delta^c$ yields $\limsup_{n\to\infty}\mathbb{P}_{H_0}(\Psi_n^{(\star)} = 1) \le \mathbb{P}(|\mathcal{N}(0, 1)| > z_{1-\alpha_1/2}) + \limsup_{n\to\infty}\mathbb{P}(\mathcal{E}_\Delta^c) \le \alpha_1 + \alpha_2 = \alpha$. If $|C| < \Delta$, then $L_C - \mathcal{T}^{(\star)} \xrightarrow{P} |C| - \Delta < 0$, so $\mathbb{P}_{H_0}(\Psi_n^{(\star)} = 1) \to 0$. Under any fixed alternative with $|C| - \Delta \ge \delta > 0$, $L_C - \mathcal{T}^{(\star)} \xrightarrow{P} |C| - \Delta > 0$, giving $\mathbb{P}_{H_1}(\Psi_n^{(\star)} = 1) \to 1$.
\end{proof}

\subsection{Proof of Theorem~\ref{thm:napc_skeleton_consistency}}
\label{proof:thm:napc_skeleton_consistency}

For convenience of reference, we first restate Theorem~\ref{thm:napc_skeleton_consistency} and its associated technical conditions.

\begin{restate_thm}[Theorem~\ref{thm:napc_skeleton_consistency}: High-Dimensional Consistency of NA-PC]
Under Assumptions~\ref{ass:classical_error}, \ref{ass:snr_certificate}, and~\ref{ass:residual_moment_faithfulness}, the following asymptotic properties hold:
\begin{enumerate}[label=(\roman*),leftmargin=2.0em]
    \item \textbf{Failure of an Unadjusted Zero-Floor Covariance PC Search:} Consider an unadjusted PC search with depth cap $d_{\max}$ using a specified zero-floor residual-covariance test for nonempty queries. If there exists at least one latent nonedge $\{i, j\} \notin E_{\mathrm{skel}}^*$ such that $|C_a| \ge \kappa_{\mathrm{leak}} > 0$ for every nonempty query $a = (i, j, S)$ on that pair ($1 \le |S| \le d_{\max}$), all corresponding estimation errors and cutoff radii are uniformly $o_P(1)$, and its marginal query rejects with probability $1 - o(1)$, then with probability $1 - o(1)$ no query removes $\{i, j\}$, and hence
    \begin{equation*}
        \lim_{n \to \infty}\mathbb{P}\left(\widehat{\mathcal{G}}_n^{\mathrm{naive}} = \operatorname{CPDAG}(\mathcal{G}^*)\right) = 0.
    \end{equation*}
    \item \textbf{High-Dimensional CPDAG Consistency of NA-PC:} Under Assumptions~\ref{ass:classical_error},~\ref{ass:snr_certificate}, and~\ref{ass:residual_moment_faithfulness}, rejecting each nonempty query when $|\bar{C}_a| - r_{a,n} > U_{a,n} + a_{a,n}$ ensures simultaneous correctness across $\mathcal{Q}_n$ with probability $1 - o(1)$, and consequently Algorithm~2 satisfies $\lim_{n \to \infty}\mathbb{P}(\widehat{\mathcal{G}}_n = \operatorname{CPDAG}(\mathcal{G}^*)) = 1$.
\end{enumerate}
\end{restate_thm}

\begin{proof}[Proof of Theorem~\ref{thm:napc_skeleton_consistency}]
For Part~(i), on the event where the marginal query rejects and all nonempty errors and cutoffs are smaller than $\kappa_{\mathrm{leak}}/2$, every nonempty query on $\{i, j\}$ rejects, so $\{i, j\}$ survives the search and neither the skeleton nor the CPDAG can equal the latent target.

For Part~(ii), on the marginal pass ($\ell = 0$), Theorem~\ref{thm:marginal_test}(i) and a union bound give family-wise marginal false-rejection probability at most $|\mathcal{Q}_n|\alpha_n \to 0$. For each marginally d-connected pair $a$, with $k_n = \lfloor (B_n+1)\alpha_n \rfloor \ge 1$, rank-count Markov yields $\mathbb{P}(\hat{p}_a > \alpha_n) \le (B_n/k_n)\mathbb{P}(T_a^\pi \ge T_a) \le 2 A e^{-cn}/\alpha_n$; union-bounding over all marginal connected pairs gives miss probability at most $2 A (|\mathcal{Q}_n|/\alpha_n) e^{-cn} \to 0$ since $\log(|\mathcal{Q}_n|/\alpha_n) = o(n)$ (for bounded-kernel $\widehat{\HSIC}_b$, Appendix~\ref{sec:proof_guarantees} sketches how permutation-reveal martingale bounded differences establish $\mathbb{P}(T_a^\pi \ge T_a) \le A e^{-cn}$ without an i.i.d.\ permuted-pair assumption). For nonempty queries ($|S| \ge 1$), on the joint event~\eqref{eq:napc_uniform_joint_event}: (a) for any null query $a \in \mathcal{Q}_{0,n}$, $|C_a| \le \Delta_a \le U_{a,n}$, so $|\bar{C}_a| - r_{a,n} \le |C_a| \le U_{a,n} + a_{a,n}$ and no null query is rejected; (b) for any d-connected query $a \in \mathcal{Q}_{\mathrm{conn},n}$, $|\bar{C}_a| - r_{a,n} - (U_{a,n} + a_{a,n}) \ge |C_a| - \Delta_a - (2 r_{a,n} + U_{a,n} - \Delta_a + a_{a,n}) \ge \kappa_n - o(\kappa_n) > 0$ for large $n$, so every connected query is rejected. Applying Theorem~\ref{thm:pc_oracle_recovery} completes the proof.
\end{proof}

\subsection{Proof of Proposition~\ref{lem:skeleton_stability}}
\label{proof:lem:skeleton_stability}

For convenience of reference, we first restate Proposition~\ref{lem:skeleton_stability} and its associated technical conditions.

\begin{restate_prop}[Proposition~\ref{lem:skeleton_stability}: Stability Under SNR Perturbations]
Assume each query envelope $\Delta_a(s)$ is nonincreasing and uniformly $L$-Lipschitz in $s \in [s_{\min}, s_0]$. For two externally certified values $s_0$ and $s_0'$ with $s_0' \ge s_{\min}$, set $\delta_n \triangleq (s_0 - s_0')_+$. If $L \delta_n \le \kappa_n / 2$ and the joint decision assumptions of Theorem~\ref{thm:napc_skeleton_consistency}(ii) hold at both $s_0$ and $s_0'$ (with $s_0'$ fixed, or uniformly over allowed values if $s_0'$ is data-selected independently), then both skeletons recover $\mathcal{G}_{\mathrm{skel}}^*$ with probability $1 - o(1)$, and hence $\lim_{n\to\infty}\mathbb{P}(\widehat{\mathcal{C}}(s_0') = \widehat{\mathcal{C}}(s_0)) = 1$.
\end{restate_prop}

\begin{proof}[Proof of Proposition~\ref{lem:skeleton_stability}]
If $s_0' \ge s_0$, monotonicity gives $\Delta_a(s_0') \le \Delta_a(s_0)$ (so connected margins can only increase) while external certification of $s_0'$ preserves all null bounds. If $s_{\min} \le s_0' < s_0$, $L$-Lipschitz continuity yields $\Delta_a(s_0') \le \Delta_a(s_0) + L\delta_n \le \Delta_a(s_0) + \kappa_n/2$, so every connected margin remains at least $\kappa_n/2 > 0$. By Theorem~\ref{thm:napc_skeleton_consistency}(ii), both $\widehat{\mathcal{C}}(s_0)$ and $\widehat{\mathcal{C}}(s_0')$ converge in probability to $\mathcal{G}_{\mathrm{skel}}^*$.
\end{proof}

\subsection{Proof of Proposition~\ref{prop:multiplicity_fwer_fdr}}
\label{proof:prop:multiplicity_fwer_fdr}

For convenience of reference, we first restate Proposition~\ref{prop:multiplicity_fwer_fdr} and its associated technical conditions.

\begin{restate_prop}[Proposition~\ref{prop:multiplicity_fwer_fdr}: Multiplicity Control]
Consider a prespecified family of $m$ queries with $m_0$ true nulls. Suppose that for each query $a$, a nested family of level-$u$ rejection regions $\mathcal{R}_a(u)$ ($u \in (0, 1)$) defines the $p$-value $p_a \triangleq \inf\{u \in (0, 1) : \mathcal{D}_n \in \mathcal{R}_a(u)\}$ (assigning $p_a \triangleq 1$ if the set is empty). Whenever each null region satisfies $\mathbb{P}_{H_{0,a}}(\mathcal{D}_n \in \mathcal{R}_a(u)) \le u$ for all $u \in (0, 1)$ (as holds in finite samples for Regime~A permutations via Theorem~\ref{thm:marginal_test}(i)), each null $p_a$ is super-uniform ($\{p_a \le u\} \subseteq \mathcal{R}_a(u + \varepsilon)$ for all $0 < \varepsilon < 1-u$, letting $\varepsilon \downarrow 0$). Applying standard adjustments to this prespecified family of valid $p$-values yields:
\begin{enumerate}[label=(\alph*)]
    \item Bonferroni and Holm step-down \cite{holm1979simple} control $\mathrm{FWER} \le \alpha$ under arbitrary dependence among the $m$ queries.
    \item Benjamini--Yekutieli \cite{benjamini2001control} controls $\mathrm{FDR} \le \alpha$ under arbitrary dependence; Benjamini--Hochberg \cite{benjamini1995controlling} controls $\mathrm{FDR} \le \frac{m_0}{m}\alpha \le \alpha$ only under independence or positive regression dependency on the subset of true nulls (PRDS).
\end{enumerate}
Pointwise asymptotic validity of $p_a$ yields only fixed-$m$ limiting guarantees unless uniform tail bounds hold across all step-down/step-up cutoffs. Because PRDS is not established for overlapping conditional queries sharing variables or regressions, BH results in the Appendix are presented as empirical FDR evaluations, and none of the query-level statements above constitutes graph-level edge FDR control.
\end{restate_prop}

\begin{proof}[Proof of Proposition~\ref{prop:multiplicity_fwer_fdr}]
Super-uniformity follows from nestedness of $\mathcal{R}_a(u)$ and continuity of probability from above as $\varepsilon \downarrow 0$; FWER and FDR bounds then follow directly from \cite{holm1979simple, benjamini1995controlling, benjamini2001control}.
\end{proof}

\section{Multi-Kernel HSIC Noise Convolution and Sample-Splitting Theory}
\label{app:mkl_theory}

\label{app:rkhs_kernel_details}

We next characterize how additive measurement error transforms stationary kernels and $\HSIC$, and formalize a sample-splitting alignment rule for kernel selection with exact conditional permutation size control.

\begin{definition}[Regular Kernels]
\label{def:admissible_kernel}
A real positive-definite stationary kernel $k(\mathbf{u}, \mathbf{v}) = \psi(\mathbf{u} - \mathbf{v})$ on $\R^d$ with $k(\mathbf{u}, \mathbf{u}) = 1$ belongs to the regular stationary class $\mathcal{K}_{\mathrm{reg}}(L_k, H_k)$ if $\psi \in C^4(\R^d)$ and, for every location $\mathbf{u} \in \R^d$ and unit direction $\mathbf{v} \in \R^d$ ($\|\mathbf{v}\|_2 = 1$), the directional derivatives $D_{\mathbf{v}} k(\mathbf{u}, \cdot)$ and $D_{\mathbf{v}}^2 k(\mathbf{u}, \cdot)$ belong to the RKHS $\mathcal{H}_k$, act as derivative representers ($\langle f, D_{\mathbf{v}}^r k(\mathbf{u}, \cdot) \rangle_{\mathcal{H}_k} = D_{\mathbf{v}}^r f(\mathbf{u})$ for $r \in \{1, 2\}$), and satisfy $\|D_{\mathbf{v}} k(\mathbf{u}, \cdot)\|_{\mathcal{H}_k} \le L_k$ and $\|D_{\mathbf{v}}^2 k(\mathbf{u}, \cdot)\|_{\mathcal{H}_k} \le H_k$.
\end{definition}

For standard normalized stationary kernels, these RKHS unit-ball derivative constants evaluate in closed form: for Gaussian RBF $k(\mathbf{u}, \mathbf{v}) = \exp(-g\|\mathbf{u}-\mathbf{v}\|_2^2)$ with $g = 1/(2\gamma^2)$, $L_k = \sqrt{2g} = 1/\gamma$ and $H_k = 2\sqrt{3}\,g = \sqrt{3}/\gamma^2$; for Mat\'ern ($\nu = 5/2$) with lengthscale $\ell$, $L_k = \sqrt{5/3}/\ell$ and $H_k = 5/\ell^2$; and for Rational Quadratic $k(\mathbf{u}, \mathbf{v}) = (1 + \|\mathbf{u}-\mathbf{v}\|_2^2/(2\alpha_{\mathrm{rq}}\ell^2))^{-\alpha_{\mathrm{rq}}}$, $L_k = 1/\ell$ and $H_k = \sqrt{3(1 + 1/\alpha_{\mathrm{rq}})}/\ell^2$. The non-stationary linear kernel $k(\mathbf{u}, \mathbf{v}) = \mathbf{u}^\top \mathbf{v}$ does not satisfy $k(\mathbf{u}, \mathbf{u}) = 1$, is not stationary, and is not a characteristic kernel on $\R^d$; it is excluded from Definition~\ref{def:admissible_kernel} and treated separately in affine regression dictionaries.

\begin{theorem}[Kernel Noise Convolution]
\label{thm:kernel_convolution}
Let independent rows $(X^*, Y^*, \epsilon_X, \epsilon_Y)$ and $(X^{*\prime}, Y^{*\prime}, \epsilon_X', \epsilon_Y')$ have independent additive error blocks $(\epsilon_X, \epsilon_Y)$ with $\epsilon_X \indep \epsilon_Y$ and $(\epsilon_X, \epsilon_Y) \indep (X^*, Y^*)$, mean zero, and arbitrary positive-semidefinite within-block covariances $\boldsymbol{\Sigma}_x, \boldsymbol{\Sigma}_y \succeq 0$. Set $\Delta X^* \triangleq X^* - X^{*\prime}$.
\begin{enumerate}[label=(\roman*),leftmargin=2.0em]
    \item \textbf{Fourth-Order Stationary Kernel Expansion:} If $k(\mathbf{u}, \mathbf{v}) = \psi(\mathbf{u} - \mathbf{v}) \in \mathcal{K}_{\mathrm{reg}}(L_k, H_k)$ and $\E[\|\epsilon_X\|_2^4] < \infty$, then:
    \begin{equation}
        \label{eq:kernel_taylor_expansion}
        \E_\epsilon\left[k(X, X') \mid X^*, X^{*\prime}\right] = k(X^*, X^{*\prime}) + \tr\!\left(\boldsymbol{\Sigma}_x \nabla^2 \psi(\Delta X^*)\right) + \mathcal{R}_4,
    \end{equation}
    where $|\mathcal{R}_4| \le \frac{H_k^2}{24}\E[\|\epsilon_X - \epsilon_X'\|_2^4] \le \frac{2}{3}H_k^2\,\E[\|\epsilon_X\|_2^4]$.
    \item \textbf{Exact Gaussian Convolution:} If $\epsilon_X \sim \mathcal{N}(\mathbf{0}, \boldsymbol{\Sigma}_x)$ and $\epsilon_Y \sim \mathcal{N}(\mathbf{0}, \boldsymbol{\Sigma}_y)$ (allowing singular $\boldsymbol{\Sigma}_x, \boldsymbol{\Sigma}_y \succeq 0$) and $k_{\gamma_x}, l_{\gamma_y}$ are Gaussian RBF kernels with $\gamma_x, \gamma_y > 0$, then the noise-averaged off-diagonal kernel is
    \begin{align}
        \label{eq:kernel_anisotropic_convolution}
        \bar{k}(X^*, X^{*\prime}) &\triangleq \E_\epsilon\left[k_{\gamma_x}(X, X') \mid X^*, X^{*\prime}\right] \notag \\
        &= \det\!\left(\mathbf{I}_{d_X} + \frac{2}{\gamma_x^2}\boldsymbol{\Sigma}_x\right)^{-1/2} \exp\!\left(-\frac{1}{2}(\Delta X^*)^\top (\gamma_x^2\mathbf{I}_{d_X} + 2\boldsymbol{\Sigma}_x)^{-1}(\Delta X^*)\right),
    \end{align}
    and similarly for $\bar{l}(Y^*, Y^{*\prime})$. Consequently, the population $\HSIC$ satisfies $\HSIC(P_{XY}; k_{\gamma_x}, l_{\gamma_y}) = \HSIC(P_{X^*Y^*}; \bar{k}, \bar{l})$. Under isotropic noise $\boldsymbol{\Sigma}_x = \sigma_x^2\mathbf{I}_{d_X}$ and $\boldsymbol{\Sigma}_y = \sigma_y^2\mathbf{I}_{d_Y}$, with dilated bandwidths $\widetilde{\gamma}_x \triangleq \sqrt{\gamma_x^2 + 2\sigma_x^2}$, $\widetilde{\gamma}_y \triangleq \sqrt{\gamma_y^2 + 2\sigma_y^2}$ and amplitude contractions $c_x \triangleq (1 + 2\sigma_x^2/\gamma_x^2)^{-d_X/2}$, $c_y \triangleq (1 + 2\sigma_y^2/\gamma_y^2)^{-d_Y/2}$:
    \begin{align}
        \label{eq:hsic_exact_convolution}
        \HSIC_{\mathrm{obs}}(\gamma_x, \gamma_y) &= c_x c_y \cdot \HSIC_{\mathrm{latent}}(\widetilde{\gamma}_x, \widetilde{\gamma}_y), \notag \\
        \E\left[\widehat{\HSIC}_{u,\mathrm{obs}}\right] &= c_x c_y\,\HSIC_{\mathrm{latent}}(\widetilde{\gamma}_x, \widetilde{\gamma}_y) \quad (\text{for } n \ge 4).
    \end{align}
    No analogous identity is asserted for the empirical V-statistic $\widehat{\HSIC}_b$, whose diagonal terms satisfy $k_{\gamma_x}(X_r, X_r) = 1 \ne c_x$.
\end{enumerate}
\end{theorem}

\begin{proof}[Proof of Theorem~\ref{thm:kernel_convolution}]
For Part~(i), set $\Delta\epsilon \triangleq \epsilon_X - \epsilon_X'$. By symmetry of $\Delta\epsilon$ and finite fourth moments, the first and third moments of $\Delta\epsilon$ exist and vanish, and $\Cov(\Delta\epsilon) = 2\boldsymbol{\Sigma}_x$. Applying Taylor's formula through degree three with Lagrange remainder to $t \mapsto \psi(\Delta X^* + t\Delta\epsilon)$ gives $\E_\epsilon[\psi(\Delta X^* + \Delta\epsilon) \mid X^*, X^{*\prime}] = \psi(\Delta X^*) + \tr(\boldsymbol{\Sigma}_x \nabla^2\psi(\Delta X^*)) + \mathcal{R}_4$. By stationarity and the derivative reproducing property of Definition~\ref{def:admissible_kernel}, for every displacement $\mathbf{w}$ and unit direction $\mathbf{v}$, $|D_{\mathbf{v}}^4 \psi(\mathbf{w})| = |\langle D_{\mathbf{v}}^2 k(\mathbf{w}, \cdot), D_{\mathbf{v}}^2 k(\mathbf{0}, \cdot) \rangle_{\mathcal{H}_k}| \le H_k^2$, yielding $|\mathcal{R}_4| \le \frac{H_k^2}{24}\E[\|\Delta\epsilon\|_2^4] \le \frac{2}{3}H_k^2\,\E[\|\epsilon_X\|_2^4]$.

For Part~(ii), write $\epsilon_X - \epsilon_X' = \sqrt{2}\boldsymbol{\Sigma}_x^{1/2}\mathbf{G}$ with $\mathbf{G} \sim \mathcal{N}(\mathbf{0}, \mathbf{I}_{d_X})$. Diagonalizing $\boldsymbol{\Sigma}_x$ orthogonally and multiplying the univariate Gaussian integrals (including zero-eigenvalue coordinates, where the factor is $1$) establishes~\eqref{eq:kernel_anisotropic_convolution} without requiring $\boldsymbol{\Sigma}_x \succ 0$. In the population identity $\HSIC(P_{XY}; k, l) = \E[k(X, X')l(Y, Y')] + \E[k(X, X')]\E[l(Y, Y')] - 2\E[k(X, X')l(Y, Y'')]$, no term reuses an error realization within the same channel, and errors across channels are independent; conditioning each term on the latent rows yields $\HSIC(P_{XY}; k_{\gamma_x}, l_{\gamma_y}) = \HSIC(P_{X^*Y^*}; \bar{k}, \bar{l})$. For $n \ge 4$, every summand of the order-4 unbiased U-statistic $\widehat{\HSIC}_u$ \cite{song2012feature} involves four distinct sample indices with independent cross-channel errors, so taking expectations channel-wise gives~\eqref{eq:hsic_exact_convolution}.
\end{proof}

\noindent\textit{Proof.} Deferred to Appendix~\ref{app:main_proofs}.

\begin{remark}
\label{rem:hsic_v_statistic}
Theorem~\ref{thm:kernel_convolution} establishes the exact amplitude contraction $c_x c_y$ for the population $\HSIC$ and the unbiased U-statistic $\widehat{\HSIC}_u$. The empirical V-statistic $\widehat{\HSIC}_b$ used in finite-sample screening retains diagonal kernel entries $k(X_r, X_r) = 1$ that induce an $\mathcal{O}(1/n)$ expectation shift. Crucially, under the permutation null, these diagonal entries remain fixed under any permutation $\pi$, ensuring that the permutation null distribution inherits the exact same $\mathcal{O}(1/n)$ structure. Consequently, finite-sample permutation size control is exact for $\widehat{\HSIC}_b$, while power curves against fixed alternatives reflect an $\mathcal{O}(1/n)$ difference relative to the U-statistic identity.
\end{remark}

Given a dictionary of $M$ candidate kernels $\{k_1, \dots, k_M\}$, consider the convex combination $k_{\boldsymbol{\mu}} \triangleq \sum_{m=1}^M \mu_m k_m$ over the simplex $\Delta^{M-1} \triangleq \{\boldsymbol{\mu} \in \R^M : \mu_m \ge 0, \sum_{m=1}^M \mu_m = 1\}$, using Euclidean simplex projection $\Pi_{\Delta^{M-1}}$ \cite{duchi2008efficient, beck2003mirror}.

\begin{lemma}
\label{lem:mkl_pgd_convergence}
Fix a training sample of size $n_1 \ge 2$ and target kernel $l$, and let $\mathbf{g} = (g_1, \dots, g_M)^\top \in \R^M$ be the empirical alignment vector computed with declared normalization $g_m \triangleq n_1^{-2}\tr(\mathbf{K}_m\mathbf{H}\mathbf{L}\mathbf{H})$ (or $(n_1-1)^{-2}\tr(\mathbf{K}_m\mathbf{H}\mathbf{L}\mathbf{H})$ with matching penalty scaling). For $\rho > 0$, define $F_\rho(\boldsymbol{\mu}) \triangleq \mathbf{g}^\top \boldsymbol{\mu} - \frac{\rho}{2}\|\boldsymbol{\mu}\|_2^2$ on $\Delta^{M-1}$. For any step size $0 < \eta \le 1/\rho$ and initialization $\boldsymbol{\mu}^{(0)} \in \Delta^{M-1}$, the projected gradient ascent iterates $\boldsymbol{\mu}^{(t+1)} = \Pi_{\Delta^{M-1}}(\boldsymbol{\mu}^{(t)} + \eta(\mathbf{g} - \rho\boldsymbol{\mu}^{(t)}))$ converge geometrically to the unique global maximizer $\widehat{\boldsymbol{\mu}} = \Pi_{\Delta^{M-1}}(\mathbf{g}/\rho)$ (attained in one step when $\eta = 1/\rho$):
\begin{equation}
    \label{eq:mkl_geometric_contraction}
    \|\boldsymbol{\mu}^{(t+1)} - \widehat{\boldsymbol{\mu}}\|_2^2 \le (1 - \eta \rho)^2 \|\boldsymbol{\mu}^{(t)} - \widehat{\boldsymbol{\mu}}\|_2^2 \le (1 - \eta \rho) \|\boldsymbol{\mu}^{(t)} - \widehat{\boldsymbol{\mu}}\|_2^2.
\end{equation}
\end{lemma}

\begin{proof}[Proof of Lemma~\ref{lem:mkl_pgd_convergence}]
Because $F_\rho(\boldsymbol{\mu}) = -\frac{\rho}{2}\|\boldsymbol{\mu} - \mathbf{g}/\rho\|_2^2 + \frac{1}{2\rho}\|\mathbf{g}\|_2^2$, its unique maximizer over the closed convex simplex $\Delta^{M-1}$ is $\widehat{\boldsymbol{\mu}} = \Pi_{\Delta^{M-1}}(\mathbf{g}/\rho)$. Writing $\mathbf{v}^{(t)} = (1-\eta\rho)\boldsymbol{\mu}^{(t)} + \eta\mathbf{g}$ and $\mathbf{v}^* = (1-\eta\rho)\widehat{\boldsymbol{\mu}} + \eta\mathbf{g}$, first-order optimality implies $\widehat{\boldsymbol{\mu}} = \Pi_{\Delta^{M-1}}(\mathbf{v}^*)$, and non-expansiveness of Euclidean projection yields $\|\boldsymbol{\mu}^{(t+1)} - \widehat{\boldsymbol{\mu}}\|_2^2 \le \|\mathbf{v}^{(t)} - \mathbf{v}^*\|_2^2 = (1-\eta\rho)^2\|\boldsymbol{\mu}^{(t)} - \widehat{\boldsymbol{\mu}}\|_2^2$.
\end{proof}

\noindent\textit{Proof.} Deferred to Appendix~\ref{app:main_proofs}.

Maximizing a raw regularized alignment objective does not by itself guarantee optimal HSIC test power, which also depends on null variance, calibration, and sample-splitting loss; rather, we evaluate this alignment rule as one tractable way to select the evaluation kernel. Keeping a jointly selected kernel frozen on the full dataset invalidates permutation calibration, whereas either (i) freezing the kernel selected on an independent training split (Algorithm~\ref{alg:split_mkl_hsic}, Appendix~\ref{sec:proof_kernel}) or (ii) rerunning the entire selection map inside every permutation on the full dataset preserves finite-sample permutation validity.

\begin{theorem}[MKL-HSIC Test]
\label{thm:mkl_split_validity}
Let $\mathcal{D}_n = \{(\mathbf{x}_r, \mathbf{y}_r)\}_{r=1}^n$ consist of i.i.d.\ rows satisfying Assumption~\ref{ass:classical_error}(1)--(2), split into $D_{\mathrm{train}}$ and $D_{\mathrm{test}}$ by indices independent of the data values as in Algorithm~\ref{alg:split_mkl_hsic}. For any kernel selection rule computed solely on $D_{\mathrm{train}}$ (including $\widehat{\boldsymbol{\mu}} = \Pi_{\Delta^{M-1}}(\mathbf{g}/\rho)$ or any finite iterate), any uniform permutation budget $B \ge 1$ drawn independently of $\mathcal{D}_n$, and any $\alpha \in (0, 1)$, under $H_0: X^* \indep Y^*$:
\begin{equation}
    \label{eq:mkl_level_guarantee}
    \mathbb{P}_{H_0}\left(\hat{p}_{\mathrm{split}} \le \alpha \;\middle|\; D_{\mathrm{train}}\right) \le \frac{\lfloor (B+1)\alpha \rfloor}{B+1} \le \alpha, \qquad \text{and} \qquad \mathbb{P}_{H_0}\left(\hat{p}_{\mathrm{split}} \le \alpha\right) \le \alpha.
\end{equation}
\end{theorem}

\begin{proof}[Proof of Theorem~\ref{thm:mkl_split_validity}]
Under $H_0: X^* \indep Y^*$, Assumption~\ref{ass:classical_error}(2) gives $X \indep Y$. Because the split indices are independent of the i.i.d.\ rows, $D_{\mathrm{test}}$ is independent of $D_{\mathrm{train}}$ and remains i.i.d.\ from $P_X \otimes P_Y$. Conditioning on $D_{\mathrm{train}}$ freezes all learned kernel weights and tuning parameters, so the conditional evaluation law is invariant under independent uniform $Y$-reassignments. Applying the orbit-conditioning rank argument of Theorem~\ref{thm:marginal_test}(i) conditional on $D_{\mathrm{train}}$ yields the first inequality in~\eqref{eq:mkl_level_guarantee}; taking expectations over $D_{\mathrm{train}}$ gives unconditional level control.
\end{proof}

\noindent\textit{Proof.} Deferred to Appendix~\ref{app:main_proofs}.

% =========================================================================
% =========================================================================

\section{Supplementary Mathematical Results and Proofs}
\label{app:proofs_paper}

Established results from prior literature—including characteristic-function independence invariance \cite{feller1971introduction, fan2022independence}, rank-based random permutation testing \cite{lehmann2005testing, phipson2010permutation, hemerik2018exact}, HSIC and distance-covariance spectral limits \cite{gretton2005measuring, gretton2008hsictest, song2012feature, szekely2007measuring}, Euclidean simplex projection \cite{duchi2008efficient}, matrix perturbation and linear errors-in-variables identities \cite{horn2012matrix, fuller1987measurement, blom2018upper}, conditional residual orthogonality \cite{dawid1979conditional, shah2020hardness}, sub-exponential Bernstein bounds \cite{vershynin2018high}, self-normalized moderate deviations \cite{jing2003self}, and Holm/BH/BY multiplicity control \cite{holm1979simple, benjamini1995controlling, benjamini2001control}—are credited directly by citation. Below we present the supplementary derivations, counterexamples, and auxiliary lemmas supporting Sections~\ref{sec:models}--\ref{sec:pc_discovery}.

\subsection{Remark on Minimax Density-Estimation Deconvolution Rates}
\label{sec:app_deconvolution}

\begin{remark}
\label{rem:deconvolution_barrier}
For known supersmooth error models whose characteristic functions decay exponentially as $|\varphi_{\epsilon_X}(\omega)| \asymp \exp(-b\|\omega\|_2^\gamma)$ ($\gamma > 0$) and corresponding Sobolev density classes $\mathcal{F}_\beta(L)$ of smoothness order $\beta > 0$, the minimax squared $L_2$ density-estimation risk satisfies $\inf_{\hat{f}_n} \sup_{f_{X^*} \in \mathcal{F}_\beta(L)} \E[\|\hat{f}_n - f_{X^*}\|_{L_2}^2] \asymp (\log n)^{-2\beta/\gamma}$ \cite{fan1991optimal, meister2009deconvolution}, so achieving root-$L_2$ risk $\varepsilon$ requires $\log n$ of order $\varepsilon^{-\gamma/\beta}$. These are density-estimation minimax rates rather than testing lower bounds for the residual-covariance testing problem studied here.
\end{remark}

\subsection{Supplementary Results for Section~\ref{sec:regime_a_main} (Regime A and Category-Level Misclassification)}
\label{sec:proof_marginal}

\begin{lemma}
\label{lem:inadmissibility_positive_marginal_floor}
Let $T_n$ be any permutation test statistic and let $Q_n$ denote its conditional orbit $(1-\alpha)$-quantile (with no nonnegativity or unbiasedness required). For any fixed buffer $b > 0$, compare the buffered rule $T_b \triangleq \mathbf{1}\{T_n > Q_n + b\}$ with the unbuffered rule $T_0 \triangleq \mathbf{1}\{T_n > Q_n\}$. Then both rules have size at most $\alpha$ under $H_0$, and under any alternative $H_1$ the exact power difference is
\begin{equation*}
    \mathbb{P}_{H_1}(T_0 = 1) - \mathbb{P}_{H_1}(T_b = 1) = \mathbb{P}_{H_1}(Q_n < T_n \le Q_n + b).
\end{equation*}
In particular, whenever $Q_n = o_P(1)$ and $T_n \xrightarrow{P} \theta$ for $0 < \theta < b$, this power loss converges to $1$. (No conclusion at the boundary $\theta = b$ follows from consistency alone.)
\end{lemma}
\begin{proof}
Because $b > 0$, $\{T_n > Q_n + b\} \subseteq \{T_n > Q_n\}$ and $\{T_n > Q_n\} = \{T_n > Q_n + b\} \cup \{Q_n < T_n \le Q_n + b\}$ as a disjoint union. Thus $\mathbb{P}_{H_0}(T_b = 1) \le \mathbb{P}_{H_0}(T_0 = 1) \le \alpha$ and $\mathbb{P}_{H_1}(T_0 = 1) - \mathbb{P}_{H_1}(T_b = 1) = \mathbb{P}_{H_1}(Q_n < T_n \le Q_n + b)$. When $Q_n = o_P(1)$ and $T_n \xrightarrow{P} \theta \in (0, b)$, $\mathbb{P}_{H_1}(Q_n < T_n \le Q_n + b) \to 1$.
\end{proof}

\begin{theorem}[Category-Level Independent Misclassification and Replicate Effect-Size Deattenuation]
\label{thm:discrete_invariance}
Let $(A^*, B^*)$ be latent categorical variables taking values in $[r] \times [c]$ ($r, c \ge 2$) with joint table $\boldsymbol{\Pi}^*$, marginals $p^*, q^*$, and defect $\mathbf{D}^* \triangleq \boldsymbol{\Pi}^* - p^*(q^*)^\top$ (so $\mathbf{1}_r^\top \mathbf{D}^* = \mathbf{0}$ and $\mathbf{D}^*\mathbf{1}_c = \mathbf{0}$). Assume a category-level independent misclassification model with row-stochastic transition matrices $\mathbf{M}_X \in \R^{r \times r}$ and $\mathbf{M}_Y \in \R^{c \times c}$:
\begin{equation}
    \label{eq:cat_misclass_model}
    \mathbb{P}(A = k, B = \ell \mid A^* = i, B^* = j) = (\mathbf{M}_X)_{ik} (\mathbf{M}_Y)_{j\ell}.
\end{equation}
(Note that~\eqref{eq:cat_misclass_model} does \emph{not} follow from bucketing continuous additive measurements $X = X^* + \epsilon_X, Y = Y^* + \epsilon_Y$: for example, if $X^* = Y^* = U \sim \mathrm{Unif}[-1, 1]$ split at $0$ with independent $\mathrm{Unif}[-1, 1]$ errors, the true joint negative probability is $\E[(1-U)^2]/4 = 1/3$ whereas the matrix product yields $5/16$. For bucketed continuous measurements, latent continuous independence still implies observed bucket independence directly by Section~\ref{sec:regime_a_main}.)
\begin{enumerate}[label=(\roman*),leftmargin=2.0em]
    \item \textbf{Category Independence Invariance:} Under~\eqref{eq:cat_misclass_model}, the observed table $\boldsymbol{\Pi}$ and marginals $p, q$ satisfy $\boldsymbol{\Pi} = \mathbf{M}_X^\top \boldsymbol{\Pi}^* \mathbf{M}_Y$, $p = \mathbf{M}_X^\top p^*$, $q = \mathbf{M}_Y^\top q^*$, and
    \begin{equation*}
        \mathbf{D} \triangleq \boldsymbol{\Pi} - p q^\top = \mathbf{M}_X^\top \mathbf{D}^* \mathbf{M}_Y.
    \end{equation*}
    Thus $A^* \indep B^* \implies A \indep B$ (and conversely if both $\mathbf{M}_X$ and $\mathbf{M}_Y$ are nonsingular), so testing category independence using the raw observed Pearson $\chi_n^2$ or permutation test requires zero buffer ($\Delta_m \equiv 0$; cf.\ \cite{bross1954misclassification, mote1965investigation}).
    \item \textbf{Replicate Agreement and Regularized Pearson Effect-Size Deattenuation:} Suppose i.i.d.\ subjects have two technical categorical measurements per channel that are conditionally independent given the latent category and share symmetric transition matrices $\mathbf{M}_X = a_X \mathbf{I}_r + \frac{1-a_X}{r}\mathbf{1}_r\mathbf{1}_r^\top$ and $\mathbf{M}_Y = a_Y \mathbf{I}_c + \frac{1-a_Y}{c}\mathbf{1}_c\mathbf{1}_c^\top$ with retentions $a_X \triangleq 1-\varepsilon_1 \in (0, 1]$ and $a_Y \triangleq 1-\varepsilon_2 \in (0, 1]$. Then the duplicate agreement probability on channel $X$ is $p_{\mathrm{agree}, X} = \frac{1}{r} + (1 - \frac{1}{r})a_X^2$ (without requiring uniform latent marginals), and
    \begin{equation}
        \label{eq:replicate_discrete_attenuation_est}
        \hat{a}_X \triangleq \sqrt{\max\!\left(0, \frac{r \hat{p}_{\mathrm{agree}, X} - 1}{r - 1}\right)} \xrightarrow{P} a_X, \qquad
        \hat{a}_Y \triangleq \sqrt{\max\!\left(0, \frac{c \hat{p}_{\mathrm{agree}, Y} - 1}{c - 1}\right)} \xrightarrow{P} a_Y.
    \end{equation}
    Assume additionally equal-probability latent marginals ($p^* = \frac{1}{r}\mathbf{1}_r, q^* = \frac{1}{c}\mathbf{1}_c$) and positive observed margins. Then $\mathbf{D} = a_X a_Y \mathbf{D}^*$ and the population Pearson divergence $J \triangleq \sum_{k,\ell} D_{k\ell}^2 / (p_k q_\ell)$ satisfies $J = a_X^2 a_Y^2 J^*$. For a regularization sequence $\tau_n \downarrow 0$ (or reporting non-resolution whenever $\hat{a}_X \hat{a}_Y = 0$), the regularized plug-in estimator
    \begin{equation}
        \label{eq:discrete_deattenuated_statistic}
        \widehat{J}^* \triangleq \frac{\chi_n^2 / n}{\max(\hat{a}_X^2 \hat{a}_Y^2, \, \tau_n)} \xrightarrow{P} J^*
    \end{equation}
    consistently estimates the latent Pearson divergence $J^*$ (and corresponding Cram\'er's effect size $\sqrt{\widehat{J}^*/\min(r-1, c-1)}$). Independence hypothesis testing must use the raw observed $\chi_n^2$ or its permutation distribution; the deattenuated effect $\widehat{J}^*$ is \emph{not} compared against central $\chi^2$ critical values.
\end{enumerate}
\end{theorem}
\begin{proof}
Part~(i) follows directly by expanding $\boldsymbol{\Pi} = \mathbf{M}_X^\top \boldsymbol{\Pi}^* \mathbf{M}_Y$ and subtracting $p q^\top = \mathbf{M}_X^\top p^*(q^*)^\top \mathbf{M}_Y$. For Part~(ii), conditioning on $A^* = i$ yields
\begin{equation*}
    \mathbb{P}(A^{(1)} = A^{(2)} \mid A^* = i) = \left(a_X + \frac{1-a_X}{r}\right)^2 + (r-1)\left(\frac{1-a_X}{r}\right)^2 = \frac{1}{r} + \left(1 - \frac{1}{r}\right)a_X^2,
\end{equation*}
independent of $i$, yielding~\eqref{eq:replicate_discrete_attenuation_est} by the law of large numbers and continuous mapping. Since $\mathbf{1}_r^\top \mathbf{D}^* = \mathbf{0}$ and $\mathbf{D}^*\mathbf{1}_c = \mathbf{0}$, $\mathbf{M}_X^\top \mathbf{D}^* \mathbf{M}_Y = a_X a_Y \mathbf{D}^*$. When $p^* = \frac{1}{r}\mathbf{1}_r$ and $q^* = \frac{1}{c}\mathbf{1}_c$, $\mathbf{M}_X^\top p^* = p^*$ and $\mathbf{M}_Y^\top q^* = q^*$, so $J = a_X^2 a_Y^2 J^*$. For i.i.d.\ subjects and fixed positive margins, $\chi_n^2 / n \xrightarrow{P} J$, and Slutsky's theorem yields~\eqref{eq:discrete_deattenuated_statistic} whenever $a_X a_Y > 0$.
\end{proof}

\begin{proposition}
\label{prop:calibration_inflation}
Suppose a nonnegative biased V-statistic $T_n$ (such as $\widehat{\HSIC}_b$ or $\mathcal{V}_n^2$) satisfies $n T_n \xrightarrow{d} Q \triangleq \sum_{k=1}^\infty \lambda_k Z_k^2$ under $H_0: X \indep Y$ for a nonzero summable nonnegative spectrum $\{\lambda_k\}_{k=1}^\infty$ and $Z_k \stackrel{\mathrm{i.i.d.}}{\sim} \mathcal{N}(0, 1)$ \cite{gretton2008hsictest, szekely2007measuring}, and suppose a variance estimator satisfies $n^2 \widehat{\Var}_0(T_n) \xrightarrow{P} V_0 \triangleq 2\sum_{k=1}^\infty \lambda_k^2 > 0$. Then:
\begin{equation*}
    Z_n \triangleq \frac{T_n}{\sqrt{\widehat{\Var}_0(T_n)}} \xrightarrow{d} Z_\infty \triangleq \frac{Q}{\sqrt{V_0}}.
\end{equation*}
The limiting variable $Z_\infty$ has mean $C_0/\sqrt{V_0} \ge 1/\sqrt{2}$ (where $C_0 \triangleq \sum_{k=1}^\infty \lambda_k$) and variance $1$ (which need not match finite-sample moments of $Z_n$). For any cutoff $z > 0$, the limiting rejection probability is $\mathbb{P}(Q > z\sqrt{V_0})$, which depends on the complete eigenvalue sequence rather than the mean-to-SD ratio alone (for instance, a rank-one spectrum gives $\mathbb{P}(\chi_1^2 > 1.96\sqrt{2}) \approx 0.096$). Even centering $T_n$ by its null mean leaves a centered weighted-$\chi^2$ limit rather than a normal limit. Numerical rejection rates reported in Section~\ref{sec:experiments} are simulation-based negative-control evaluations.
\end{proposition}
\begin{proof}
Immediate from Slutsky's theorem applied to $n T_n \xrightarrow{d} Q$ and $\sqrt{n^2 \widehat{\Var}_0(T_n)} \xrightarrow{P} \sqrt{V_0}$, together with $\E[Q] = \sum_k \lambda_k = C_0$ and $\Var(Q) = 2\sum_k \lambda_k^2 = V_0$.
\end{proof}

\subsection{Supplementary Results for Section~\ref{sec:rkhs_kernel} (Fixed-Law HSIC Power and Attenuation)}
\label{sec:proof_mdla}
\label{sec:proof_kernel}

\begin{table}[!htbp]
\centering
\caption{Sample-Splitting Alignment-Selected MKL-HSIC Test Algorithm.}
\label{alg:split_mkl_hsic}
\footnotesize\begin{tabular}{@{}p{\linewidth}@{}}
\toprule
\textit{Algorithm 1: Sample-Splitting MKL-HSIC Test} \\
\midrule
\textit{Input:} I.i.d.\ sample $\mathcal{D}_n = \{(\mathbf{x}_r, \mathbf{y}_r)\}_{r=1}^n$, candidate kernels $\{k_1, \dots, k_M\}$, target kernel $l$, split fraction $s_{\mathrm{split}} \in (0, 1)$, ridge penalty $\rho > 0$, permutation budget $B \ge 1$, sampling mode (with replacement from $\mathcal{S}_{n_2}$ or without replacement from $\mathcal{S}_{n_2}\setminus\{\mathrm{id}\}$). \newline
\textit{Output:} Calibrated $p$-value $\hat{p}_{\mathrm{split}}$ and selected weights $\widehat{\boldsymbol{\mu}}$. \\
\midrule
1. \textit{Independent Sample Split:} Partition row indices independently of the data values into training split $D_{\mathrm{train}}$ ($n_1 = \lfloor s_{\mathrm{split}} n \rfloor \ge 2$) and evaluation split $D_{\mathrm{test}}$ ($n_2 = n - n_1 \ge 4$). \newline
2. \textit{Training Selection:} On $D_{\mathrm{train}}$ only, choose all paired-data tuning parameters, compute the alignment vector $\mathbf{g}$ (declaring the normalization $n_1^{-2}$ or $(n_1-1)^{-2}$), and compute the exact maximizer $\widehat{\boldsymbol{\mu}} = \Pi_{\Delta^{M-1}}(\mathbf{g}/\rho)$ via \cite{duchi2008efficient}. Freeze $k_{\widehat{\boldsymbol{\mu}}} = \sum_{m=1}^M \hat{\mu}_m k_m$ and $l$. \newline
3. \textit{Evaluation Permutation Test on $D_{\mathrm{test}}$:} Compute $T_{\mathrm{obs}} = \widehat{\HSIC}_u(D_{\mathrm{test}}; k_{\widehat{\boldsymbol{\mu}}}, l)$ (or $\widehat{\HSIC}_b$). Independently of $\mathcal{D}_n$, draw $B$ uniform permutations $\pi_1, \dots, \pi_B$ of $[n_2]$ under the declared sampling mode, compute $T^{(b)}$ using the same HSIC statistic on permuted $Y$ assignments, and set:
\[
    \hat{p}_{\mathrm{split}} \triangleq \frac{1 + \sum_{b=1}^B \mathbf{1}\{T^{(b)} \ge T_{\mathrm{obs}}\}}{1 + B}.
\] \\
\midrule
\textit{Return:} Valid $p$-value $\hat{p}_{\mathrm{split}}$ and frozen weights $\widehat{\boldsymbol{\mu}}$. \\
\bottomrule
\end{tabular}
\end{table}

\begin{proposition}
\label{thm:mdla}
Fix an observed law $P$ with independent isotropic Gaussian error blocks $\boldsymbol{\Sigma}_x = \sigma_x^2\mathbf{I}_{d_X}$, $\boldsymbol{\Sigma}_y = \sigma_y^2\mathbf{I}_{d_Y}$, fixed positive Gaussian RBF bandwidths $\gamma_x, \gamma_y > 0$, and a dependent latent law $P_{X^*Y^*}$. Put $\theta^* \triangleq \HSIC_{\mathrm{latent}}(\widetilde{\gamma}_x, \widetilde{\gamma}_y) > 0$, $a \triangleq c_x c_y = (1 + 2\sigma_x^2/\gamma_x^2)^{-d_X/2}(1 + 2\sigma_y^2/\gamma_y^2)^{-d_Y/2} > 0$, and let $\sigma_1^2(P) \triangleq 16\,\Var_P\{\E_P[h_{\mathrm{obs}}(W_1, W_2, W_3, W_4) \mid W_1]\} > 0$ denote the first-order U-statistic projection variance (which depends on the full joint law $P$ and bandwidths). Then $\sqrt{n}(\widehat{\HSIC}_{u,\mathrm{obs}} - a\theta^*)/\sigma_1(P) \xrightarrow{d} \mathcal{N}(0, 1)$, so any level-$\alpha$ test with critical value $q_{n,\alpha} = o_P(1)$ is consistent whenever $\theta^* > 0$. Moreover, if an alternative sequence $P_n$ independently satisfies this standardized CLT with variance $\sigma_{1,n}^2$ and a deterministic upper bound $b_{n,\alpha}$ covers $q_{n,\alpha}$ with probability $1 - o(1)$, then
\begin{equation*}
    a_n \theta_n^* \ge b_{n,\alpha} + \frac{z_{1-\beta}\sigma_{1,n}}{\sqrt{n}} + o(n^{-1/2})
\end{equation*}
eventually suffices for $\liminf_{n\to\infty}\mathbb{P}_{P_n}(\text{reject}) \ge 1 - \beta$. (We do not claim a local minimax detection boundary, nor point identification of oracle latent-variance bandwidths from a single noisy sample.)
\end{proposition}
\begin{proof}
By Theorem~\ref{thm:kernel_convolution}(ii), $\E[\widehat{\HSIC}_{u,\mathrm{obs}}] = a\theta^*$. Hoeffding's non-degenerate U-statistic CLT gives $\sqrt{n}(\widehat{\HSIC}_{u,\mathrm{obs}} - a\theta^*)/\sigma_1(P) \xrightarrow{d} \mathcal{N}(0, 1)$, and the sufficient power inequality follows directly by standardizing $\{\widehat{\HSIC}_{u,\mathrm{obs}} > q_{n,\alpha}\}$ on $\{q_{n,\alpha} \le b_{n,\alpha}\}$.
\end{proof}

\subsection{Supplementary Results for Section~\ref{sec:regime_b_main} (Auxiliary Regime B Lemmas)}
\label{sec:proof_conditioning}

\begin{lemma}
\label{lem:precision_perturbation}
Assume $\Cov(\mathbf{V}^*, \boldsymbol{\epsilon}) = \mathbf{0}$, $\boldsymbol{\Sigma}_* \triangleq \Var(\mathbf{V}^*) \succ 0$, and $\mathbf{E} \triangleq \Var(\boldsymbol{\epsilon}) \succeq 0$ (permitting noiseless components). Let $\boldsymbol{\Omega}_* \triangleq \boldsymbol{\Sigma}_*^{-1}$, $\boldsymbol{\Omega} \triangleq (\boldsymbol{\Sigma}_* + \mathbf{E})^{-1}$, and $\mathbf{F} \triangleq \mathbf{E}^{1/2}$. Then:
\begin{equation}\label{eq:woodbury_precision}
    \boldsymbol{\Omega} = \boldsymbol{\Omega}_* - \boldsymbol{\Omega}_*\mathbf{F}\left(\mathbf{I} + \mathbf{F}\boldsymbol{\Omega}_*\mathbf{F}\right)^{-1}\mathbf{F}\boldsymbol{\Omega}_*.
\end{equation}
If $(\boldsymbol{\Omega}_*)_{ij} = 0$, then $\Omega_{ij} = -[\boldsymbol{\Omega}_*\mathbf{F}(\mathbf{I} + \mathbf{F}\boldsymbol{\Omega}_*\mathbf{F})^{-1}\mathbf{F}\boldsymbol{\Omega}_*]_{ij}$. (This algebraic identity does not assert nonvanishing for a given graph path, nor interpret precision entries as conditional independence outside jointly Gaussian models.)
\end{lemma}
\begin{proof}
Immediate from the Woodbury matrix identity applied to $(\boldsymbol{\Sigma}_* + \mathbf{F}\mathbf{F})^{-1}$ \cite[Section~0.7.4]{horn2012matrix}, since $\mathbf{I} + \mathbf{F}\boldsymbol{\Omega}_*\mathbf{F} \succeq \mathbf{I} \succ 0$.
\end{proof}

\begin{proposition}
\label{prop:eiv_leakage}
Consider the linear structural equations $X^* = \boldsymbol{\beta}_X^\top \mathbf{Z}^* + U_X$ and $Y^* = \beta_{XY}X^* + \boldsymbol{\beta}_Y^\top \mathbf{Z}^* + U_Y$. Assume finite second moments, disturbance orthogonality $\Cov(U_X, U_Y) = 0$ and $\Cov(U_j, \mathbf{Z}^*) = \mathbf{0}$ for $j \in \{X, Y\}$, and uncorrelated classical errors ($\Cov(\mathbf{V}^*, \boldsymbol{\epsilon}) = \mathbf{0}$ with diagonal $\boldsymbol{\Sigma}_\epsilon$). Set $\mathbf{b}_Y \triangleq \boldsymbol{\beta}_Y + \beta_{XY}\boldsymbol{\beta}_X$ and $\boldsymbol{\Sigma} \triangleq \boldsymbol{\Sigma}_* + \mathbf{E} \succ 0$. Then the affine projection residuals of $X$ and $Y$ onto $\mathbf{Z}$ satisfy \cite{fuller1987measurement, blom2018upper}:
\begin{equation*}
    C_{\mathrm{aff}} = \beta_{XY}\Var(U_X) + \boldsymbol{\beta}_X^\top \left(\boldsymbol{\Sigma}_* - \boldsymbol{\Sigma}_*\boldsymbol{\Sigma}^{-1}\boldsymbol{\Sigma}_*\right)\mathbf{b}_Y = \beta_{XY}\Var(U_X) + \boldsymbol{\beta}_X^\top \mathbf{E}\boldsymbol{\Sigma}^{-1}\boldsymbol{\Sigma}_*\mathbf{b}_Y.
\end{equation*}
If $\Var(U_X) > 0$, then $\beta_{XY}\Var(U_X) = 0 \iff \beta_{XY} = 0$. When $\beta_{XY} = 0$ and $\mathbf{Z}^*$ is scalar, $C_{\mathrm{aff}} = \beta_X\beta_Y\Var(Z^*)/(1 + \SNR_Z)$, vanishing in the noiseless limit. No equivalence to conditional independence is claimed without further distributional assumptions.
\end{proposition}
\begin{proof}
Direct expansion of $\Cov(X, Y) - \Cov(X, \mathbf{Z})\boldsymbol{\Sigma}^{-1}\Cov(\mathbf{Z}, Y)$ using $\Cov(\mathbf{Z}, X) = \boldsymbol{\Sigma}_*\boldsymbol{\beta}_X$, $\Cov(\mathbf{Z}, Y) = \boldsymbol{\Sigma}_*\mathbf{b}_Y$, and $\boldsymbol{\Sigma}_* - \boldsymbol{\Sigma}_*\boldsymbol{\Sigma}^{-1}\boldsymbol{\Sigma}_* = \mathbf{E}\boldsymbol{\Sigma}^{-1}\boldsymbol{\Sigma}_*$.
\end{proof}

\begin{theorem}[Conditional Residuals Are Not Generally Marginally Independent]
\label{thm:residual_vanishing_scope}
Let $X, Y \in L_2$ satisfy $X \indep Y \mid T$ with residuals $R_X = X - \E[X \mid T]$ and $R_Y = Y - \E[Y \mid T]$. Then $R_X \indep R_Y \mid T$ and $\E[R_X R_Y] = 0$ \cite{dawid1979conditional, shah2020hardness}. If at least one of the conditional distributions $P(R_X \in \cdot \mid T)$ or $P(R_Y \in \cdot \mid T)$ is almost surely constant in $T$, then $R_X \indep R_Y$ marginally. In general, however, $R_X$ and $R_Y$ need not be marginally independent: for example, with independent $U, V \sim \mathcal{N}(0, 1)$ and independent uniform $A \in \{1, 2\}$, setting $X = AU, Y = AV, T = A$ gives $X \indep Y \mid T$ and $C = 0$, yet $\Cov(R_X^2, R_Y^2) = \Var(A^2) > 0$.
\end{theorem}
\begin{proof}
Subtracting $T$-measurable conditional means preserves conditional independence, so $\E[R_X R_Y \mid T] = 0$ and $\E[R_X R_Y] = 0$. If $P(R_X \in B_1 \mid T) = \mu_X(B_1)$ a.s., then $\mathbb{P}(R_X \in B_1, R_Y \in B_2) = \E[\mu_X(B_1)\mathbb{P}(R_Y \in B_2 \mid T)] = \mu_X(B_1)\mathbb{P}(R_Y \in B_2)$. The heteroscedastic counterexample follows by direct computation since $\E[X \mid A] = \E[Y \mid A] = 0$, $R_X^2 = A^2 U^2$, and $R_Y^2 = A^2 V^2$.
\end{proof}

\begin{lemma}
\label{lem:noise_gradient_contraction_paper}
Assume $\boldsymbol{\epsilon}_Z \indep \mathbf{Z}^*$ with $\E[\boldsymbol{\epsilon}_Z] = \mathbf{0}$, finite covariance $\boldsymbol{\Sigma}_{\boldsymbol{\epsilon}_Z}$, $\E[\|\mathbf{Z}^*\|_2^2] < \infty$, $\E[g(\mathbf{Z}^*)^2] < \infty$, and let $h: \R^q \to \R$ be differentiable with $M_H$-Lipschitz gradient $\nabla h$. Then for every $\tau \in [0, 1]$ and $\widetilde{\mathbf{Z}} \triangleq \mathbf{Z}^* + \tau\boldsymbol{\epsilon}_Z$:
\begin{equation*}
    \left|\E\left[g(\mathbf{Z}^*)\nabla h(\widetilde{\mathbf{Z}})^\top \boldsymbol{\epsilon}_Z\right]\right| \le \tau M_H \|g(\mathbf{Z}^*)\|_{L_2}\,\tr(\boldsymbol{\Sigma}_{\boldsymbol{\epsilon}_Z}).
\end{equation*}
\end{lemma}
\begin{proof}
Because $\nabla h$ is $M_H$-Lipschitz, $\|\nabla h(\mathbf{z})\|_2 \le \|\nabla h(\mathbf{0})\|_2 + M_H\|\mathbf{z}\|_2$; combined with $\E[\|\mathbf{Z}^*\|_2^2] < \infty$, $\E[g(\mathbf{Z}^*)^2] < \infty$, and finite noise covariance, $g(\mathbf{Z}^*)\nabla h(\mathbf{Z}^*)^\top \boldsymbol{\epsilon}_Z$ is integrable and has mean zero by independence. Subtracting it and applying $\|\nabla h(\widetilde{\mathbf{Z}}) - \nabla h(\mathbf{Z}^*)\|_2 \le \tau M_H \|\boldsymbol{\epsilon}_Z\|_2$ with Cauchy--Schwarz gives the stated bound.
\end{proof}

\begin{lemma}
\label{lem:gcm_bernstein_concentration}
Let $D_2$ be an i.i.d.\ evaluation sample of size $n_2$ independent of the training split $D_1$, and assume a certified null bound $|C| \le \Delta$ for the true conditional-mean residuals $R_X, R_Y$. Conditional on $D_1$, assume the fitted residuals $\hat{R}_X, \hat{R}_Y$ have conditional $\psi_2$ norms bounded by $K_X, K_Y$ and the fitted mean errors have conditional $L_2(P_Z)$ norms bounded by $d_X, d_Y$. Then for any $\delta \in (0, 1)$, under $H_0: X^* \indep Y^* \mid \mathbf{Z}^*$ there exists a universal constant $c' > 0$ such that:
\begin{equation*}
    \mathbb{P}_{H_0}\left(|\bar{C}| > \Delta + d_X d_Y + c' K_X K_Y \left(\sqrt{\frac{\ln(2/\delta)}{n_2}} + \frac{\ln(2/\delta)}{n_2}\right) \;\middle|\; D_1\right) \le \delta.
\end{equation*}
(If $K_X, K_Y, d_X, d_Y$, or $\Delta$ are estimated, their failure probabilities must be budgeted separately.)
\end{lemma}
\begin{proof}
Conditional on $D_1$, conditional-mean orthogonality gives $|\E[\hat{R}_X \hat{R}_Y \mid D_1]| \le |C| + d_X d_Y \le \Delta + d_X d_Y$. Without requiring within-row independence of $\hat{R}_X$ and $\hat{R}_Y$, the centered product $\hat{R}_X\hat{R}_Y - \E[\hat{R}_X\hat{R}_Y \mid D_1]$ has conditional $\psi_1$ norm at most $c K_X K_Y$; applying Bernstein's sub-exponential inequality \cite[Theorem~2.8.1]{vershynin2018high} across the i.i.d.\ evaluation rows of $D_2$ yields the result.
\end{proof}

\begin{lemma}
\label{lem:sample_splitting_level}
Let $D_2$ be independent of $D_1$ and put $\theta_c \triangleq \E[\bar{C} \mid D_1]$. Suppose $D_1$ (augmented with auxiliary replicates, validation subsamples, or instruments) produces a directional certificate $\hat{s}(D_1)$ such that $G \triangleq \{\hat{s}(D_1) \le s_{\mathrm{true}}\}$ satisfies $\mathbb{P}(G^c) \le \gamma < \alpha$. Assume $B(D_1, s)$ is nonincreasing in $s$ and satisfies $B(D_1, s_{\mathrm{true}}) \ge |\theta_c|$ on $G$ under $H_0$ (including nuisance bias), and let $Q(D_1)$ be a conditional quantile satisfying $\mathbb{P}_{H_0}(|\bar{C} - \theta_c| > Q(D_1) \mid D_1) \le \alpha - \gamma$ on $G$. Then rejecting when $|\bar{C}| > B(D_1, \hat{s}(D_1)) + Q(D_1)$ satisfies $\mathbb{P}_{H_0}(\text{Reject}) \le \alpha$.
\end{lemma}
\begin{proof}
On $G$, monotonicity gives $B(D_1, \hat{s}(D_1)) \ge B(D_1, s_{\mathrm{true}}) \ge |\theta_c|$, so by the triangle inequality $\{|\bar{C}| > B(D_1, \hat{s}) + Q(D_1)\} \subseteq \{|\bar{C} - \theta_c| > Q(D_1)\}$, which has conditional probability at most $\alpha - \gamma$ given $D_1$. Union-bounding with $\mathbb{P}(G^c) \le \gamma$ gives $\mathbb{P}_{H_0}(\text{Reject}) \le (\alpha - \gamma) + \gamma = \alpha$.
\end{proof}

\begin{proposition}
\label{prop:replicate_eiv_comparator}
For the separately labeled duplicate-data linear comparator in Section~\ref{sec:experiments}, let independent subjects $r = 1, \dots, n$ have two noisy records $\mathbf{V}_r^{(m)} = \mathbf{V}_r^* + \boldsymbol{\epsilon}_r^{(m)}$ ($m = 1, 2$), where each error vector is uncorrelated with $\mathbf{V}_r^*$ and $\Cov(\boldsymbol{\epsilon}_r^{(1)}, \boldsymbol{\epsilon}_r^{(2)}) = \mathbf{0}$. Then $\Cov(\mathbf{V}^{(1)}, \mathbf{V}^{(2)}) = \boldsymbol{\Sigma}_*$. Estimating $\boldsymbol{\Sigma}_*$ by the symmetrized subject-level empirical cross-covariance $\widehat{\boldsymbol{\Sigma}}_*$ and, for $\boldsymbol{\Sigma}_{*,SS} \succ 0$, estimating the latent partial covariance $C_* \triangleq \Sigma_{*,ij} - \boldsymbol{\Sigma}_{*,iS}\boldsymbol{\Sigma}_{*,SS}^{-1}\boldsymbol{\Sigma}_{*,Sj}$ by plug-in (with a declared regularization/non-resolution rule if $\widehat{\boldsymbol{\Sigma}}_{*,SS}$ is singular) yields a studentized root-$n$ test of $C_* = 0$ via the subject-level covariance CLT and delta method under finite fourth moments, fixed dimension, and nonzero influence-function variance. Under a jointly Gaussian latent law this tests latent CI; subjects, not individual duplicate records, are the inferential units.
\end{proposition}
\begin{proof}
Immediate from $\Cov(\mathbf{V}^{(1)}, \mathbf{V}^{(2)}) = \Var(\mathbf{V}^*) = \boldsymbol{\Sigma}_*$, the multivariate central limit theorem across i.i.d.\ subjects, and the delta method applied to the smooth map $\mathbf{A} \mapsto A_{ij} - \mathbf{A}_{iS}\mathbf{A}_{SS}^{-1}\mathbf{A}_{Sj}$ on $\{\mathbf{A}_{SS} \succ 0\}$.
\end{proof}

\subsection{Supplementary Results for Section~\ref{sec:pc_discovery} (Local Power and High-Dimensional Uniform Lemmas)}
\label{sec:proof_guarantees}

\begin{lemma}
\label{lem:pitman_power}
(a) For a signed covariance statistic satisfying $\sqrt{n}(\bar{C} - C_n)/\hat{\sigma}_C \xrightarrow{d} \mathcal{N}(0, 1)$ with $\hat{\sigma}_C \xrightarrow{P} \sigma > 0$ under $C_n = c/\sqrt{n}$, the zero-floor two-sided normal rule has limiting power $1 - \Phi(z_{1-\alpha/2} - |c|/\sigma) + \Phi(-z_{1-\alpha/2} - |c|/\sigma)$ (this is not the local power formula for degenerate HSIC or distance covariance). (b) For the sensitivity-adjusted rule of Theorem~\ref{thm:conditional_test} with $\Delta_n \to \Delta > 0$, $U_{\Delta, n} - \Delta_n = o_P(1)$, and $C_n = \mathcal{O}(n^{-1/2})$, the rejection probability converges to zero.
\end{lemma}
\begin{proof}
Part~(a) follows from Slutsky's theorem and the continuous mapping theorem. For Part~(b), $\sqrt{n}\bar{C} = \mathcal{O}_P(1)$ implies $|\bar{C}| = \mathcal{O}_P(n^{-1/2})$, while $U_{\Delta, n} \xrightarrow{P} \Delta > 0$, so $\mathbb{P}(|\bar{C}| - z_{1-\alpha_1/2}\hat{\sigma}_C/\sqrt{n} > U_{\Delta, n}) \to 0$.
\end{proof}

\begin{remark}
\label{lem:collider_orientation}
In an isolated scalar linear collider $X^* \to Z^* \leftarrow Y^*$ with marginally independent endpoints ($X^* \indep Y^*$) and classical measurement error, the observable affine partial covariance is $C_{\mathrm{aff}} = -\Cov(X^*, Z^*)\Cov(Y^*, Z^*)/\Var(Z)$, which equals its latent counterpart multiplied by $\Var(Z^*)/\Var(Z) = \SNR_Z / (1 + \SNR_Z)$. For general conditioning sets and multivariate graphs, collider orientation is governed directly by the deterministic separator proof of Theorem~\ref{thm:pc_oracle_recovery}. Likewise, for the bounded-kernel biased $\widehat{\HSIC}_b$ option in Assumption~\ref{ass:residual_moment_faithfulness}(ii), the uniform permuted-comparison bound $\mathbb{P}(T_a^\pi \ge T_a) \le A e^{-cn}$ follows without an i.i.d.\ permuted-pair assumption: conditional on the paired data, the average of permuted $\widehat{\HSIC}_b$ equals $n^{-2}(n-1)^{-1}\tr(\mathbf{H}\mathbf{K}\mathbf{H})\tr(\mathbf{H}\mathbf{L}\mathbf{H}) = \mathcal{O}(n^{-1})$; for centered feature maps with uniformly bounded norms, swapping two assignments changes the empirical cross-covariance operator norm by $\mathcal{O}(n^{-1})$, so a permutation-reveal martingale (coupling any two next labels by one transposition) yields an exponentially small upper tail above its $\mathcal{O}(n^{-1/2})$ mean, while bounded differences give the corresponding lower tail for the observed operator at a fixed HSIC gap.
\end{remark}

\begin{lemma}
\label{lem:uniform_moderate_deviation}
Let $W_{r,a} \triangleq R_{r,i}(S) R_{r,j}(S) - C_a$ for i.i.d.\ rows $r = 1, \dots, n$ and nonempty queries $a = (i, j, S) \in \mathcal{Q}_n$, with oracle sample mean $\bar{C}_a^0 \triangleq \frac{1}{n}\sum_{r=1}^n R_{r,i}(S)R_{r,j}(S)$ and sample SD $\hat{\sigma}_a^0 \triangleq (\frac{1}{n}\sum_{r=1}^n (R_{r,i}(S)R_{r,j}(S) - \bar{C}_a^0)^2)^{1/2}$. Assume uniformly over $a \in \mathcal{Q}_n$ that $\Var(W_{r,a}) \ge v_0 > 0$ and $\E[|W_{r,a}|^3] \le M < \infty$, and set $L_n \triangleq (d_{\max} + 2)\log p_n + \log n = o(n^{1/3})$.
\begin{enumerate}[label=(\alph*),leftmargin=2.0em]
    \item \textit{Oracle Self-Normalized Moderate Deviations:} For any deterministic $b_n = o(n^{1/6})$, uniformly in $a \in \mathcal{Q}_n$ and $0 \le t \le b_n$ (defining the statistic conservatively as $0$ when $\hat{\sigma}_a^0 = 0$):
    \begin{equation*}
        \sup_{a \in \mathcal{Q}_n}\sup_{0 \le t \le b_n}\left|\frac{\mathbb{P}\left(\sqrt{n}|\bar{C}_a^0 - C_a|/\hat{\sigma}_a^0 > t\right)}{2(1 - \Phi(t))} - 1\right| = o(1).
    \end{equation*}
    \item \textit{Family Null Error Under a Joint Nuisance and Coverage Event:} Fix $\alpha_{1,n}/\alpha_n \in (0, 1)$ independent of $n$ with $\alpha_n = p_n^{-(d_{\max}+2)}n^{-\gamma}$ ($\gamma > 0$), so $t_n \triangleq z_{1-\alpha_{1,n}/2} = o(n^{1/6})$. Assume a joint event $N_n$ with $\mathbb{P}(N_n^c) = o(1)$ on which $\sup_a \sqrt{n}|\bar{C}_a - \bar{C}_a^0|/\hat{\sigma}_a^0 = o(L_n^{-1/2})$ and $\sup_a |\hat{\sigma}_a/\hat{\sigma}_a^0 - 1| = o(L_n^{-1})$, and separately assume simultaneous envelope coverage failure $\mathbb{P}(\exists a \in \mathcal{Q}_{0,n}, |S|\ge 1 : U_{\Delta, a} < \Delta_a) \le |\mathcal{Q}_{0,n}^{\mathrm{cond}}|\alpha_{2,n}(1 + o(1)) + o(1)$. Then the unconditional family null error satisfies
    \begin{equation*}
        \mathbb{P}(\exists a \in \mathcal{Q}_{0,n} : \text{reject } H_{0,a}) \le \mathbb{P}(N_n^c) + \mathbb{P}(\exists a : U_{\Delta,a} < \Delta_a) + |\mathcal{Q}_{0,n}^{\mathrm{cond}}|\alpha_{1,n}(1+o(1)) + |\mathcal{Q}_{0,n}^{\mathrm{marg}}|\alpha_n \to 0.
    \end{equation*}
    (Such nuisance and extreme-coverage events are explicit high-level assumptions and do not follow from entrywise moment consistency alone.)
    \item \textit{Connected Nonempty Query Retention:} Suppose on a joint event of probability $1 - o(1)$ that the fitted score error $|\bar{C}_a - \bar{C}_a^0|$, operational envelope excess $(U_{\Delta, a} + a_{a,n} - \Delta_a)_+$, and critical radius $t_n \hat{\sigma}_a/\sqrt{n}$ sum to at most $\kappa_n / 2$ for every connected nonempty query $a \in \mathcal{Q}_{\mathrm{conn},n}$ ($|S| \ge 1$), and that the centered oracle products have uniformly bounded $\psi_1$ norm $K$. Then there exists a universal constant $c > 0$ such that:
    \begin{equation*}
        \mathbb{P}(\text{any connected nonempty query missed}) \le o(1) + 2|\mathcal{Q}_n|\exp\!\left(-c n \min\!\left\{\frac{\kappa_n^2}{K^2}, \, \frac{\kappa_n}{K}\right\}\right),
    \end{equation*}
    which vanishes whenever the displayed exponential term tends to zero.
\end{enumerate}
\end{lemma}
\begin{proof}
For~(a), writing $S_a \triangleq \sum_{r=1}^n W_{r,a}$ and $V_a^2 \triangleq \sum_{r=1}^n W_{r,a}^2$, the studentized event $\{\sqrt{n}|\bar{C}_a^0 - C_a|/\hat{\sigma}_a^0 > t\}$ coincides on $\{\hat{\sigma}_a^0 > 0\}$ with $\{|S_a|/V_a > t / \sqrt{1 + t^2/n}\}$. Applying the self-normalized moderate deviation theorem \cite[Theorem~2.1]{jing2003self} to $W_{r,a}$ and $-W_{r,a}$ for $0 \le t \le b_n = o(n^{1/6})$ (where $(1 + t)^3 / \sqrt{n} \to 0$ and the cutoff shift $t - t/\sqrt{1+t^2/n} = \mathcal{O}(t^3/n)$ changes $1 - \Phi(t)$ by $1 + o(1)$) yields~(a). For~(b), on $N_n$ and simultaneous envelope coverage, any false conditional rejection implies $\sqrt{n}|\bar{C}_a^0 - C_a|/\hat{\sigma}_a^0 > t_n - o(L_n^{-1/2})$; applying~(a) and union-bounding over disjoint marginal and conditional null counts yields the stated family bound. For~(c), on the joint event where the three error terms sum to at most $\kappa_n/2$, any missed connected query ($|C_a| - \Delta_a \ge \kappa_n$) implies $|\bar{C}_a^0 - C_a| \ge \kappa_n / 2$; applying sub-exponential Bernstein's inequality \cite[Theorem~2.8.1]{vershynin2018high} and a union bound over $|\mathcal{Q}_n|$ completes the proof.
\end{proof}

\section{Additional Empirical Evaluations and Diagnostic Benchmarks}
\label{app:extra_empirical}

This appendix collects the supplementary Tier~1 unit tests, single-realization Tier~2 structure-learning stress tests, and descriptive Sachs marginal screening plot summarized in Section~\ref{sec:experiments}.

\subsection{Pearson Correlation Attenuation Under Classical Additive Noise}
\begin{figure}[!htbp]
\centering
\includegraphics[width=0.65\textwidth]{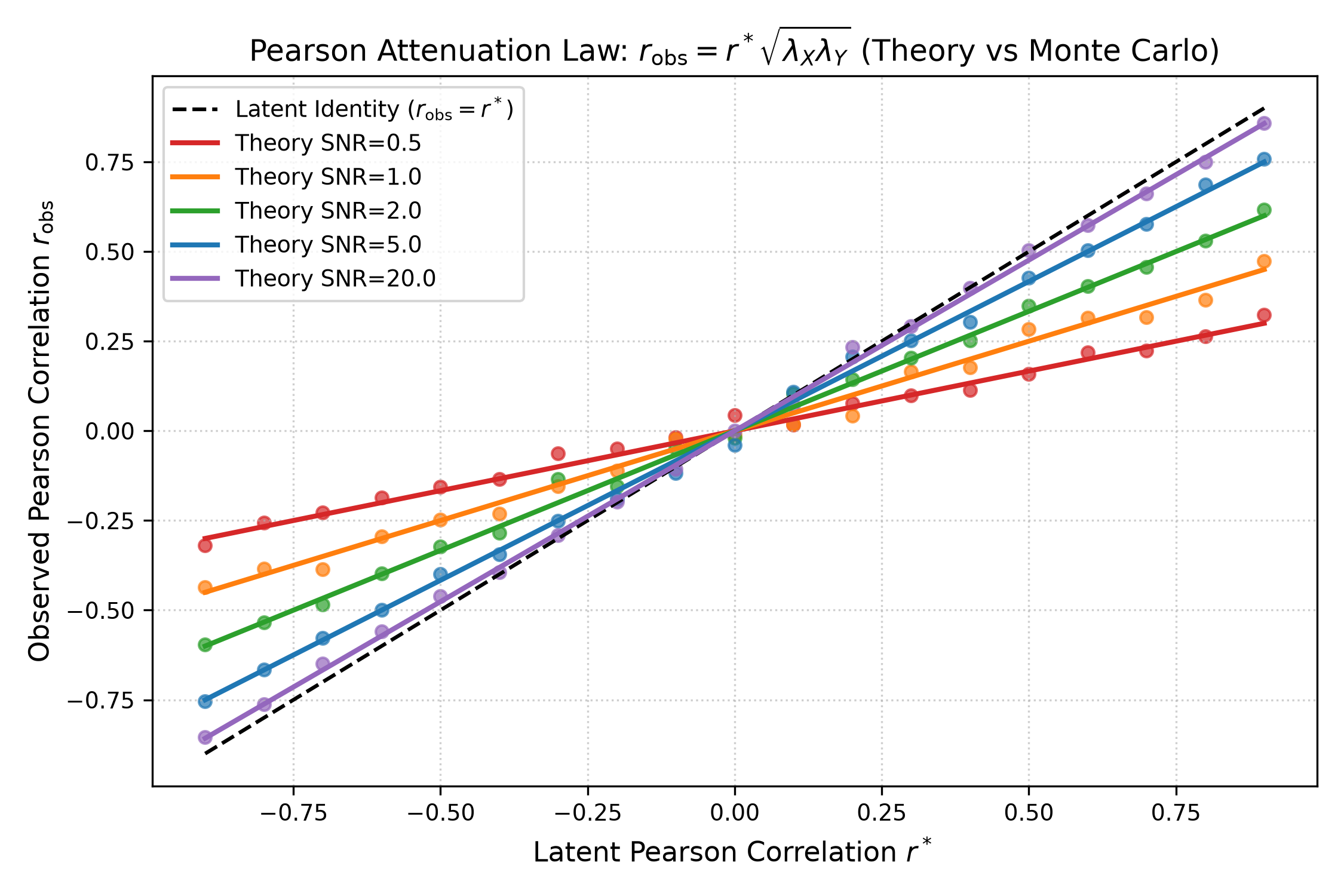}
\caption{Population Pearson attenuation and simulated sample correlations. Under classical additive errors and positive latent variances, $\rho_{\mathrm{obs}} = \rho^* / \sqrt{(1+\SNR_X^{-1})(1+\SNR_Y^{-1})}$. This is a population identity, not an equality for individual sample correlations.}
\label{fig:pearson_atten}
\end{figure}

Under independent classical additive measurement errors with positive latent variances, the population Pearson correlation obeys $\rho_{\mathrm{obs}} = \rho^* / \sqrt{(1+\SNR_X^{-1})(1+\SNR_Y^{-1})}$. Figure~\ref{fig:pearson_atten} illustrates this population attenuation curve alongside simulated sample correlations across noise levels.

\subsection{Split-Sample MKL-HSIC and Category-Level Misclassification Deattenuation}
We evaluate: (i) sample-splitting MKL-HSIC (Table~\ref{alg:split_mkl_hsic}, Theorem~\ref{thm:mkl_split_validity}, $N_{\mathrm{mc}} = 200$, $n = 400$, 50/50 split, $B = 99$ permutations, seed $2026$) under $H_0$ ($X^* \sim \mathcal{U}[-\pi, \pi], Y^* \sim \mathcal{N}(0, 1)$) and $H_1$ ($Y^* = \cos(3X^*) + U_Y$ with $U_Y \sim \mathcal{N}(0, 0.25^2)$), compared descriptively against median-bandwidth HSIC and a negative control (\emph{Frozen Full-Sample MKL}) that selects kernel weights on the full dataset and keeps them frozen during permutation; and (ii) category-level symmetric misclassification on a $4 \times 4$ table ($N = 600$, $N_{\mathrm{mc}} = 500$, seed $2026$) generated by discretizing standard bivariate normal latent draws ($\rho^* \in \{0, 0.65\}$) at population normal quartiles $\Phi^{-1}(0.25, 0.50, 0.75)$ and applying the symmetric category transition $\mathbf{M}(\varepsilon) = (1-\varepsilon)\mathbf{I}_4 + \frac{\varepsilon}{4}\mathbf{1}_4\mathbf{1}_4^\top$ ($\varepsilon \in \{0.10, 0.25, 0.40\}$) independently to two replicates per channel. Raw Pearson $\chi^2$ testing is evaluated on replicate~1, while $\hat{\varepsilon} \triangleq 1 - \frac{1}{2}(\hat{a}_X + \hat{a}_Y)$ and deattenuated bias-corrected Cram\'er's $\widehat{V}^*$ (with denominator floor $\tau = 0.02$) measure effect-size estimation under Theorem~\ref{thm:discrete_invariance}.

\begin{figure}[!htbp]
\centering
\includegraphics[width=0.92\textwidth]{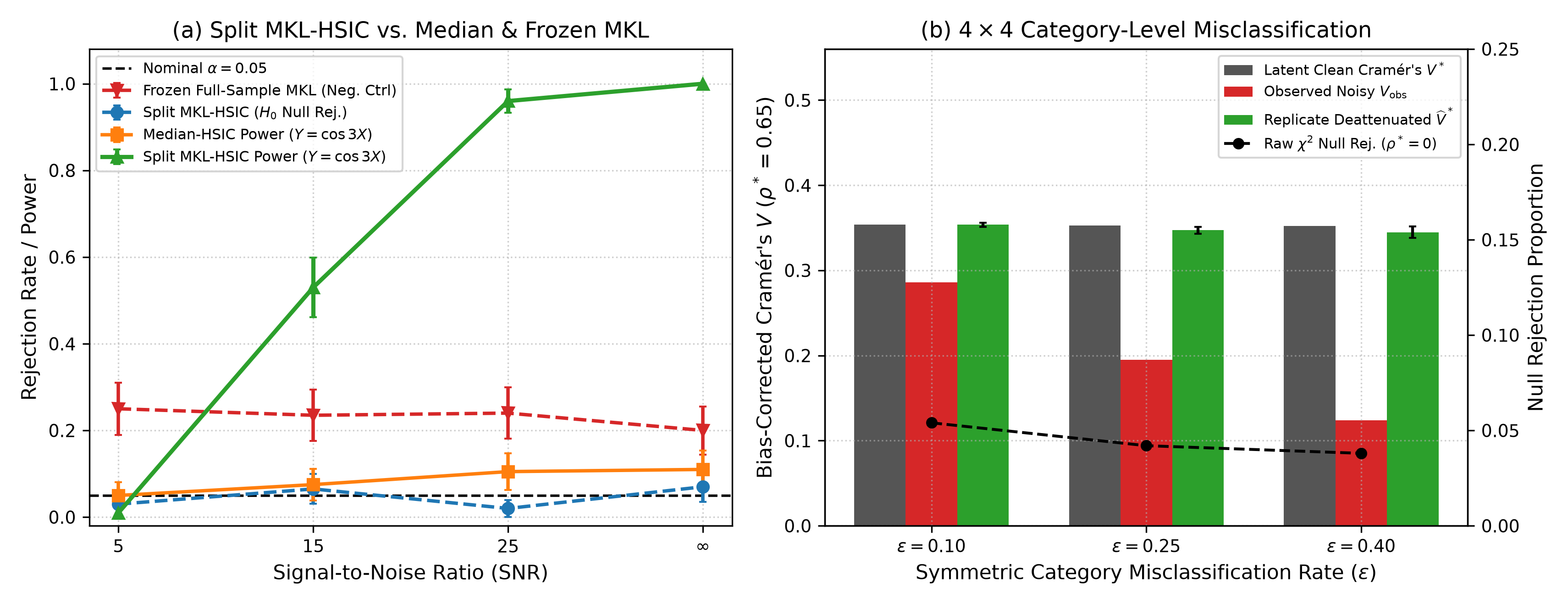}
\caption{Empirical evaluation of split-sample MKL-HSIC ($N_{\mathrm{mc}}=200$, $n=400$, $B=99$) and $4\times 4$ category-level symmetric misclassification ($N_{\mathrm{mc}}=500$, $N=600$). Panel~(a) contrasts split-sample MKL against median-HSIC and frozen full-sample selection; Panel~(b) displays raw Pearson null rejection proportions ($\rho^*=0$) alongside replicate-deattenuated Cram\'er's $V$ estimates ($\rho^*=0.65$).}
\label{fig:mkl_discrete}
\end{figure}

\begin{table}[!htbp]
\centering
\caption{Split MKL-HSIC ($N_{\mathrm{mc}}=200$, $n=400$, $B=99$, plug-in SEs where shown) and $4\times 4$ Category-Level Symmetric Misclassification ($N_{\mathrm{mc}}=500$, $N=600$, raw $\chi^2$ null rejection on replicate~1 and deattenuated Cram\'er's $V$ effect size).}
\label{tab:mkl_discrete}
\resizebox{\linewidth}{!}{%
\footnotesize\setlength{\tabcolsep}{4pt}%
\begin{tabular}{lccccc|lccccc}
\toprule
\multicolumn{6}{c|}{(a) Split MKL-HSIC vs.\ Median \& Frozen Full-Sample MKL} & \multicolumn{6}{c}{(b) $4\times 4$ Category-Level Symmetric Misclassification} \\
\cmidrule(lr){1-6}\cmidrule(lr){7-12}
$\SNR$ & Frozen MKL FP & Split MKL FP & Med FP & Med Pow & Split MKL Pow & $\varepsilon$ ($\hat{\varepsilon}$) & Raw $\chi^2$ Null Rej & $V^*_{\mathrm{lat}}$ ($\rho^*=0.65$) & $V_{\mathrm{obs}}$ & $\widehat{V}^*$ (Deatt.) & Raw Alt Rej \\
\midrule
$5.0$ & $0.250\,(0.031)$ & $0.030\,(0.012)$ & $0.045\,(0.015)$ & $0.050$ & $0.010\,(0.007)$ & $0.10\,(0.101)$ & $0.054\,(0.010)$ & $0.354$ & $0.286$ & $0.354\,(0.001)$ & $1.000$ \\
$15.0$ & $0.235\,(0.030)$ & $0.065\,(0.017)$ & $0.080\,(0.019)$ & $0.075$ & $0.530\,(0.035)$ & $0.25\,(0.251)$ & $0.042\,(0.009)$ & $0.353$ & $0.195$ & $0.347\,(0.002)$ & $1.000$ \\
$25.0$ & $0.240\,(0.030)$ & $0.020\,(0.010)$ & $0.040\,(0.014)$ & $0.105$ & $0.960\,(0.014)$ & $0.40\,(0.400)$ & $0.038\,(0.009)$ & $0.352$ & $0.124$ & $0.345\,(0.003)$ & $0.980$ \\
$\infty$ & $0.200\,(0.028)$ & $0.070\,(0.018)$ & $0.065\,(0.017)$ & $0.110$ & $1.000\,(0.000)$ & \multicolumn{6}{c}{\textit{Note: Category-level symmetric channel; raw $\chi^2$ uses $\Delta_m \equiv 0$.}} \\
\bottomrule
\end{tabular}%
}
\end{table}

In Table~\ref{tab:mkl_discrete}(a), the split method's null rejection proportions range from $0.020$ to $0.070$ across $200$ runs per cell, while its alternative rejection proportions are $0.010$, $0.530$, $0.960$, and $1.000$ (at $\SNR=5.0$, split MKL rejects less frequently than median-HSIC, $2/200$ vs.\ $10/200$). Finite-sample level validity of the split procedure follows from Theorem~\ref{thm:mkl_split_validity}, not from these Monte Carlo proportions; the frozen full-sample comparator illustrates post-selection inflation ($0.200$--$0.250$) when kernel selection is not rerun per permutation. In Table~\ref{tab:mkl_discrete}(b), the categorical simulation applies symmetric misclassification directly at the category level (not continuous additive bucketing). Raw $\chi^2$ null rejection proportions on replicate~1 range from $0.038$ to $0.054$ across $500$ runs per cell, while the mean deattenuated effect-size estimates $\widehat{V}^*$ ($0.345$--$0.354$) closely track the clean-table Cram\'er's $V^*$ ($0.352$--$0.354$); we do not compare the deattenuated statistic against central $\chi^2$ cutoffs.

\subsection{Quadratic Confounding: Fitted Global and Interior Curvature Heuristics}
To examine non-monotone confounding, we generate $Z^* \sim \mathcal{N}(0, 1)$, independent $U_X, U_Y \sim \mathcal{N}(0, 0.75^2)$, $X^* = 0.65(Z^{*2} - 1) + 0.35 Z^* + U_X$, $Y_0^* = 0.65(Z^{*2} - 1) + 0.35 Z^* + U_Y$, and $Y_1^* = Y_0^* + 0.85 X^*$, with independent Gaussian measurement errors $\epsilon_Z \sim \mathcal{N}(0, 1/\SNR)$, $\epsilon_X \sim \mathcal{N}(0, \Var(X^*)/\SNR)$, and $\epsilon_Y \sim \mathcal{N}(0, \Var(Y^*)/\SNR)$ for $\SNR \in \{3.0, 5.0, 10.0\}$ ($n = 500$, $N_{\mathrm{mc}} = 250$ independent trials, seed $3033$, 2-fold cross-fitting with GCV-tuned RBF KRR). As established in Remark~\ref{rem:signed_interval_and_rkhs_caveat}, the fitted RKHS Hessian norm $\widehat{M}_{H,\mathrm{RKHS}} = 2\sqrt{3}\,g\,\|\hat{m}\|_{\mathcal{H}_k}$ bounds only the fitted surrogate regression $\hat{m}$, not the latent regression $m^*$ (whose exact generating curvatures here are $H_X^* = H_{Y_0}^* = 1.30$ and $H_{Y_1}^* = 2.405$), even after rescaling by $((1+\SNR_0)/\SNR_0)^2$. Accordingly, both the global-fit rule $\widehat{M}_{H,\mathrm{RKHS}}$ and the interior 80\%-quantile median Hessian rule $\widehat{M}_{H,\mathrm{med}}$ are reported transparently as \emph{sensitivity heuristics}.

\begin{figure}[!htbp]
\centering
\includegraphics[width=0.90\textwidth]{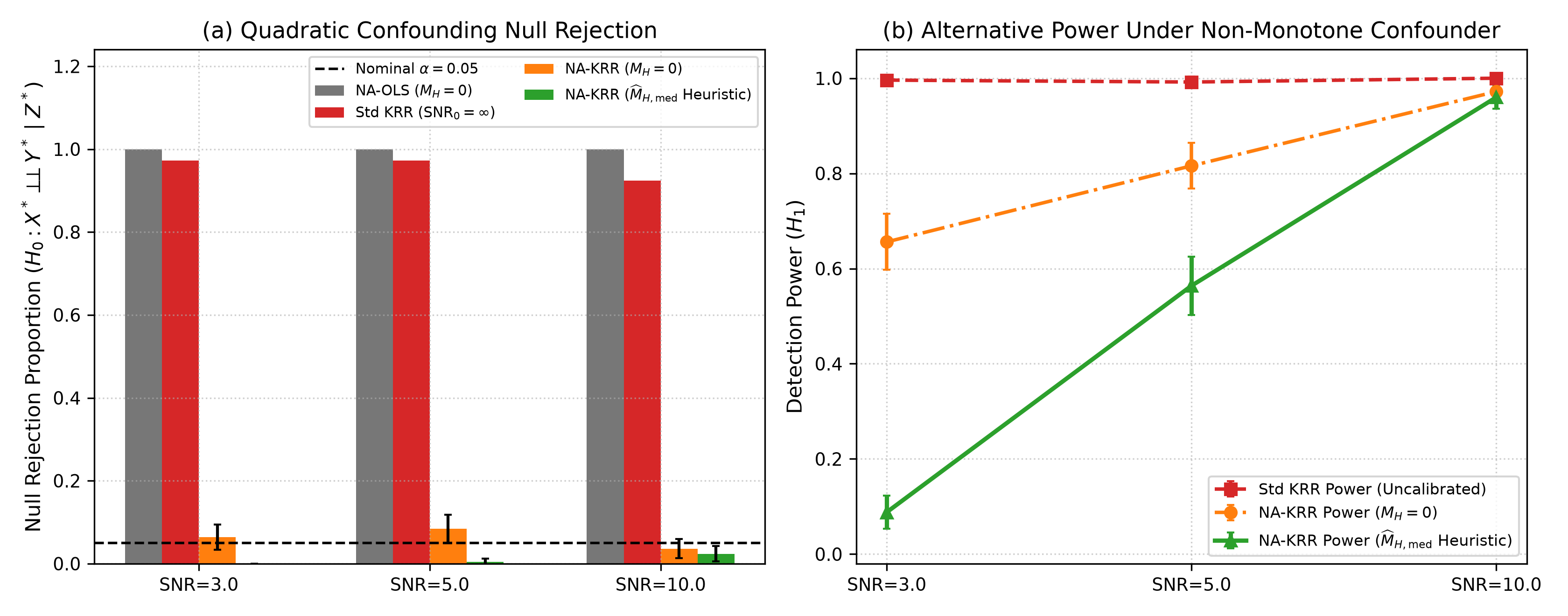}
\caption{Quadratic-confounding simulation ($n=500$, $250$ Monte Carlo datasets per cell). The global RKHS bound is for the fitted surrogate regression, not the latent regression. In Table~\ref{tab:nonlin_curvature}, the global-fit rule produced $0/250$ null rejections (exact one-sided 95\% upper bound $0.0119$) with power $0/250$ to $0.068$; the interior heuristic reached $0.960$ empirical power at $\SNR = 10$.}
\label{fig:nonlinear_curvature}
\end{figure}

\begin{table}[!htbp]
\centering
\caption{Quadratic Confounding Simulation ($n=500$, $N_{\mathrm{mc}}=250$, $\alpha=0.05$): NA-OLS, Standard KRR, and Fitted Global vs.\ Interior Curvature Heuristics. Parentheses report plug-in Monte Carlo SEs where nonzero; $0/250$ carries exact one-sided 95\% upper bound $0.0119$.}
\label{tab:nonlin_curvature}
\resizebox{\linewidth}{!}{%
\footnotesize\setlength{\tabcolsep}{3.5pt}%
\begin{tabular}{lcccccccc}
\toprule
$\SNR$ & NA-OLS FP & Std-KRR FP & KRR ($M_H=0$) FP & KRR ($\widehat{M}_{H,\mathrm{med}}$) FP & KRR ($\widehat{M}_{H,\mathrm{RKHS}}$) FP & Pow ($M_H=0$) & Pow ($\widehat{M}_{H,\mathrm{med}}$) & Pow ($\widehat{M}_{H,\mathrm{RKHS}}$) \\
\midrule
$3.0$ & $1.000$ & $0.972$ & $0.064\,(0.015)$ & $0/250$ & $0/250$ & $0.656\,(0.030)$ & $0.088\,(0.018)$ & $0/250$ \\
$5.0$ & $1.000$ & $0.972$ & $0.084\,(0.018)$ & $0.004\,(0.004)$ & $0/250$ & $0.816\,(0.025)$ & $0.564\,(0.031)$ & $0.004\,(0.004)$ \\
$10.0$ & $1.000$ & $0.924$ & $0.036\,(0.012)$ & $0.024\,(0.010)$ & $0/250$ & $0.972\,(0.010)$ & $0.960\,(0.012)$ & $0.068\,(0.016)$ \\
\bottomrule
\end{tabular}%
}
\end{table}

In Table~\ref{tab:nonlin_curvature}, NA-OLS rejected every simulated null ($1.000$) and standard KRR rejected $0.924$--$0.972$ of null datasets, while omitting the curvature term ($M_H=0$) produced $0.064$ and $0.084$ null rejections at $\SNR=3$ and $\SNR=5$. The fitted global RKHS heuristic produced $0/250$ null rejections across all displayed SNRs (one-sided exact 95\% upper bound $1 - 0.05^{1/250} \approx 0.0119$) with power $0/250$ to $0.068$, illustrating severe conservatism at $n=500$, whereas the interior heuristic yielded $0/250$ to $0.024$ null rejections with $0.088$ to $0.960$ empirical power. Both curvature rules remain empirical heuristics unless latent curvature $H_j^*$ and an envelope UCB are separately certified.

\subsection{Direct D-Separation Evaluation Across Conditioning Cardinalities \texorpdfstring{$|Z|$}{|Z|}}
We evaluate direct d-separation queries across $N_{\mathrm{DAG}} = 20$ independent linear-Gaussian random DAGs ($p = 14$, edge probability $0.25$, $n = 1{,}500$, coordinate generating $\SNR = 2.0$, $\alpha = 0.05$, $|Z| \in \{1, 2, 3, 4\}$). The counts of sampled null queries at depths one through four are $2{,}739$, $2{,}982$, $3{,}000$, and $3{,}000$, with $3{,}000$ sampled d-connected queries at each depth. Reported means and standard errors are unweighted DAG-level summaries across the $20$ independent DAGs ($\bar{r} = \frac{1}{20}\sum_{g=1}^{20} r_g$, $\mathrm{SE} = s(r_1, \dots, r_{20})/\sqrt{20}$). Because the simulation supplies the coordinate generating $\SNR = 2.0$ rather than a separate multivariate directional certificate $s_S$ for correlated conditioners, single-sample NA-GCM here uses the linear spectral envelope with $s_0 = 2.0$ and influence-function UCB; duplicate-based Replicate GCM (Proposition~\ref{prop:replicate_eiv_comparator}) uses additional technical replicate measurements.

\begin{figure}[!htbp]
\centering
\includegraphics[width=0.90\textwidth]{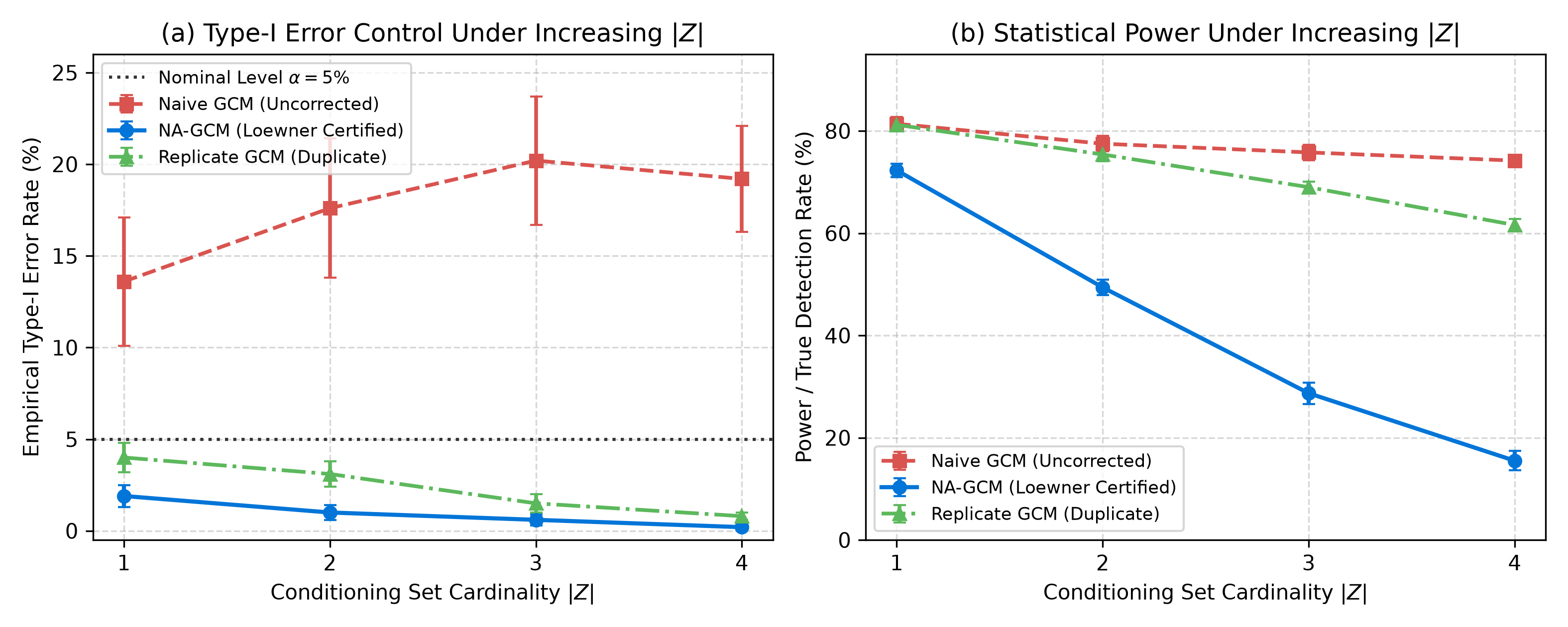}
\caption{Direct d-separation query evaluation across conditioning cardinalities $|Z| \in \{1, 2, 3, 4\}$ ($20$ linear-Gaussian DAGs, $n=1{,}500$, coordinate $\SNR=2.0$, DAG-level mean $\pm$ SE).}
\label{fig:dsep_multiz}
\end{figure}

\begin{table}[!htbp]
\centering
\caption{Direct d-Separation Query Performance Across Conditioning Cardinalities ($20$ linear-Gaussian DAGs, $p=14$, $n=1{,}500$, coordinate generating $\SNR=2.0$, $\alpha=0.05$; DAG-level mean and SE across 20 DAGs; sampled null totals $2{,}739/2{,}982/3{,}000/3{,}000$, d-connected totals $3{,}000$ each).}
\label{tab:dsep_benchmark}
\resizebox{\linewidth}{!}{%
\footnotesize\setlength{\tabcolsep}{4pt}%
\begin{tabular}{lccc|ccc}
\toprule
Cardinality & \multicolumn{3}{c|}{Null Rejection Proportion ($\alpha = 0.05$)} & \multicolumn{3}{c}{Alternative Rejection Proportion} \\
\cmidrule(lr){2-4} \cmidrule(lr){5-7}
$|Z|$ & Naive GCM & Noise-Aware GCM & Replicate GCM (Dup) & Naive GCM & Noise-Aware GCM & Replicate GCM (Dup) \\
\midrule
$|Z| = 1$ & $0.136\,(0.035)$ & $\mathbf{0.019\,(0.006)}$ & $0.040\,(0.008)$ & $0.814\,(0.013)$ & $0.723\,(0.013)$ & $0.812\,(0.012)$ \\
$|Z| = 2$ & $0.176\,(0.038)$ & $\mathbf{0.010\,(0.004)}$ & $0.031\,(0.007)$ & $0.775\,(0.015)$ & $0.494\,(0.015)$ & $0.754\,(0.013)$ \\
$|Z| = 3$ & $0.202\,(0.035)$ & $\mathbf{0.006\,(0.003)}$ & $0.015\,(0.005)$ & $0.758\,(0.015)$ & $0.287\,(0.021)$ & $0.690\,(0.011)$ \\
$|Z| = 4$ & $0.192\,(0.029)$ & $\mathbf{0.002\,(0.001)}$ & $0.008\,(0.002)$ & $0.742\,(0.009)$ & $0.155\,(0.019)$ & $0.616\,(0.012)$ \\
\bottomrule
\end{tabular}%
}
\end{table}

In Table~\ref{tab:dsep_benchmark}, Naive GCM exhibits $0.136$--$0.202$ null rejection proportions, whereas single-sample NA-GCM yields $0.002$--$0.019$ null rejections while its alternative rejection proportions decline sharply with conditioning cardinality from $0.723$ ($|Z|=1$) to $0.155$ ($|Z|=4$). Duplicate Replicate GCM (which has additional replicate measurements) retains $0.616$--$0.812$ alternative rejection proportions at $0.008$--$0.040$ null rejection proportions. This steep power attenuation for single-sample NA-GCM directly reflects the mathematical decay of the conditioning block's minimum eigenvalue $\lambda_{\min}(\mathbf{P}_Z)$ as correlated conditioning variables accumulate, which mechanically expands the certified worst-case leakage bound $U_{a,n}$ in the absence of technical replicate measurements.

\begin{figure}[!htbp]
\centering
\includegraphics[width=0.96\textwidth]{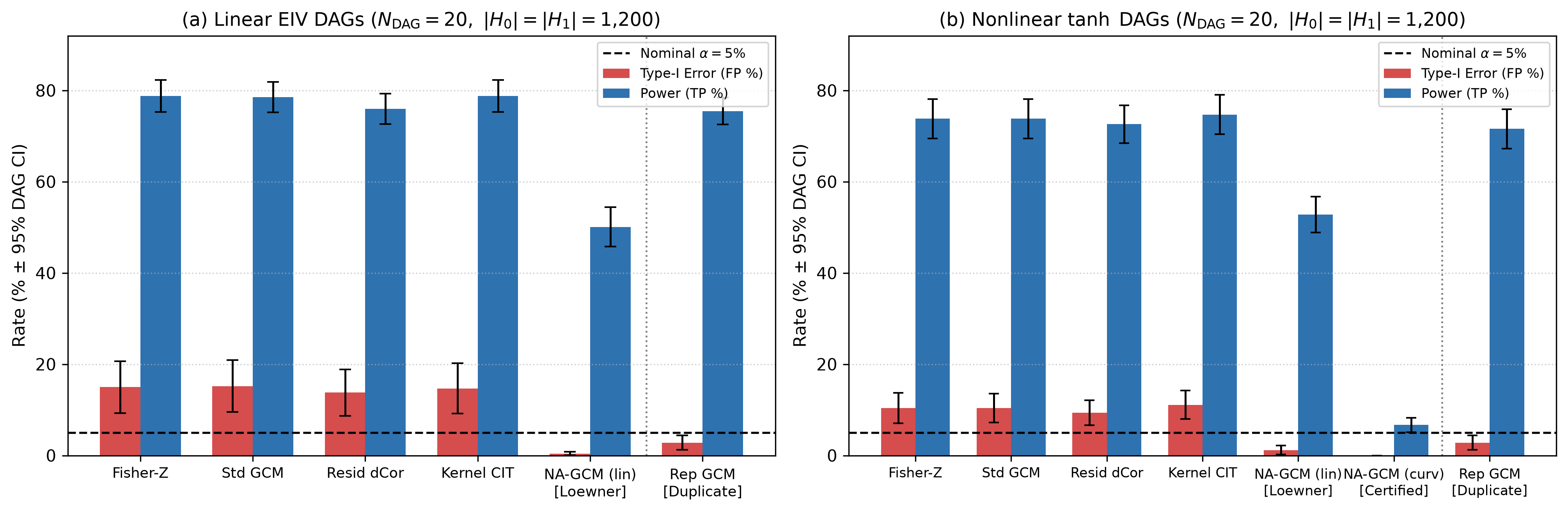}
\caption{Head-to-head surrogate tests and duplicate-measurement comparator in linear and nonlinear $\tanh$ simulation designs ($20$ DAGs per regime, coordinate $\SNR=2.0$, showing DAG-level means with $1.96\times$ DAG-level SE intervals). Plot axis labels ``Kernel CIT'' and ``[Certified]'' correspond to the 20-feature RFF approximation and fitted-curvature heuristic documented in Table~\ref{tab:competing_ci_tests}.}
\label{fig:competing_ci_benchmark}
\end{figure}

\subsection{Single-Realization Simulated DREAM4-Style Hill Network}
We evaluate a single realization of a synthetic 10-gene nonlinear Hill-kinetics simulation ($p = 10$, $n = 1{,}000$, $15$ acyclic reference regulatory edges) modeled on the 10-gene DREAM4 topology \cite{marbach2010revealing, marbach2012wisdom}, corrupted with independent coordinate Gaussian noise at generating $\SNR = 3.0$. Because the graph search uses linear projections under nonlinear Hill kinetics and a coordinate generating SNR (rather than certified multivariate directional bounds for the linear class), the sensitivity settings $s_0$ in Table~\ref{tab:dream4_results} are stratified solely by whether $s_0 \le 3.0$ or $s_0 > 3.0$, not by full certification of all leakage hypotheses.

\begin{figure}[!htbp]
\centering
\includegraphics[width=0.90\textwidth]{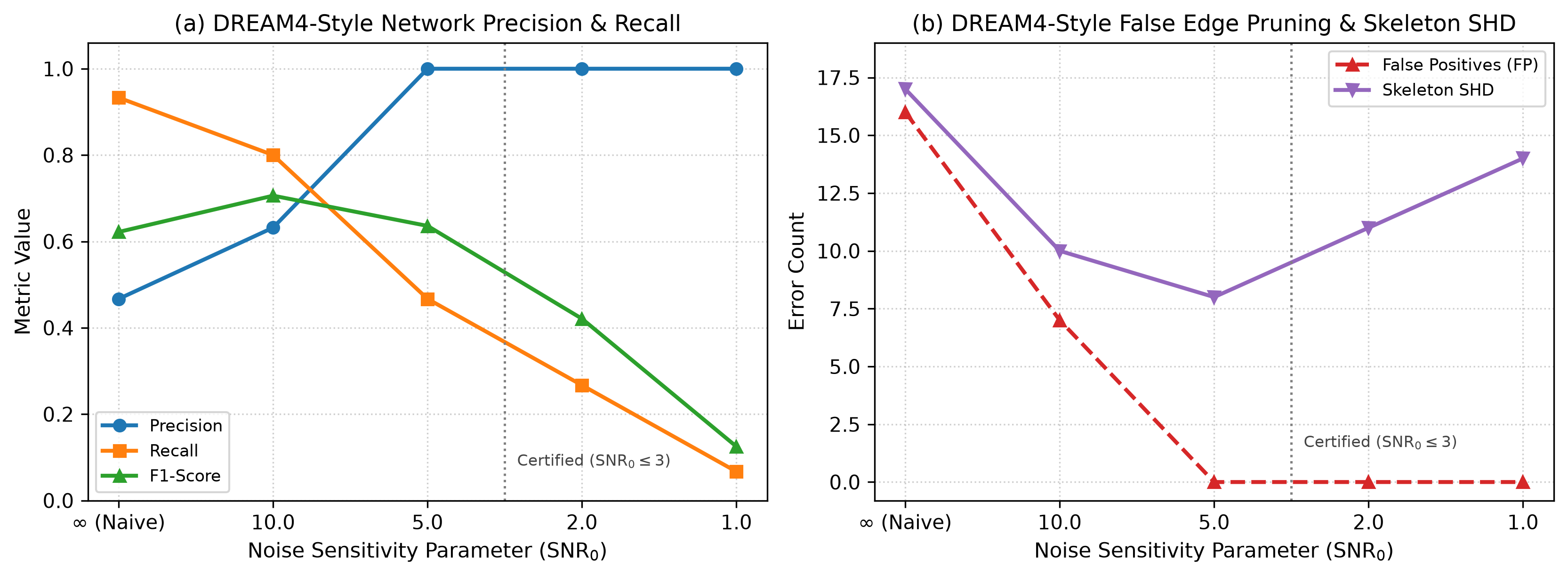}
\caption{Reference-graph agreement in one synthetic DREAM4-style Hill-network realization ($p=10$, $n=1{,}000$, $15$ reference regulatory edges, generating coordinate $\SNR = 3.0$).}
\label{fig:dream4_bench}
\end{figure}

\begin{table}[!htbp]
\centering
\caption{Reference-graph agreement in one synthetic DREAM4-style Hill-network realization ($p=10$, $n=1{,}000$, $15$ reference regulatory edges). Values of $s_0$ are sensitivity settings stratified by whether they exceed the generating coordinate SNR of $3.0$, not by whether all leakage hypotheses hold.}
\label{tab:dream4_results}
\resizebox{\linewidth}{!}{%
\footnotesize\setlength{\tabcolsep}{3.5pt}%
\begin{tabular}{llcccccccccc}
\toprule
Setting & Configuration ($s_0$) & Pred & TP & FP & Precision & Recall & F1 & Skel.\ SHD & Coll.\ F1 & CPDAG SHD & Time (ms) \\
\midrule
Baseline & Standard PC ($\infty$) & $30$ & $14$ & $16$ & $0.467$ & $0.933$ & $0.622$ & $17$ & $0.167$ & $23$ & $1058$ \\
Baseline & PC-stable ($\alpha=0.005$) & $26$ & $14$ & $12$ & $0.538$ & $0.933$ & $0.683$ & $13$ & $\mathbf{0.231}$ & $19$ & $789$ \\
$s_0 > 3$ & Covariance-corrected PC ($5.0$) & $17$ & $9$ & $8$ & $0.529$ & $0.600$ & $0.562$ & $14$ & $0.071$ & $18$ & $272$ \\
$s_0 \le 3$ & Covariance-corrected PC ($3.0$) & $21$ & $12$ & $9$ & $0.571$ & $0.800$ & $0.667$ & $12$ & $0.080$ & $15$ & $165$ \\
\midrule
$s_0 > 3$ & Noise-Aware PC ($10.0$) & $19$ & $12$ & $7$ & $0.632$ & $0.800$ & $0.706$ & $10$ & $0.167$ & $14$ & $838$ \\
$s_0 > 3$ & Noise-Aware PC ($5.0$) & $7$ & $7$ & $0$ & $1.000$ & $0.467$ & $0.636$ & $8$ & $0.182$ & $11$ & $423$ \\
\midrule
$s_0 \le 3$ & Noise-Aware PC ($2.0$) & $4$ & $4$ & $0$ & $1.000$ & $0.267$ & $0.421$ & $11$ & $0.222$ & $12$ & $116$ \\
$s_0 \le 3$ & Noise-Aware PC ($1.0$) & $1$ & $1$ & $0$ & $1.000$ & $0.067$ & $0.125$ & $14$ & $0.000$ & $14$ & $44$ \\
\bottomrule
\end{tabular}%
}
\end{table}

Table~\ref{tab:dream4_results} highlights the sharp recall cost in a single nonlinear realization. At $s_0 = 2.0$, NA-PC predicts $4$ reference edges and $0$ outside-reference edges, missing $11$ reference edges; its skeleton F1 ($0.421$) and collider F1 ($0.222$) are both lower than the low-$\alpha$ PC-stable ($\alpha = 0.005$) baseline ($0.683$ and $0.231$, respectively), while $s_0 = 1.0$ retains only $1$ of $15$ edges.

\subsection{Single-Realization Synthetic Atmospheric Sensor Simulation}
As a second single-realization Tier~2 synthetic example, we generate $n = 2{,}000$ observations across $8$ coupled atmospheric variables (Temperature, Humidity, Dew Point, Pressure, Wind Speed, CO, $\mathrm{NO}_2$, and $\mathrm{PM}_{2.5}$) from deterministic thermodynamic and dispersion equations with $9$ reference physical associations and independent coordinate noise $\sigma_\epsilon = 0.35\,\sigma_{\mathrm{clean}}$ (coordinate $\SNR = 1/0.35^2 \approx 8.16$). Sensitivity inputs $s_0$ do not certify multivariate directional SNRs across correlated channels or certify latent regression classes.

\begin{figure}[!htbp]
\centering
\includegraphics[width=0.90\textwidth]{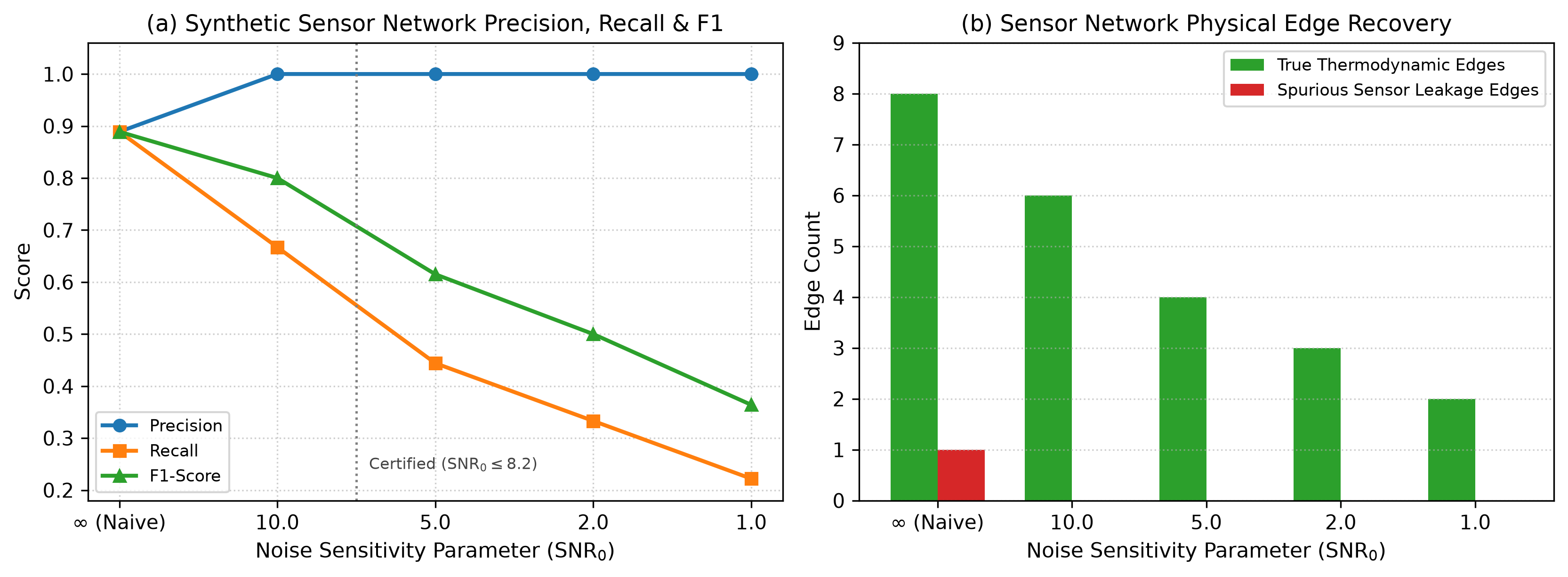}
\caption{One synthetic eight-channel atmospheric example ($n=2{,}000$, nine reference associations, coordinate $\SNR \approx 8.16$).}
\label{fig:climate_bench}
\end{figure}

\begin{table}[!htbp]
\centering
\caption{One synthetic eight-channel atmospheric example with nine reference associations ($n=2{,}000$). The noise SD is $0.35$ times each clean marginal SD (coordinate $\SNR \approx 8.16$). Sensitivity inputs do not certify multivariate directional SNRs or latent regression classes.}
\label{tab:climate_results}
{\footnotesize\setlength{\tabcolsep}{4.5pt}%
\begin{tabular}{llcccccccc}
\toprule
Setting & Configuration & $s_0$ & Pred & TP & FP & Precision & Recall & F1 & Skel.\ SHD \\
\midrule
Baseline & Standard PC & $\infty$ & 9 & 8 & 1 & 0.889 & 0.889 & 0.889 & 2 \\
$s_0 > 8.16$ & Noise-Aware PC & 10.0 & 6 & 6 & 0 & 1.000 & 0.667 & 0.800 & 3 \\
\midrule
$s_0 \le 8.16$ & Noise-Aware PC & 5.0 & 4 & 4 & 0 & 1.000 & 0.444 & 0.615 & 5 \\
$s_0 \le 8.16$ & Noise-Aware PC & 2.0 & 3 & 3 & 0 & 1.000 & 0.333 & 0.500 & 6 \\
$s_0 \le 8.16$ & Noise-Aware PC & 1.0 & 2 & 2 & 0 & 1.000 & 0.222 & 0.364 & 7 \\
\bottomrule
\end{tabular}%
}
\end{table}

In Table~\ref{tab:climate_results}, Standard PC achieves the highest displayed skeleton F1 ($0.889$) and lowest skeleton SHD ($2$). At $s_0 = 5.0$, the sensitivity filter retains $4$ reference edges and $0$ edges outside the reference (SHD $5$). We report agreement with a simulated association reference rather than field validation, causal direction recovery, or attribution of individual disagreements to sensor noise.

\subsection{Descriptive Pairwise Marginal Screening and Sensitivity Sweep on Sachs Flow Cytometry}
\begin{figure}[!htbp]
\centering
\includegraphics[width=0.90\textwidth]{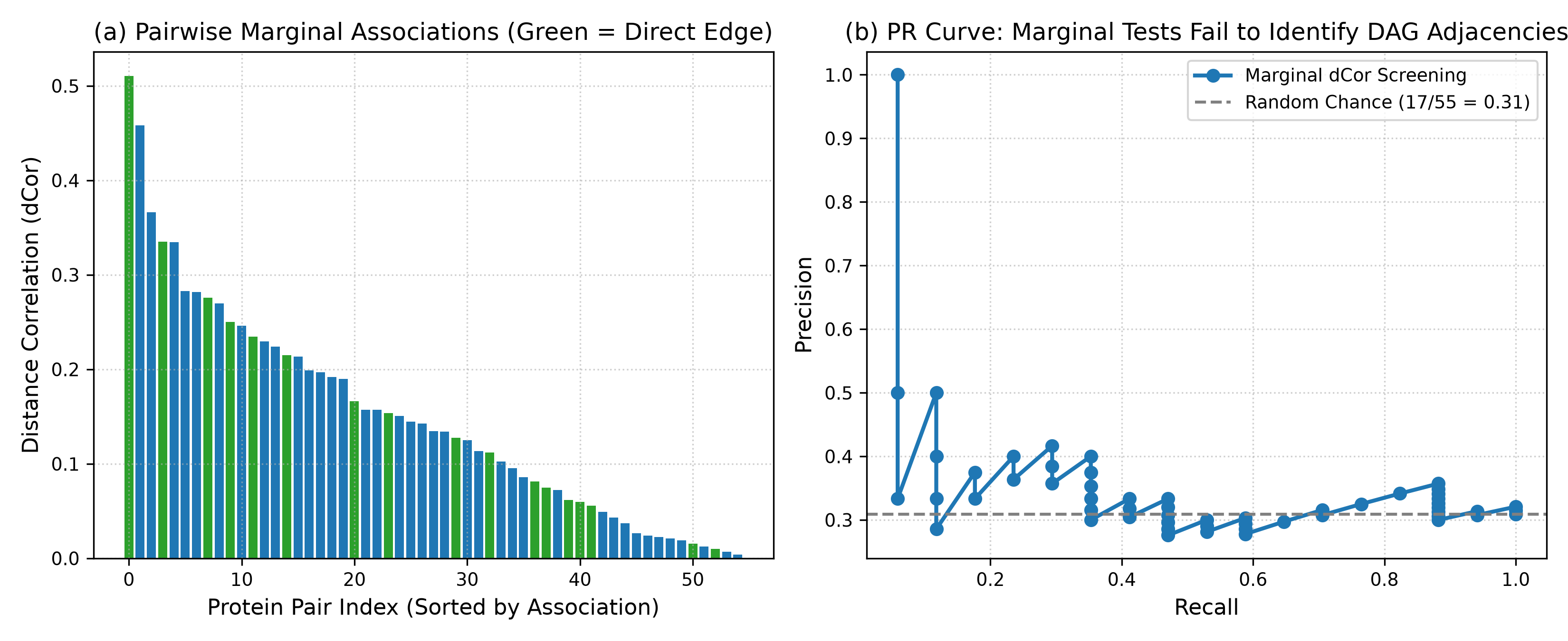}
\caption{Pairwise marginal association screening across the 55 signaling pairs in the Sachs cytometry dataset relative to the 17-edge biological reference network (marginal association is descriptive and is not a conditional edge test).}
\label{fig:sachs_screening}
\end{figure}

\begin{figure}[!htbp]
\centering
\includegraphics[width=0.90\textwidth]{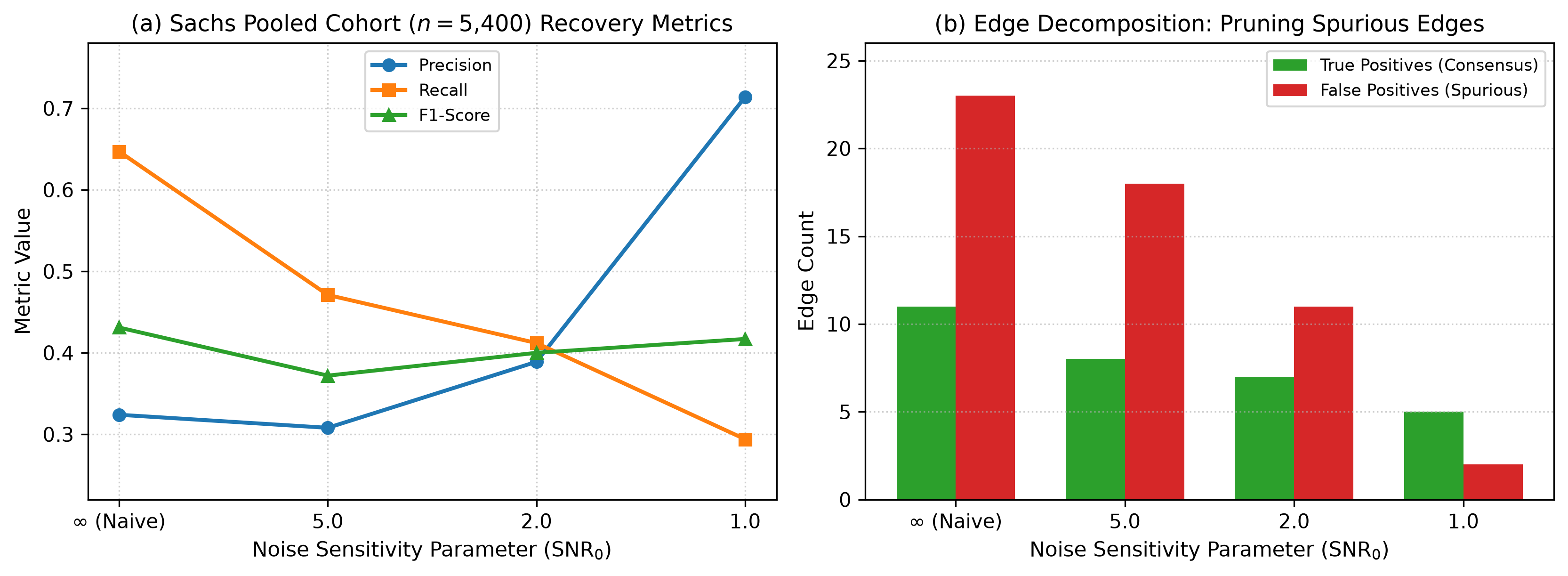}
\caption{Pooled Sachs sensitivity sweep ($n=5{,}400$). The plot shows changes in retained edges as the user-supplied threshold $s_0$ varies; it does not identify measurement error or eliminate intervention heterogeneity.}
\label{fig:sachs_pc}
\end{figure}

Figure~\ref{fig:sachs_screening} displays descriptive pairwise marginal association screening across the $55$ signaling pairs in the Sachs cytometry dataset relative to the $17$-edge biological reference network discussed in Section~\ref{sec:sachs_sensitivity}. In addition, Figure~\ref{fig:sachs_pc} illustrates the changes in retained reference and non-reference edges across the user-supplied threshold $s_0$ for the pooled cohort ($n = 5{,}400$) documented in Table~\ref{tab:sachs_results}.

\section{Marginal Multiplicity Simulation Under Gaussian Measurement Noise}
We evaluate empirical FWER and FDR across dimensions $p \in \{5, 10, 15, 20, 25, 30\}$ ($m = \binom{p}{2} \in [10, 435]$ pairwise marginal hypotheses, $n = 100$ observations per dataset, $200$ independent simulated datasets per dimension, seed $808$, target $\alpha = 0.05$). In each dataset, $n_{\mathrm{active}} \triangleq \max(2, \lfloor 0.3 p \rfloor)$ latent Gaussian variables share a common standard normal factor with loading $0.65$ (so even at $p = 5$, two variables are active and $m_0 = 9 < 10$), while all remaining pairs are independent Gaussian nulls, corrupted by independent Gaussian noise at generating coordinate $\SNR = 2.0$. Because the marginal Gaussian null pairs are tested via the exact two-sided Pearson Student-$t$ $p$-value ($df = n - 2$) rather than random permutations, the permutation-grid resolution constraint $B + 1 \ge \lceil m/\alpha \rceil$ does not apply to this table. (For random-permutation marginal $p$-values, a Bonferroni cutoff of $\alpha/m$ is reachable only if $B + 1 \ge \lceil m/\alpha \rceil$.)

\begin{figure}[!htbp]
\centering
\includegraphics[width=0.65\textwidth]{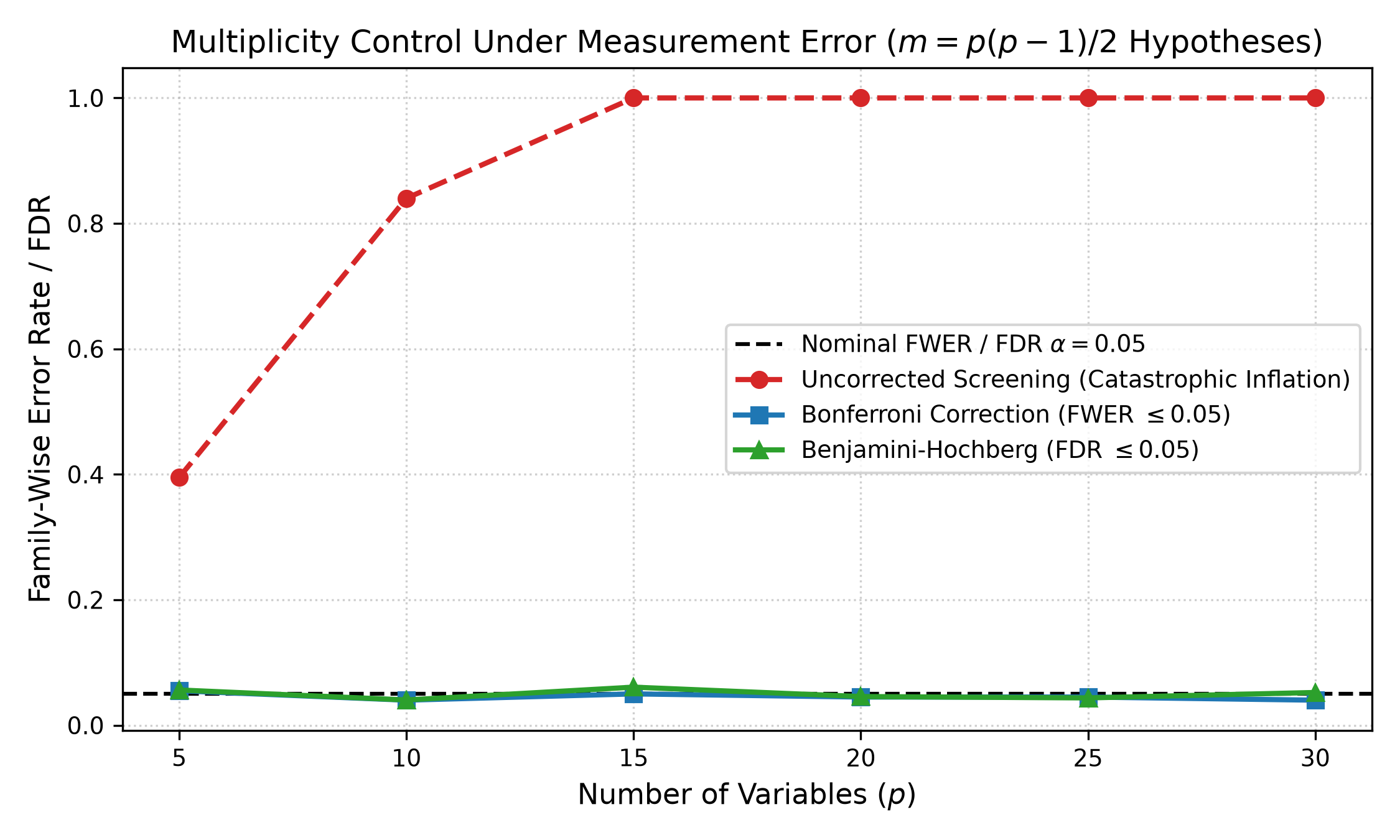}
\caption{Estimated FWER and FDR in a marginal Gaussian screening simulation ($n=100$, $200$ datasets per dimension, generating coordinate $\SNR=2.0$, exact Student-$t$ $p$-values, target level $0.05$).}
\label{fig:multiple_testing}
\end{figure}

\begin{table}[!htbp]
\centering
\caption{Estimated FWER and FDR in a marginal screening simulation ($n=100$ observations per dataset, $200$ datasets per dimension, generating coordinate $\SNR=2.0$, exact Student-$t$ $p$-values, target level $0.05$). Empirical estimates may exceed $0.05$ due to Monte Carlo variability.}
\label{tab:mult_testing}
{\footnotesize\setlength{\tabcolsep}{6pt}%
\begin{tabular}{ccccc}
\toprule
Dim $p$ & Hyp $m$ ($m_0$) & Raw FWER & Bonferroni FWER (SE) & BH FDR (SE) \\
\midrule
$5$ & $10$ ($9$) & $0.395\,(0.035)$ & $0.055\,(0.016)$ & $0.056\,(0.016)$ \\
$10$ & $45$ ($42$) & $0.840\,(0.026)$ & $0.040\,(0.014)$ & $0.040\,(0.014)$ \\
$15$ & $105$ ($99$) & $1.000\,(0.000)$ & $0.050\,(0.015)$ & $0.061\,(0.017)$ \\
$20$ & $190$ ($175$) & $1.000\,(0.000)$ & $0.045\,(0.015)$ & $0.046\,(0.015)$ \\
$25$ & $300$ ($279$) & $1.000\,(0.000)$ & $0.045\,(0.015)$ & $0.044\,(0.014)$ \\
$30$ & $435$ ($399$) & $1.000\,(0.000)$ & $0.040\,(0.014)$ & $0.052\,(0.016)$ \\
\bottomrule
\end{tabular}%
}
\end{table}

In Table~\ref{tab:mult_testing}, FWER is the empirical mean of $\mathbf{1}\{V > 0\}$ and FDR is the empirical mean of $V / \max(R, 1)$ across the $200$ simulated families (where $V$ and $R$ denote false and total rejections per family). Unadjusted screening exhibits severe FWER inflation ($0.395$ at $p=5$, $0.840$ at $p=10$, and $200/200$ at $p \ge 15$). Bonferroni estimates range from $0.040$ to $0.055$ and BH estimates range from $0.040$ to $0.061$; these are empirical estimates under the stated factor model, not a verification of PRDS or a general proof of BH control across arbitrary dependence structures.

% =========================================================================

\section{Synthetic Bivariate Experiments Inspired by the T\"{u}bingen Cause--Effect Benchmark}
\label{app:tuebingen}
We evaluate marginal association and classical correlation attenuation on six synthetic bivariate data-generating mechanisms inspired by physical and socioeconomic examples in the T\"{u}bingen Cause--Effect database \cite{mooij2016distinguishing} ($n = 500$ observations per pair, seed $709$). No original T\"{u}bingen observations are used, arrow labels in Table~\ref{tab:tuebingen_results} specify the simulation design rather than inferred causal directions, and this experiment evaluates marginal dependence preservation and linear correlation attenuation rather than causal direction identification. Each standardized synthetic pair $(X^*, Y^*)$ is corrupted by independent Gaussian noise $\epsilon_X, \epsilon_Y \sim \mathcal{N}(0, 0.1)$ (generating coordinate $\SNR_X = \SNR_Y = 10.0$). Under classical additive error, the population identity is $\rho^* = \rho_{\mathrm{obs}} / \sqrt{\lambda_X \lambda_Y}$ with $\lambda_j = \Var(V_j^*)/\Var(V_j)$; in Table~\ref{tab:tuebingen_results}, $r^*$ is the clean-sample Pearson correlation and $\hat{r}^* \triangleq r_{\mathrm{obs}} / \sqrt{\hat{\lambda}_X \hat{\lambda}_Y}$ uses the oracle realized-noise reliability plug-ins $\hat{\lambda}_X = (\widehat{\Var}(X) - \widehat{\Var}(\epsilon_X))/\widehat{\Var}(X)$ and $\hat{\lambda}_Y = (\widehat{\Var}(Y) - \widehat{\Var}(\epsilon_Y))/\widehat{\Var}(Y)$ (which is a finite-sample estimate, not exact recovery, and can exceed $[-1, 1]$ in finite samples).

\begin{figure}[!htbp]
\centering
\includegraphics[width=0.88\textwidth]{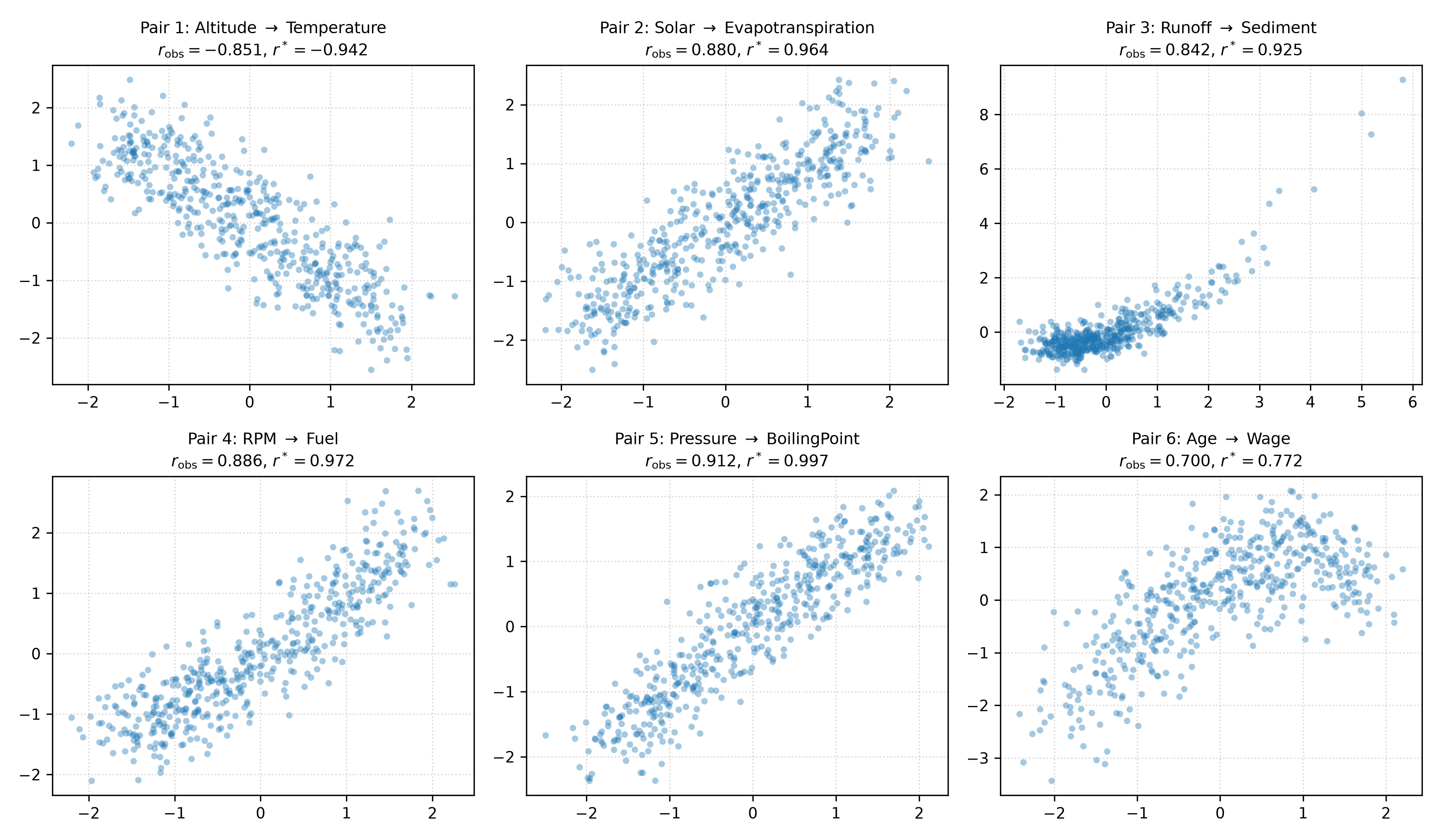}
\caption{Marginal association in six synthetic examples inspired by cause--effect benchmarks ($n=500$ per pair, generating coordinate $\SNR_X = \SNR_Y = 10.0$).}
\label{fig:tuebingen_bench}
\end{figure}

\begin{table}[!htbp]
\centering
\caption{Marginal association and oracle-reliability Pearson adjustment on six synthetic examples inspired by cause--effect benchmarks ($n=500$ per pair, $B=999$ random permutations for $\widehat{\dCor}$). Arrow labels specify the simulation design, not inferred directions.}
\label{tab:tuebingen_results}
{\footnotesize\setlength{\tabcolsep}{6pt}%
\begin{tabular}{clccccc}
\toprule
ID & Synthetic Design & $r^*$ (Clean Sample) & $r_{\mathrm{obs}}$ & $\hat{r}^*$ (Oracle Plug-in) & $\widehat{\dCor}$ & $p$-val \\
\midrule
1 & Altitude $\to$ Temperature & $-0.946$ & $-0.851$ & $-0.942$ & $0.708$ & $0.001^*$ \\
2 & Solar $\to$ Evapotranspiration & $+0.968$ & $+0.880$ & $+0.964$ & $0.763$ & $0.001^*$ \\
3 & Runoff $\to$ Sediment Load & $+0.930$ & $+0.842$ & $+0.925$ & $0.632$ & $0.001^*$ \\
4 & Engine RPM $\to$ Fuel Rate & $+0.975$ & $+0.886$ & $+0.972$ & $0.785$ & $0.001^*$ \\
5 & Pressure $\to$ Boiling Point & $+0.999$ & $+0.912$ & $+0.997$ & $0.818$ & $0.001^*$ \\
6 & Age $\to$ Wage Rate & $+0.775$ & $+0.700$ & $+0.772$ & $0.504$ & $0.001^*$ \\
\bottomrule
\end{tabular}}
\par\smallskip
{\footnotesize\noindent\textit{Note:} $^*$With $B=999$ random permutations, $p = 0.001$ indicates that no sampled permutation statistic matched or exceeded the observed $\widehat{\dCor}$ value, not that the exact tail probability equals $0.001$.\par}
\end{table}

\end{document}